\documentclass{article}
\usepackage[utf8]{inputenc}

\title{Screening Methods for Classification Based on Non-parametric Bayesian Tests}
\author{Naveed Merchant and Jeffrey D.~Hart}
\date{January 2023}
\usepackage{wrapfig}
\usepackage{amsthm}
\usepackage{amsmath}
\usepackage{ulem}
\usepackage{url}
\usepackage[colorlinks,citecolor=blue,urlcolor=blue,filecolor=blue,backref=page]{hyperref}
\usepackage{graphicx}
\usepackage{ulem}
\usepackage{amsthm}
\usepackage[colorlinks,citecolor=blue,urlcolor=blue,filecolor=blue,backref=page]{hyperref}
\usepackage{graphicx}

\usepackage{color}
\usepackage{graphicx}
\usepackage{verbatim}
\usepackage{amsmath}
\usepackage{caption}
\usepackage{subfig}
\usepackage{enumitem}
\usepackage{float}
\usepackage{amsmath}
\usepackage{amssymb}
\usepackage{setspace}
\usepackage[a4paper,
            bindingoffset=0.2in,
            left=1in,
            right=1in,
            top=1in,
            bottom=1in,
            footskip=.25in]{geometry}
\def\beqn{\begin{eqnarray*}}
\def\eeqn{\end{eqnarray*}}
\def\beq{\begin{eqnarray}}
\def\eeq{\end{eqnarray}}

\def\bi{\begin{itemize}}
\def\ei{\end{itemize}}

\def\X{\mbox{\boldmath $X$}}

\def\Y{\mbox{\boldmath $Y$}}

\def\x{\mbox{\boldmath $x$}}

\definecolor{mygreen}{rgb}{0.1,0.6,0.0}

\definecolor{myblue}{cmyk}{0.95,0.5,0,0.3}
\definecolor{mygreen}{cmyk}{1,0,0,0.74}
\definecolor{mygreen2}{rgb}{0.1,0.6,0.0}
\definecolor{myred}{rgb}{0.9,0,0.2}
\definecolor{purp}{rgb}{0.35,0,0.65}
\def\tcb#1{\textcolor{blue}{#1}}
\def\tcr#1{\textcolor{myred}{#1}}

\newcommand\myeq{\mathrel{\stackrel{\makebox[0pt]{\mbox{\normalfont\tiny def}}}{=}}}

\newtheorem{theorem}{Theorem}
\begin{document}

\maketitle
\noindent
{\bf Abstract:} Feature or variable selection is a problem inherent to large data sets. While many methods have been proposed to deal with this problem, some can scale poorly with the number of predictors in a data set. Screening methods scale linearly with the number of predictors by checking each predictor one at a time, and are a tool used to decrease the number of variables to consider 
%look through 
before further analysis or variable selection. For classification, there is a variety of techniques. There are parametric based screening tests, such as 
%like 
$t$-test or SIS based screening, and non-parametric based screening tests, such as Kolmogorov distance based screening \cite{mai2012kolmogorov}, and MV-SIS \cite{cui2015model}. We propose a method for variable screening 
%for classification 
that uses Bayesian-motivated tests, compare it to SIS based screening, and provide example applications of the method on simulated and real data. It is shown that our screening method can lead to improvements in classification rate. This is so even when our method is used in conjunction with a classifier, such as DART, which is designed to select a sparse subset of variables. Finally, we propose a classifier based on kernel density estimates that in some cases can produce dramatic improvements in classification rates relative to DART. 


%\bigskip\noindent
%\tcr{\it A thought comes to mind. Ultimately, a referee may ask for examples where the variables are dependent, but for now I don't think you need to address this.}
%\tcb{\it{We can comment on this more in the conclusion}}

\section{Introduction}

Classification involves predicting a class label for an observation, given a set of predictor variables. The techniques for doing so are wide ranging, including support vector machines \cite{boser1992training}, tree based methods \cite{brieman1984classification}, Bayesian trees \cite{linero2018bayesian} and gradient boosting trees \cite{friedman2002stochastic}.

It is common to encounter a data set with many features, but rarely are all of them important. Picking a subset of these features quickly is a task that is desired, but can be tricky for very large data sets. Removing unimportant variables can result in dramatic improvement for some of the previously mentioned classification methods. Feature selection is not a new field, and can be divided into three categories: screening or filter based methods, wrapper methods, and embedded methods \cite{tang2014feature}. Screening methods examine each variable one at a time to see if it provides useful information, and as a result scale linearly with the number of predictors. However, examining each variable individually can cause information on joint behavior of variables to be lost, or can cause collinear variables to be selected \cite{tang2014feature}. Wrapper methods fit different models, and then evaluate each one according to some criterion. The model that does best according to this criterion is selected. While this tends to select good variable subsets, fitting every model can be extremely time-consuming, especially if the data set is very large \cite{tang2014feature}. Embedding methods produce a model that has some sparsity built into it, producing a set of useful variables and a model built with them, simultaneously. The speed of different embedding methods varies with the strategy used to obtain sparsity, but these are typically slower than screening methods \cite{tang2014feature}. 

Our focus in this paper will be on screening methods. It is a common strategy to employ a screening method and then employ an embedding method (such as LASSO) afterwards. Fan \cite{fan2008sure} employs this strategy and improves both the time it takes to run LASSO and the accuracy of the model as a whole. Since then several filter methods have popped up. For classification in particular, maximum marginal likelihood screening \cite{fan2010sure}, MV-SIS \cite{cui2015model}, and Kolmogorov distance screening \cite{mai2012kolmogorov} are some of the screening tests that have been published. Most of these methods are applied to linear discriminant analysis and show improvement in applying these methods to a data set after the screening has been performed. This paper proposes a new screening method when the number of classes is known to be two, and show that it results in improved classification accuracy in settings where the simple model underlying linear discriminant analysis does not hold. Our screening method identifies informative features by using a two-sample Bayesian test that checks whether two data sets share the same distribution\cite{leaveoneout}. 
%\tcr{\it Explain to me what you mean by previous sentence. Also, in the introduction you should have at least a brief description of the type of methodology you're proposing.}
One of our goals is to show that classification methods, including BART \cite{linero2018bayesian}, DART \cite{linero2018bayesian} and SVM \cite{boser1992training}, can be improved when preceded by our screening procedure.

\section{Methodology}

Our screening method is based on computing a statistic for each individual feature. We will use kernel density estimates of the two distributions corresponding to the two classes. The idea is similar to that of Kolmogorov distance screening: if it seems likely that the two classes have different distributions for a feature, then we will keep the feature. We define a test statistic that can make this determination. 

%\medskip\noindent
%\tcr{\it I feel like the notation below is confusing. Why not just use $X_{i1},\ldots,X_{ip}$, $i=1,\ldots,n$ and $Y_{j1},\ldots,X_{jp}$, $j=1,\ldots,m$? At least I think you should change as indicated below.}
%\medskip

Suppose we observe the data $\X$, %and $\Y$,
an $(m+n)\times p$ matrix whose $i$th row, $(X_{i1},\ldots,X_{ip})$, contains the values of the $p$ variables for one subject.  The $i$th element, $Y_i$, of vector  $\Y$ is 0 or 1 and indicates the class to which the $i$th subject belongs, $i=1,\ldots,m+n$.
\begin{comment}
$(X^{1:p},Y)_{1},...,(X^{1:p},Y)_{n + m} $, where $Y \in \{0,1\}$, and $X^{1:p}$ denotes a $p$-dimensional variable. The variable $Y$ determines which class the data come from.
\end{comment}
For now, suppose that $Y_1,\ldots,Y_n$ are $0$ and $Y_{n+1},\ldots,Y_{n+m}$ are $1$.  We have $n + m$ data vectors and $p$ features. Consider the data $\X_j=(X_{1j},\ldots,X_{(m+n)j})$ for feature $j$ and define $\hat{h}_{i}(\cdot | \X_j,b)$ to be a kernel density estimate that has bandwidth $b$ and uses all the data in $\X_j$ except that of the $i$th subject:
$$ \hat{h}_{i}(x | \X_j, b) = \frac{1}{nb} \sum_{\substack{r=1 \\ r\neq i}}
^{n + m}  K\left(\frac{x - X_{rj}}{b} \right) .$$
We also compute kernel estimates from the data sets consisting of observations where $Y = 0$ and $Y = 1$. These are 
%Consider $\hat{g}_{i}(\cdot | X^j)$ and %\hat{h}_{i}(\cdot | X^j)$:
%
$$\hat{f}_{i}(\cdot | \X_j, b) = \frac{1}{nb} \sum_{\substack{r=1 \\ r\neq i}}^n K\left(\frac{x - X_{rj}}{b}\right)$$
and
$$\hat{g}_{i}(\cdot | \X_j, b) = \frac{1}{nb} \sum_{\substack{r=n+1 \\ r\neq i}}^{n+m} K\left(\frac{x - X_{rj}}{b}\right).$$

Then we define the test statistics $ALB_j$, $j=1,\ldots,p$:
\begin{eqnarray*}
(m+n)ALB_j &=&  \sum_{i = n + 1}^{m} \text{log}\left(\frac{\hat{g}_{i}(X_{ij} | \X_j, b)}{\hat{h}_{i}(X_{ij} | \X_j, b)}\right) + \sum_{i = 1}^n \text{log}\left(\frac{\hat{f}_{i}(X_{ij} | \X_j, b)}{\hat{h}_{i}(X_{ij} | \X_j, b)}\right) \\
&\myeq&\sum_{i = n+1}^{m} \text{log}(B_{1ji})+\sum_{i = 1}^{n} \text{log}(B_{2ji}).
\end{eqnarray*}
The notation $ALB$ stands for {\it Average Log-Bayes factor}, since, as argued by \cite{leaveoneout}, the statistics $B_{1jr}$ and $B_{2js}$ are Bayes factors based on the data ``sets" $X_{rj}$ and $X_{sj}$, respectively. We refer to \cite{leaveoneout} for a thorough exploration of such average log-Bayes factors.

Another motivation for $ALB_j$ is in terms of Kullback-Leibler divergence, which for densities $h_1$ and $h_2$ is 
$$
KL(h_1,h_2)=\int_{-\infty}^\infty h_1(x)\log\left(\frac{h_1(x)}{h_2(x)}\right)\,dx.
$$
Define $f_{j,{\rm mix}}=(nf_j+mg_j)/(n+m)$, where $f_j$ and $g_j$ are the densities of feature $j$ for classes 0 and 1, respectively.
Then the statistic $ALB_j$ is an approximately unbiased estimator of the following quantity: 
$$\left(\frac{n}{m + n}\right)KL\left(f_j,f_{j,{\rm mix}}\right) + \left(\frac{m}{m + n}\right)KL\left(g_j, f_{j,{\rm mix}}\right).
$$
%\tcr{\it Ends up being odd that $\hat g$ and $\hat h$ estimate $f_j$ and $g_j$, respectively. How about changing $\hat f$, $\hat g$ and $\hat h$ to $\hat h$, $\hat f$ and $\hat g$, respectively?} 
%\tcb{\it{I changed some of the labels to reflect this}} 
In the case $f \equiv g$, $ALB_j$ converges to 0 in probability as $m$ and $n$ tend to $\infty$.  The null distribution of $ALB_j$ can be assessed using a permutation-based procedure to determine if two sets of observations arise from a common distribution.  

\cite{leaveoneout} propose selecting the bandwidth $b$ by leave-one-out cross validation, a proposal that tacitly assumes the null hypothesis to be true. While this strategy has potential, we believe it to be too computationally expensive to use for every variable separately.
%using this procedure. 
We only have to do this $p$ times, suggesting linear scaling with the variable length, but this introduces quadratic scaling with $n$, which is prohibitive for large data sets. Instead, we opt to use a normal plug-in bandwidth in conjunction with the heavy tailed Hall kernel, which is:
$$ K_0(z) = \frac{1}{\sqrt{8 \pi e}\,\Phi(1)}\exp\left[-\frac{1}{2}(\log(1+|z|))^2\right]. $$
Simulation results show that the constant for the plug-in is $0.162$ to 3 decimal places, resulting in the following plug-in rule for variable $j$:
$$ b_{\text{plug-in},j}(\X_j) = 0.162 (m+n)^{-1/5} s_j,$$
where $s_j$ is an estimate of the underlying (null) standard deviation.
%Here $s$ is a scale. 
One possibility is to take $s_j$ to be the sample standard deviation for $\X_j$, but we prefer the more robust choice $s_{Rj}=IQR(\X_j)/1.35$. 
\begin{comment}
For normal data, we use the standard deviation as a default for $s$. For heavier tailed data where the standard deviation is unstable, the rule of $\frac{IQR(X_1,X_2,...,X_n)}{1.35}$ can be used instead.
\end{comment}

Suppose we have computed $ALB_1, ALB_2, ALB_3, ..., ALB_p$. 
%\tcr{\it Decide if you want ALB to be italicized or not. A journal would say that when it is a statistic (as in previous sentence) use italics, but if you are just referring to the ALB procedure, use Roman.} \tcb{I'll go ahead and use it as italics if refering to statistic and in roman if referring to procedure. I think math mode in latex uses italics so I'll just use that.} 
The matter still remains in choosing a cutoff for the $ALB$s such that we select all variables with $ALB$ larger than the cutoff. Below are some possible ways of doing so. 

\begin{itemize}
    \item [A1.] Choose the cutoff to be some percentile of $ALB_1, ALB_2, ALB_3, ..., ALB_p$. This is in line with what some of the authors in SIS and SIRS propose. Suppose we expect $d$ features to be important in the data set, then we can set our cutoff to be the $100(1-d/p)$th percentile. Clearly, the largest $d$ test statistics are the most likely to be significant. A problem with this approach is selecting an appropriate value for $d$. The authors in SIS \cite{fan2008sure} and SIRS \cite{zhu2011model} argue that a conservative choice for $d$ is $n$ or $n\log(n)$, although these choices seem somewhat arbitrary.
    \item [A2.] An empirical but computationally daunting way to approach this problem is to proceed with two training sets and a classification method. We can choose the cutoff that minimizes the error rate of the classification method when it is trained on one of the training sets and then applied to the other. To keep the computational scaling of the procedure linear with $p$, we recommend restricting the number of candidate cutoff values to be fairly small, say no more than ten.
    \item [A3.] Randomly select a covariate, say $\X_j$. For this covariate, permute the labels, and compute the test statistic, call it $ALB_1^*$. Using the same feature $\X_j$, repeat this procedure $d$ times, resulting in  $ALB_{1}^*,\ldots,ALB_{d}^*$. We randomly select $B-1$ more covariates without replacement, and repeat the previous steps for each of them, resulting in a total of $Bd$ values of $ALB$.
    %$ALB* = (ALB^1*,  ... , ALB^{Bd}*)$.
    Once this is done, we choose the cutoff to be a percentile of these $Bd$ values. This method also approximates the null conditional distribution of $ALB$ for a randomly selected feature, but potentially has the advantage of requiring fewer  statistics to be computed than in A3.
    \item [A4.] The Bayes factor interpretation of $ALB_j$ entails that variable $j$ should never  survive screening when $ALB_j<0$, suggesting that we use 0 as a cutoff.
    %Picking 0 as a }cutoff corresponds to using $T = 1$ in A5.  
    %and is a decent idea if the sample sizes are large. 
    While this choice may seem liberal, \cite{leaveoneout} show that an $ALB$  cutoff of 0 produces a test whose type I error probability %corresponds to a test where the alpha value 
    tends to 0 as $m + n$ tends to $\infty$.
\end{itemize}
\begin{comment}
We note the following in regards to each procedure. Simulating i.i.d. random normal variables and performing a permutation based procedure both result in ALBs that offer insight on how ALBs behave in the case where the data from classes do not differ in distribution. Simulating normal random variables is a smaller computational burden than repeatedly permuting the labels, but the distributions sampled from permutation will often be closer to those of the observed features, and hence more relevant.
\end{comment}

It should be noted that a particular screening method 
%an empirical cutoff made in conjunction with a classification method 
will work differently with different classifiers, although this is perhaps to be expected. We encourage the use of a classifier that can take advantage of differences in distribution other than location differences.  A popular classifier is linear discriminant analysis (LDA), by which we mean the version that assumes equal covariance matrices for the two classes. Features identified as important due to a scale difference between classes will usually be of no use to LDA.
%as our screening in this case should be a worse variation of the SIS procedure. \tcr{\it It should be worse than SIS? Then why use it?} \tcb{\it{The intention is to say if you use ALB screening with something like linear discriminant analysis you'll probably have a worse classifier than if t-test screening is used. It can do better but it requires a classifier that leverages differences other than location differences.}}

Of the methods A1-A4, the least computationally expensive ones are A1 and A4. 
%involve (a) choosing a cutoff based on the interpretability of ALB, and (b) choosing the cutoff to be one of the top percentiles of the $ALB$s. As previously discussed, two interpretable ALB cutoffs are 0 and one resulting from choosing $T=2$. 
Use of these cutoffs also has the advantage of producing a test with power tending to 1 as $m+n$ tends to $\infty$, since $ALB_j$ converges to 0 in probability when $f_j\equiv g_j$  \cite{leaveoneout}. Method A1 
is recommended if there is a strong idea as to how many variables are expected to be relevant or if there is a critical number of variables that are needed for another classifier to work well. 
%\tcr{\it I thought that selecting $T$ {\bf was} selecting a cutoff.  If not, then you need to define what a ``cutoff" is.}  

Lastly we wish to note that each $ALB$ has a finite upper bound, as it is easily shown that
$$ALB_j\le \log(2)\cdot\max\left(\frac{m}{(m-1)},\frac{n}{(n-1)}\right),$$
which implies that $ALB_j$ is essentially bounded by $\text{log}(2)$ so long as $m$ and $n$ are not too small. 
%An alternative strategy for choosing a cutoff is to specify $q$ in $\text{log}(2q)$, where $.5 \leq q \leq 1$. We recommend $q = .6$ in this approach. Choosing $q = .5$ results in an ALB cutoff of 0, and choosing $q = 1$ results in a cutoff of $\text{log}(2)$. The larger the $q$ value, the harsher the screening method and the less variables survive the cutoff.

\section{Consistency results}

We begin with the assumption that every variable satisfies conditions A1-A5 in \cite{leaveoneout}. We will also assume that the numbers, $m$ and $n$, of samples for the two classes tend to infinity, and the number of variables, $p$, is fixed. Suppose that a variable belongs to class $D$ if the variable marginally offers information, which means that the variable has a different distribution for one class than it does for the other. Finally, we assume that the variables are independent.
%Furthermore, the kernel
%  conditions of \cite{Hall87} are satisfied by k%ernels other than
%  $K_0$, but notably the Gaussian kernel does not %satisfy his conditions.
  
\begin{theorem}
  If the above assumptions hold, then   
  \beq\label{DistanceResult}
\lim_{n,m \rightarrow \infty} P(\max_{i \in D^c}ALB_i < \min_{j \in D}ALB_j) \rightarrow 1.
\eeq
Suppose we use a cutoff as in A4 (in our Section 3). Then we also have the following result:
\beq\label{ScreeningResult}
\lim_{n,m \rightarrow \infty} P(\min_{j \in D}ALB_j > 0 \cap \max_{j \in D^c}ALB_j < 0) \rightarrow 1.
\eeq
\end{theorem}

\noindent
{\it Proof.} \ 
Result (\ref{ScreeningResult}) implies (\ref{DistanceResult}), so we only prove the former. Using the fact that $P(A\,\cap\, B)\ge 1-P(A^c)-P(B^c)$, it is enough to show that both $P(\min_{j \in D}ALB_j > 0)$ and $P(\max_{j \in D^c}ALB_j < 0)$ tend to 1.

We only consider $P(\max_{j \in D^c}ALB_j < 0)$ as the proof for the other probability is similar. We have
\begin{eqnarray}\label{Pbound}
P(\max_{j \in D^c}ALB_j < 0)&=&P\left(\bigcap_{j\in D^c} \{ALB_j<0\}\right)\notag\\
&=&\prod_{j\in D^c} P(ALB_j<0)\notag\\
&\ge&\left(\min_{j\in D^c}P(ALB_j<0)\right)^N,
\end{eqnarray} 
where $N$ is the number of elements in $D^c$. For each $j\in D^c$, \cite{leaveoneout} show that $P(ALB_j<0)\rightarrow 1$ as $m$ and $n$ tend to $\infty$. Since $N$ is finite, this implies that the quantity on the right-hand side of (\ref{Pbound}) tends to 1, from which the result follows. $\ \blacksquare$

It is clear that if $p$ tends to $\infty$ at a sufficiently slow rate, then Theorem 1 remains true. However, determining the precise rate at which $p$ can increase relative to $m$ and $n$ requires stronger results than provided by \cite{leaveoneout}, and we will not pursue this direction further. 
%Even though we will not prove this rigorously, it stands to reason that you can let the number of variables grow at a slow enough rate relative to n and m and have the result hold in that case as well.

\begin{figure}[h!]
\centering
\includegraphics[scale=.5]{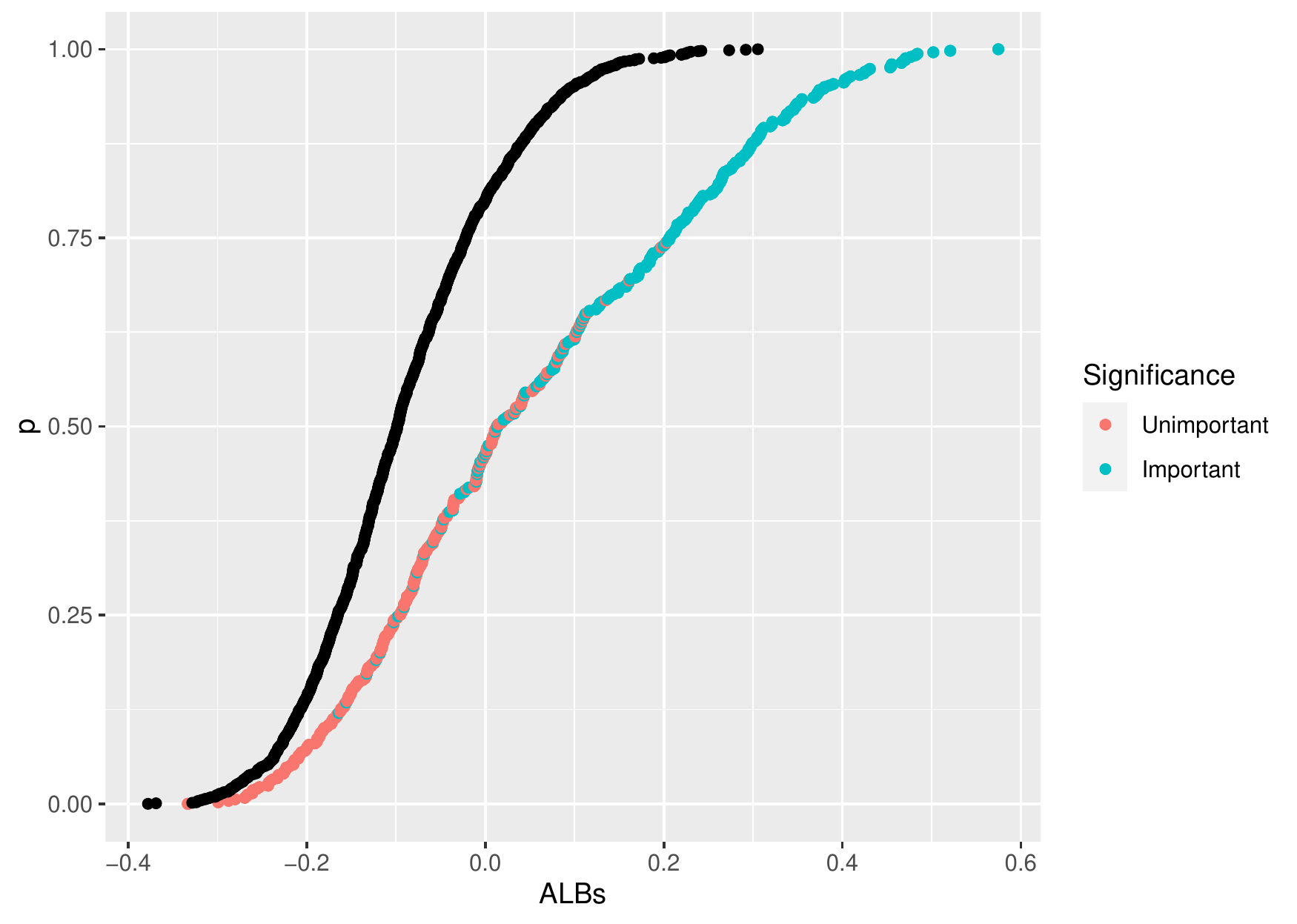}
\caption{{\it Comparison of ALB CDFs when the training set sizes are equal to 10 and important variables have a shape difference.} The black curve denotes the CDF of $ALB$s generated from data where the classes are permuted.} 
\label{fig:mn10}
\end{figure}

\begin{figure}[h!]
\centering
\includegraphics[scale=.5]{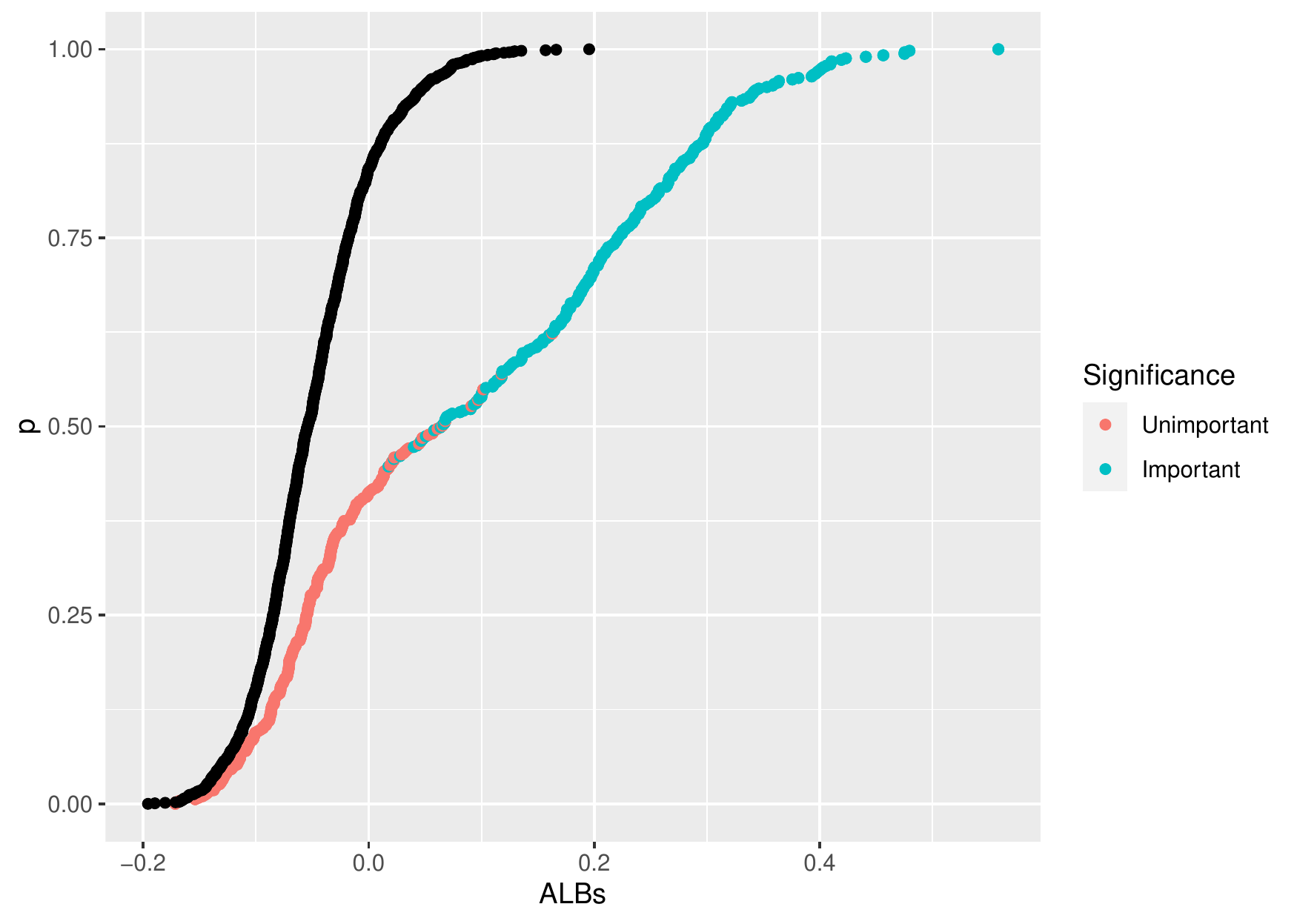}
\caption{{\it Comparison of ALB CDFs when the training set sizes are equal to 20 and important variables have a shape difference.} The black curve denotes the CDF of $ALB$s generated from data where the classes are permuted.}
\label{fig:mn20}
\end{figure}

\begin{figure}[h!]
\centering
\includegraphics[scale=.5]{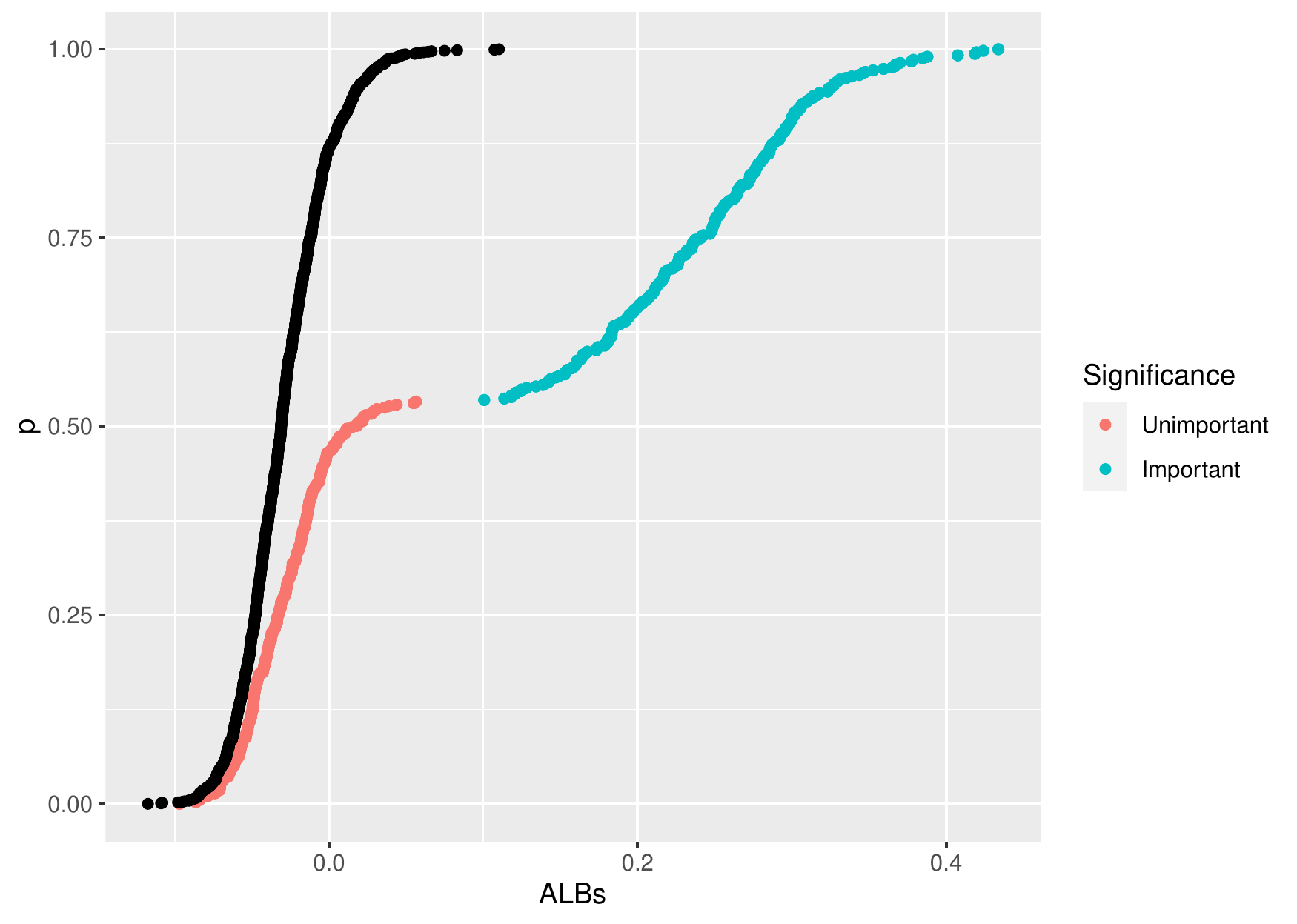}
\caption{{\it Comparison of ALB CDFs when the training set sizes are equal to 40 and important variables have a shape difference.} The black curve denotes the CDF of $ALB$s generated from data where the classes are permuted.}
\label{fig:mn40}
\end{figure}

We now show simulation results for various values of $m=n$, $p$ and $r$, where $r$ represents the proportion of important variables, i.e., variables for which the class distributions are different. 
%$n$ denotes the number of observations, and $p$ denotes the number of variables.
We generate 500 variables for a binary classification problem in the following fashion. If it is important, the variable is drawn for one class from a $t$-distribution with 4 degrees of freedom, and drawn from the other class from a mixture of two normal distributions, where the mixing parameter is 1/2, the standard deviation of both normal distributions is 1, and the means are -2.5 and 2.5. If instead the variable is unimportant, then it is drawn from a standard normal. We will name the method of generating variables in this setting ``a shape difference." Finally, each variable is determined to be important or not by performing a binomial trial with success probability $r=1/2$. Figures \ref{fig:mn10}, \ref{fig:mn20}, and \ref{fig:mn40} show how the cdfs of the $ALB$s change depending on $m$ and $n$. 

As the sample sizes $m$ and $n$ increase, the $ALB$s for important variables gradually increase. Even when the total number of observations is only eight percent of the total number of variables, we achieve the property that the largest $ALB$ of the unimportant variables is smaller than the smallest $ALB$ of the important variables. The black curve shows the cdf of $ALB$s computed by permuting the labels for each variable three times and computing the $ALB$ each time. A cutoff of 0 is not larger than the largest unimportant 
%\tcr{\it Don't you mean unimportant?} \tcb{Your correct, that was a mistake}
variable for any $n$, but is still useful for discarding a large portion of the unimportant variables.  On the other hand, using a large percentile of the permuted variables can result in discarding almost all of the unimportant variables, and at the largest sample size, choosing the cutoff to be the maximum of the permuted $ALB$s does indeed almost perfectly separate the important and unimportant variables.

\section{Discussion of classification methods}

Ideally, one should choose the screening method and classifier that work best together. 
%classification method that works best with the screening method should be tailored to work with the screening method. 
Good examples of this principle are provided by the relationship that classification methods such as logistic regression, support vector machines, and linear discriminant analysis have with $t$-test based screening. Discriminant analysis, support vector machines without a kernel trick, and logistic regression are designed to take advantage of location differences between classes. It is therefore natural to precede them with $t$-test screening, which, of course, is designed to detect differences between means.
\begin{comment}
A variable is more useful to these classifiers if there is a location difference between the classes. Linear discriminant analysis utilizes differences in means to determine its classification rule.
%on whether the class is located in different areas.
\end{comment}
On the other hand, support vector machines that use a kernel trick create a hyper-plane that best separates the two classes essentially after a transformation is performed, and can therefore deal effectively with many types of differences between distributions. To take advantage of this ability, it is thus best to use a screening method that can detect non-location differences. \begin{comment}Logistic regression seeks to choose variables that make the log odds, on average, differ. Screening via $t$-tests naturally works well with linear discriminant analysis as all variables that survive stage 1 of the screening have differences in means with high probability. \tcr{Variables with nonlocation differences can produce  
%have different means in many dimensions
clusters in multiple dimensions that require a hyper plane to be separated, thus making support vector machines valuable tools.} Finally, logistic regression that models the log odds ratio as a linear function of covariates can fail to produce a good classifier when the covariates have distributions that differ only with respect to covariance.
\end{comment}
In summary, $t$-test screening is a natural method to use when linear discriminant analysis or logistic regression are deemed to be appropriate classifiers, but is not necessarily a good method when a support vector machine with a kernel trick is required. By linear SVM, we mean a method that can classify by separating the two groups by a plane.

%For simplicity, we refer to using a support vector machine from hereforth without the kernel trick, or classification that separates based on a plane, as using "linear SVM". 
%These can do poorly if the difference does not arise from a difference in means.
%\tcr{\it We can discuss changes in the previous paragraph, but having thought about it a lot, the distinction between the methods seems to boil down to the relatively simple description above.}

\begin{figure}[h!]
\centering
\includegraphics[scale=.5]{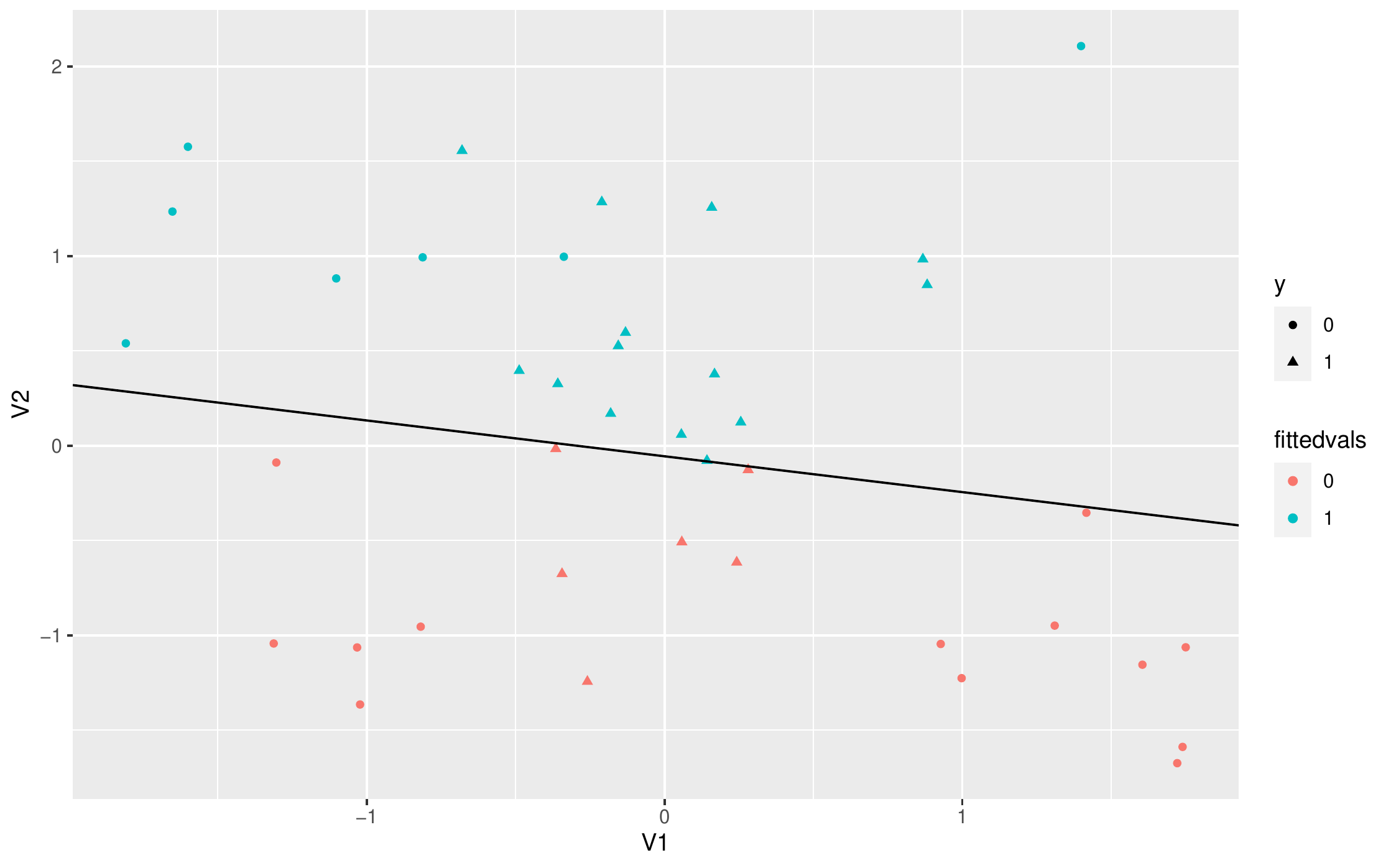}
\caption{ \textit{Prediction accuracy of two important variables in the setting of ``a shape difference" using a linear SVM.} The colors represent the predictions that the SVM produces. The line represents the discriminator that a linear SVM produces to discriminate the classes. The triangle and circle represent which class the observation arises from. }
\label{fig:normtmixlinpreds}
\end{figure}

\begin{figure}[h!]
\centering
\includegraphics[scale=.5]{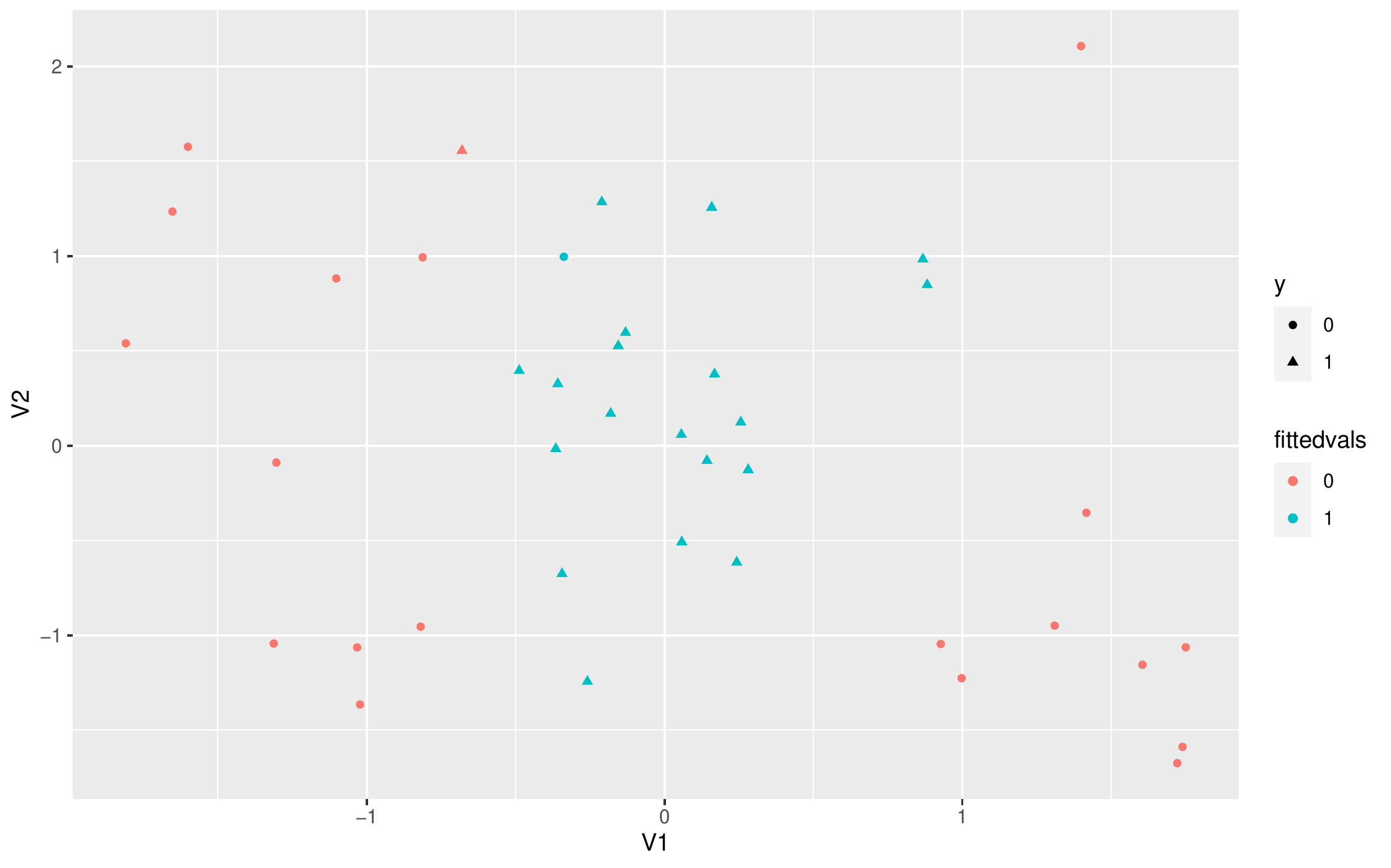}
\caption{\textit{Prediction accuracy of two important variables in the setting of ``a shape difference" with an SVM that uses a kernel trick.} The colors represent the predictions that the SVM produces when a kernel trick is applied. The triangle and circle represent which class the observation arises from. Classification is much better in this case because the trick enables the classifier to capture differences other than location shifts.}
\label{fig:normtmixkernpreds}
\end{figure}

In Figures \ref{fig:normtmixlinpreds} and \ref{fig:normtmixkernpreds} two (important) variables are generated according to a shape difference. The bimodality of one of the two class distributions makes the classes hard to separate with a plane. Figure \ref{fig:normtmixlinpreds} shows the performance of a support vector machine when only two relevant variables are used for classification, while Figure  \ref{fig:normtmixkernpreds} shows the improvement in the same situation when the kernel trick is applied to detect non-location based differences. 
%\tcr{\it (a) Is SVM the same as linear discriminant analysis when only two variables are used? The previous two sentences seem to imply that this is so.} \tcb{ \it{Its not the exact same method, but there are similar problems that occur with both of them.}}
%\tcr{\it (b) You can convey the info in Figures 4-6 more effectively with just two plots. In both plots use colors to represent the true classes.  Indicate classification with either a solid point or an x. One plot is for linear SVM and the other for SVM.}

Our method seeks to outperform $t$-test screening by considering differences other than ones of location type. Of course, this performance is not free. It comes with the cost that we lose some power in detecting differences of means. For our method to work better with a classifier, the classifier must have the ability to distinguish classes that display non-location differences. 
%\tcr{\it Do you mean the {\bf classifier} must be able to detect non-location differences?} \tcb{This is correct, I explicitly mentioned classifier} 
For example, a set of variables whose classes differ only with respect to scale would not be useful to support vector machines without the kernel trick and LDA, as there would be no hyperplane that nicely separates the classes. 
%and LDA itself assumes that there is no scale difference across classes. %\tcr{\it So, SVM can't detect scale differences?  Depending on how it's defined, discriminant analysis can detect scale differences.  So, in this sense discriminant analysis is better?  Also, you say that LDA assumes no differences across categories.  I don't know what the categories are in your context. Let's discuss.} \tcb{I replaced categories with classes, I also meant to say LDA supposes that there aren't differences in scale (they assume homogeneity across classes).} 
A kernel trick or increasing the number of variables by considering interactions and squared terms can sidestep this issue. 
%\tcr{\it I'm confused about SVM.  I thought it always used the kernel trick.  But you seem to be saying that SVM can be tricked by something as innocent as scale differences. Explain to me.} \tcb{\it{"Linear SVM" or "simple svm" can be tricked. The kernel trick replaces the distance calculation in linear svm with a difference with two kernels. My argument is along the lines that linear svm can't be used but}}
But adding variables is not ideal, as a goal of our methodology is to decrease computational complexity. 

\begin{figure}
\subfloat[Boxplots of true positives]{\includegraphics[width=.5\linewidth]{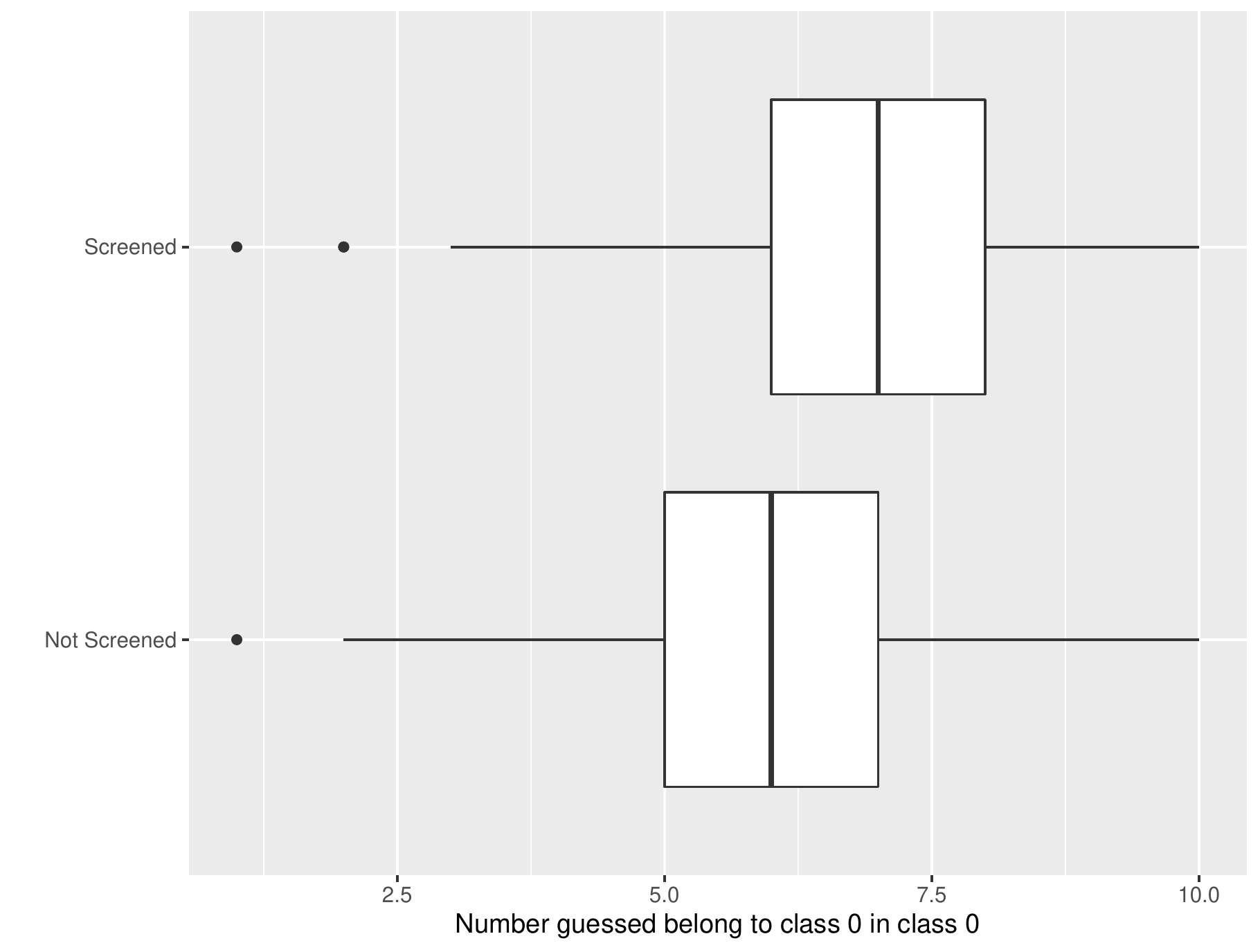}} 
\subfloat[Boxplots of false positives]{ \includegraphics[width=.5\linewidth]{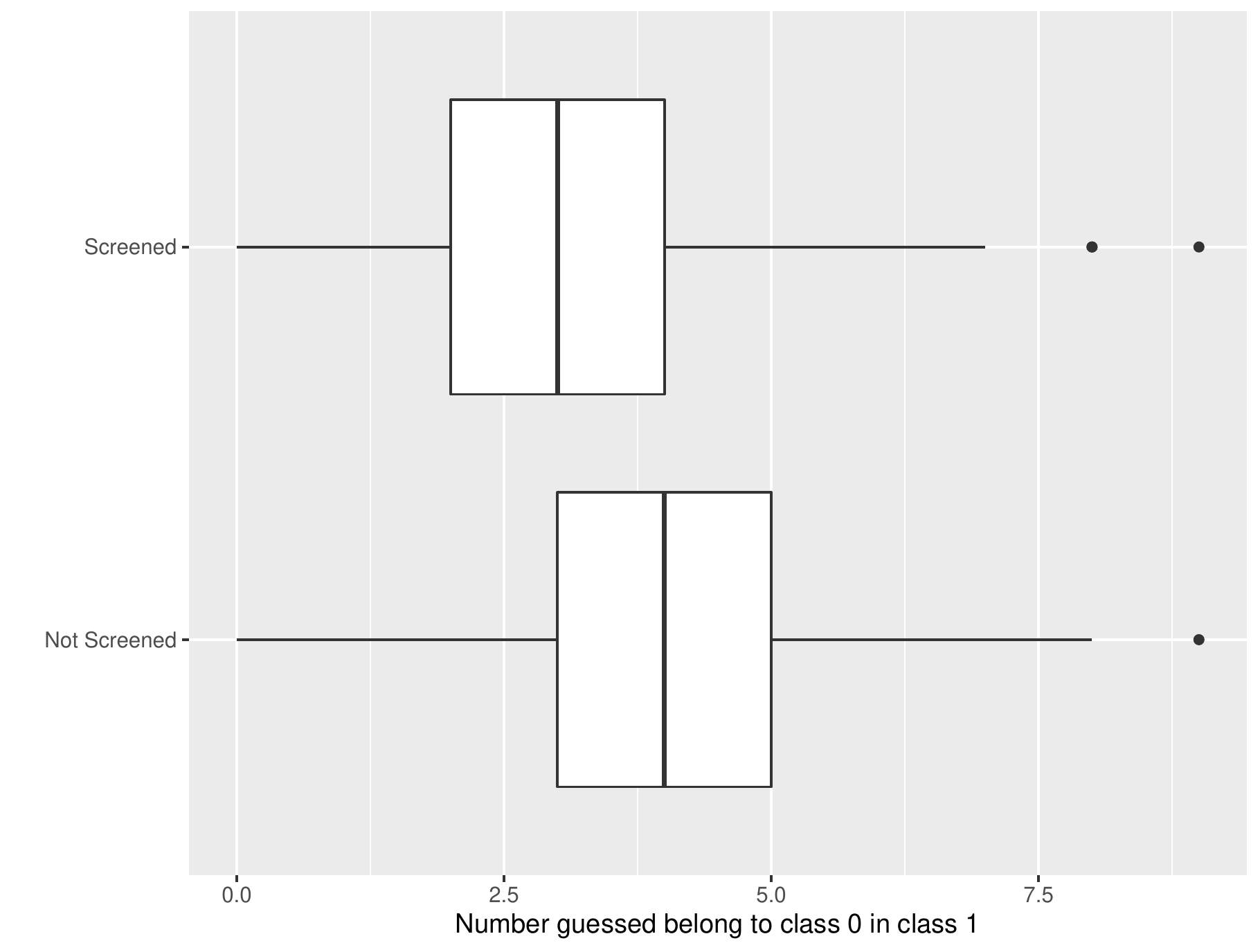}}\\
\subfloat[Boxplots of false negatives]{\includegraphics[width=.5\linewidth]{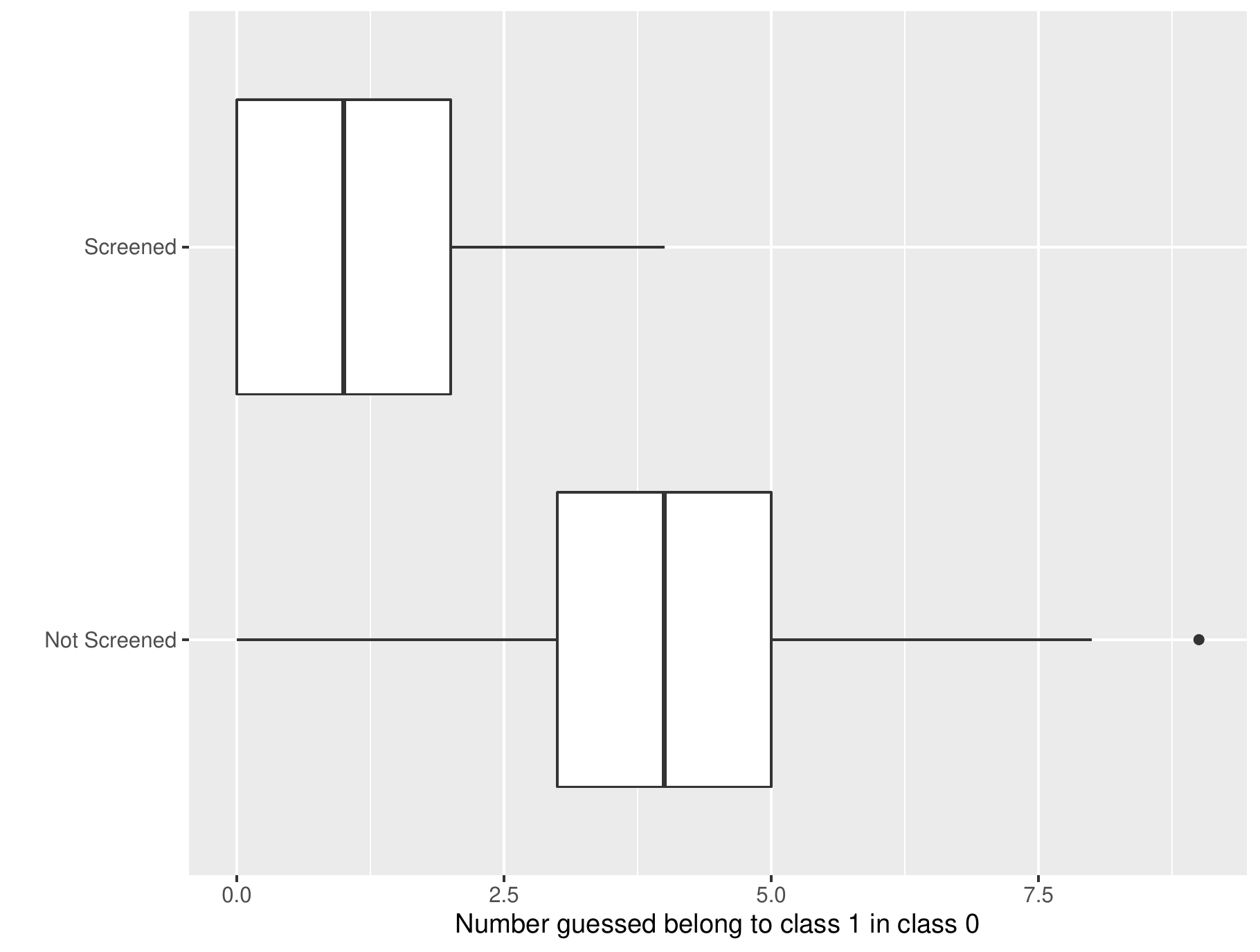}}
\subfloat[Boxplots of true negatives]{\includegraphics[width=.5\linewidth]{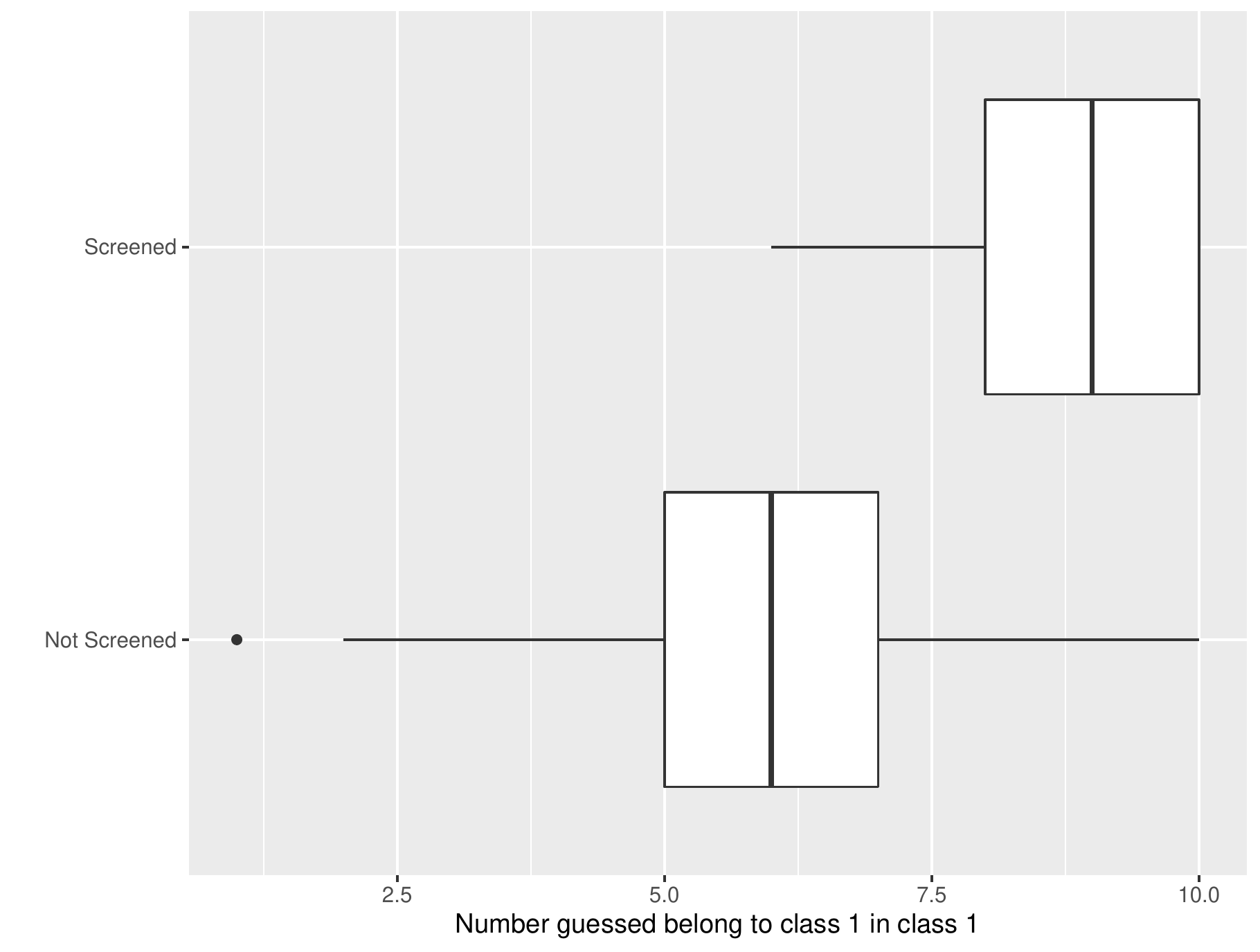}} 
\caption{\textit{Box plots displaying the accuracy of SVM when the data are generated from model where $10\%$ of the variables are important and have a shape difference.} 
%\tcr{\it In the previous section, define the model from which the data are generated for Figure 2.  Then you can just refer to this model in later sections.  If there are other models that are used repeatedly, do the same thing for them.
%as for Figure \ref{fig:mn20}}. 
To illustrate the accuracy of the methods, we do the following. First, suppose a positive case corresponds to an observation being in class 1 and a negative case corresponds to an observation being in the other class. Then to show the accuracy of the SVMs, we show boxplots on the number of ``True Positives," ``True Negatives," ``False Positives" and ``False Negative" occurrences. }
%(b) You haven't said which boxplot is for all data and which is for screened data.} \tcb{\it{I think this is an issue with the size of the plot, one of them is for screened and the other is for not screened.}}}
\label{fig:ConfusionBoxPlotScreenvNoScreen}
\end{figure}

Finally, we would like to add that even though SVM with the kernel trick is a fine classification method that uncovers many different types of differences between variables, its performance can be degraded harshly by the presence of noisy variables. To illustrate this, we use data in the same setting as the shape difference. We constructed a  training and testing set such that both consist of 10 observations from each class. Five hundred variables were used, with only 10\% on average being important. All data for unimportant variables have a standard normal distribution. 
%\tcr{\it Is the previous sentence correct?} \tcb{\it{Yes}} 
If a variable is important, then its distribution in one class is a bimodal mixture of two normals and in the other class a $t$-distribution with 4 degrees of freedom. The normal distributions in the mixture both have standard deviation 1 and means of -2.5 and 2.5. We trained an SVM with the radial basis kernel on all of the observations, and trained another SVM with the radial basis kernel but used only variables whose ALB value was larger than the interpretative cutoff of 0. We repeat this procedure 100 times, and report on its accuracy in Figure \ref{fig:ConfusionBoxPlotScreenvNoScreen}. 
%\tcr{\it Describe the important takeaway from the figure.} 
In general, classification accuracy is greatly improved, false negatives rarely happen after screening and the number of false positives is reduced.

\begin{comment}

\begin{table}
\begin{center}
\begin{tabular}{||c c c c||} 
 \hline
  &  & Actual Class & \\  
 \hline\hline
  &  & Class 1 & Class2 \\ 
 \hline
  SVM Prediction & Class1 & 9 & 1 \\
 \hline
  & Class2 & 0 & 10 \\
 \hline
\end{tabular}
\end{center}
\caption{Confusion Matrices for data given to a support vector machine similar to \ref{fig:mn20}. The data set is split in half evenly across both classes. One half is used to train the support vector machine, while the other half is used to evaluate the data. We screen this data with our screening method so that only variables with an ALB larger than $\text{log}(2) + \text{log}(.6)$ are chosen are used for training. A total of 30 variables are chosen.}
\label{table:ConfusionScreen}
\end{table}

\begin{table}
\begin{center}
\begin{tabular}{||c c c c||} 
 \hline
  &  & Actual Class &  \\ 
 \hline\hline
  &  & Class 1 & Class2 \\ 
 \hline
  SVM Prediction & Class1 & 7 & 3 \\
 \hline
  & Class2 & 7 & 3 \\
 \hline
\end{tabular}
\end{center}
\caption{Confusion Matrices for a support vector machine in the same context as \ref{table:ConfusionScreen}, the difference between the two methods is that this support vector machine uses all variables rather than variables that survive ALB screening.}
\label{table:ConfusionNoScreen}
\end{table}

\end{comment}

The ALB screening method need not be the final say as to which variables to include. Two variables that are individually important but highly correlated might be selected, although this may not be ideal for some classifiers. 
%The screening method need not do all of the variable selection for the classification scheme that is ultimately used. 
%\tcr{Does SVM do any variable selection, or does it use every variable it's given? If it's the latter, you might want to provide some context for the remarks in this paragraph.} \tcb{\it{SVM does not do variable selection, I think the kernel trick can be tricky to apply sometimes for this reason if a small subset of variables are truly responsible. I added a little bit of context. There is sparse support vector machine which is similar to the difference between lasso logistic regression and logistic regression.}} 
Screening can simply be a precursor that simplifies the job of a classifier, which does further variable selection.  Even methods that can perform variable selection and modeling simultaneously can benefit from having the number of variables reduced dramatically by screening. This is observed in SIRS \cite{zhu2011model}, SIS \cite{fan2008sure}, and is also true in our case.

\section{Interaction with BART and a tailored classification method}

We have recommended using classification methods that can take advantage of features for which the classes have non-location differences.
%or applying a kernel method to the features that remain relevant post screening. 
Methods that do further variable selection or that can handle sparse data sets can also fare quite well with our screening methods. BART and DART are methods having few parametric assumptions and that are able to capture a large variety of features from the data. BART has issues as the number of predictors grow, and DART has been proposed as a solution for this issue \cite{linero2018bayesian}. While DART can handle the case where many predictors are irrelevant, there is a cost. Mixing times of the chains for DART are increased compared to BART, and a prior that encourages sparsity may cause DART to get trapped in a posterior mode when the MCMC procedure to estimate it is run \cite{hill2020bayesian}. While we cannot directly mitigate these problems, decreasing the number of variables helps speed up the MCMC procedure. Our screening method can decrease the number of variables at a faster rate than DART can. DART is resilient against correlated nuisance variables and can therefore eliminate variables that survive ALB screening but are irrelevant due to collinearity. We provide simulations showing that use of our screening method before BART or DART can result in improved misclassification rates and computing speeds. 
%\tcr{\it Make sure my change in previous sentence is accurate.} \tcb{\it {This is fine}}

%To show this result, we look at simulations. 
%\tcr{\it See my previous comments about referring back to previous model.}
We generate data in the same context as Figure \ref{fig:mn20}, but instead roughly 10\% of the variables are relevant. If a variable is irrelevant, the distribution of the variable for both classes is standard normal. 
%\tcr{\it Define Rand index.} 
To assess the performance of a classifier, we computed the Rand index, or the percentage of correct decisions the classifier has made.  We compute the Rand index for the BART and DART procedures applied to all variables, 
%classifying both of the data sets, 
and the Rand index of the same procedures applied to variables that survive ALB screening. We consider different training set sizes that vary from 5 to 20. The testing set size for each simulation is the same as the training set size. We repeat this 100 times for each sample size. 
%Roughly 10\% of the variables are relevant. 
%To examine how accurate each method is, we will examine the rand index between the accuracy of the labels and the predicted values for each class. 
%\tcr{\it Move this definition above.}\ \tcb{This is implemented}.

\begin{figure}[h!]
\centering
\includegraphics[scale=.5]{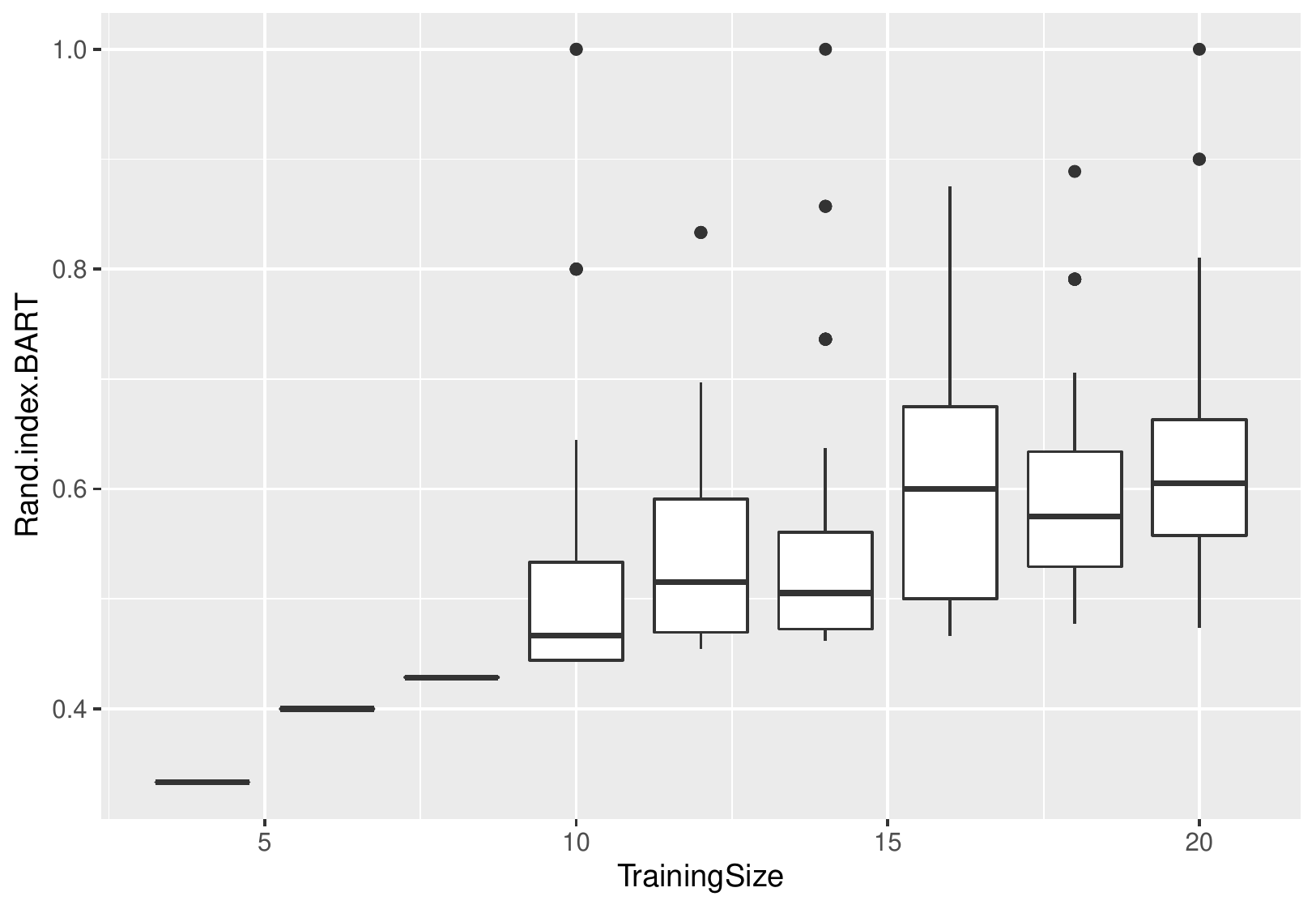}
\caption{\textit{Box plots of the Rand index of BART in the setting of a shape difference}. We vary $m$ with $m = n$, and the training size in the plot denotes $m + n$. We repeat each simulation 100 times for each sample size. Roughly 10\% of the variables are relevant.}
\label{fig:BARTacc}
\end{figure}

\begin{figure}[h!]
\centering
\includegraphics[scale=.5]{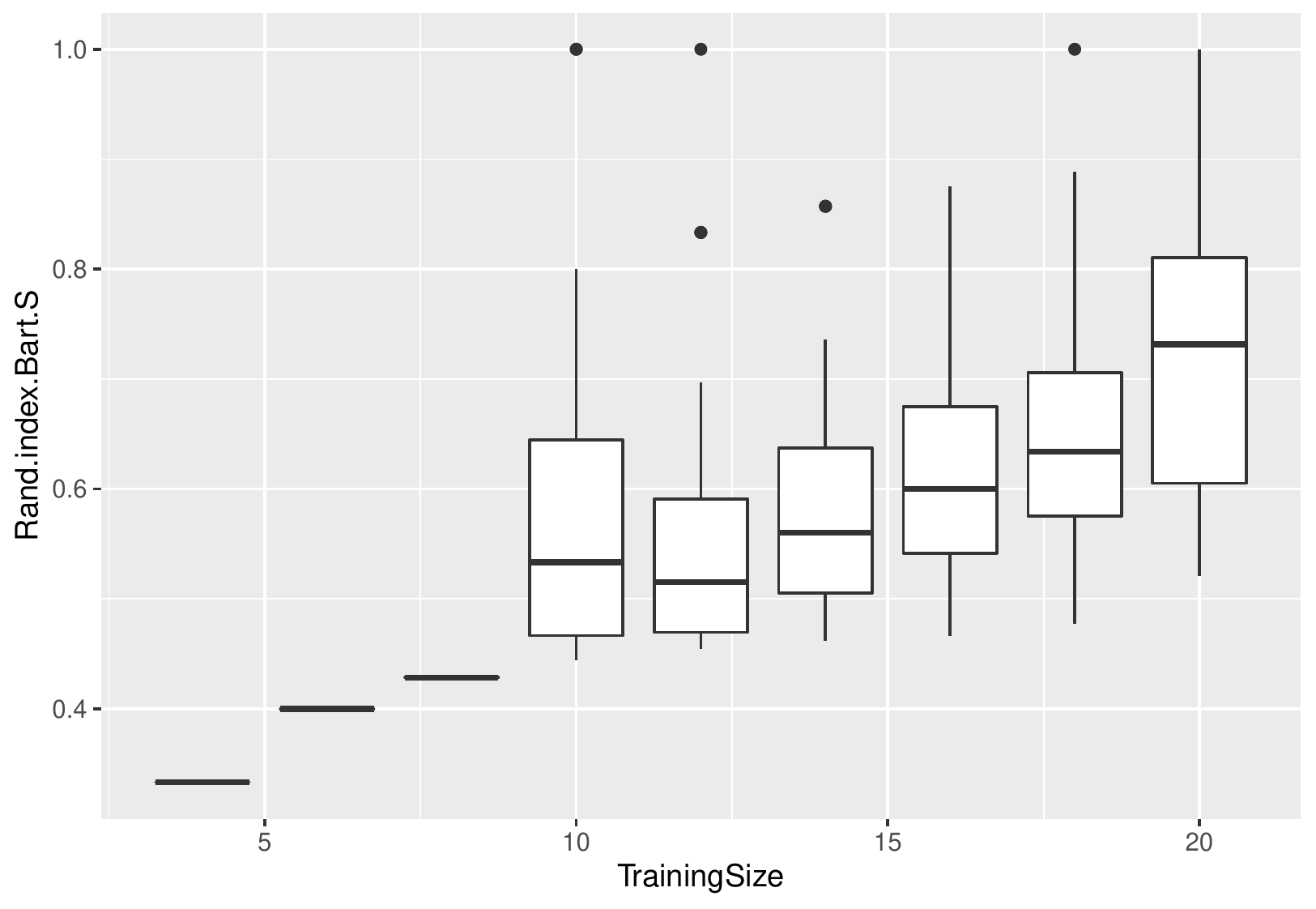}
\caption{\textit{Box plots of the Rand index of BART when BART is improved by screening.} This is in the same setting as that used in Figure \ref{fig:BARTacc}. The difference between the two plots is that we screened the variables with the ALB procedure before applying BART. We choose variables such that all generated ALBs are larger than the interpretable cutoff of 0. The power of the classification method grows large when enough data is accrued. The interpretive cutoff tends to be conservative, and power of the approach is likely to be even larger if a permutation-based cutoff is used instead.}
\label{fig:BARTScreenacc}
\end{figure}

\begin{figure}[h!]
\centering
\includegraphics[scale=.5]{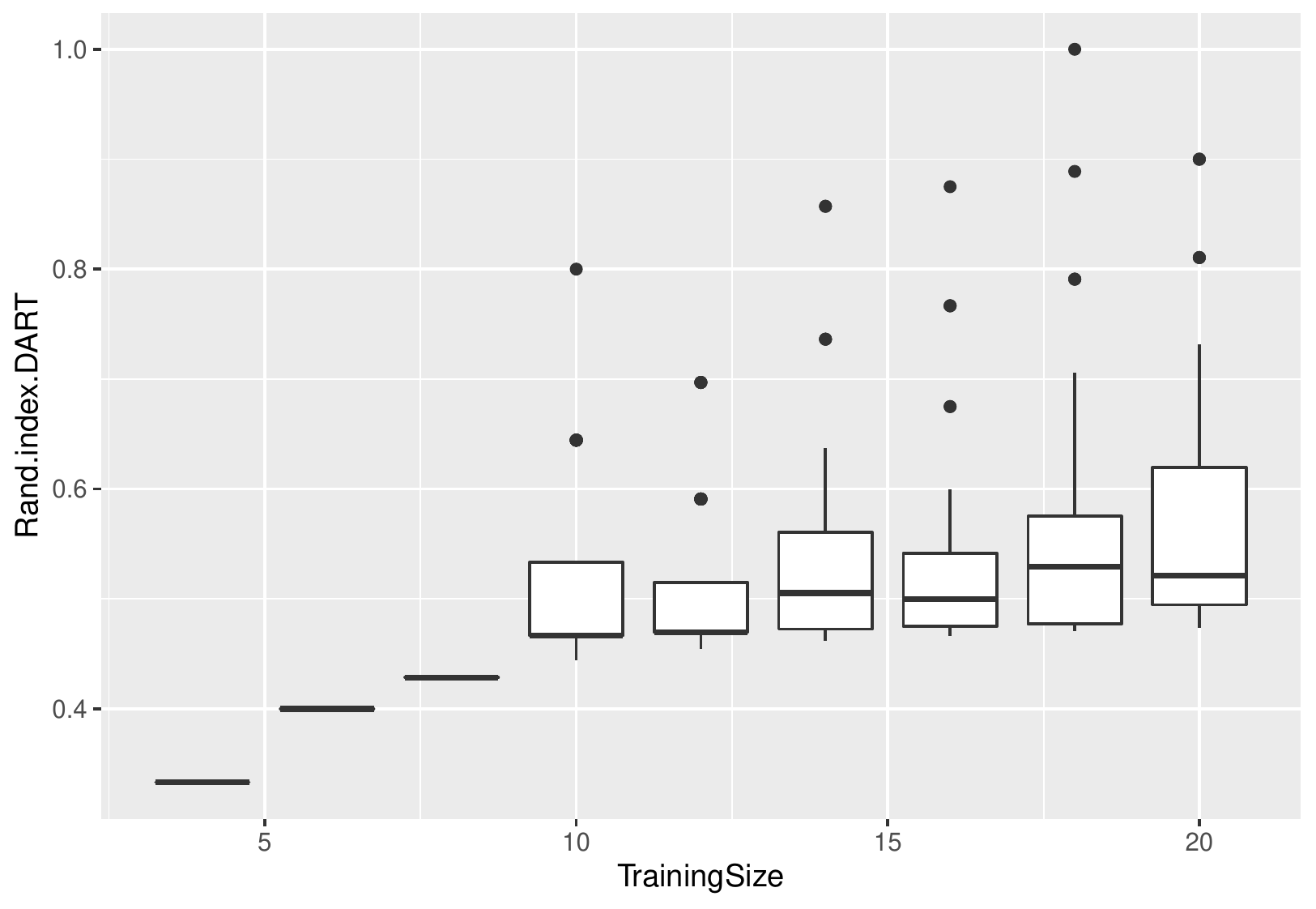}
\caption{\textit{Box plots of the Rand index of DART}. This is in the same setting as in Figure \ref{fig:BARTacc}, with the difference that we use DART instead of BART as it is capable of automatically performing variable selection.}
\label{fig:DARTacc}
\end{figure}

\begin{figure}[h!]
\centering
\includegraphics[scale=.5]{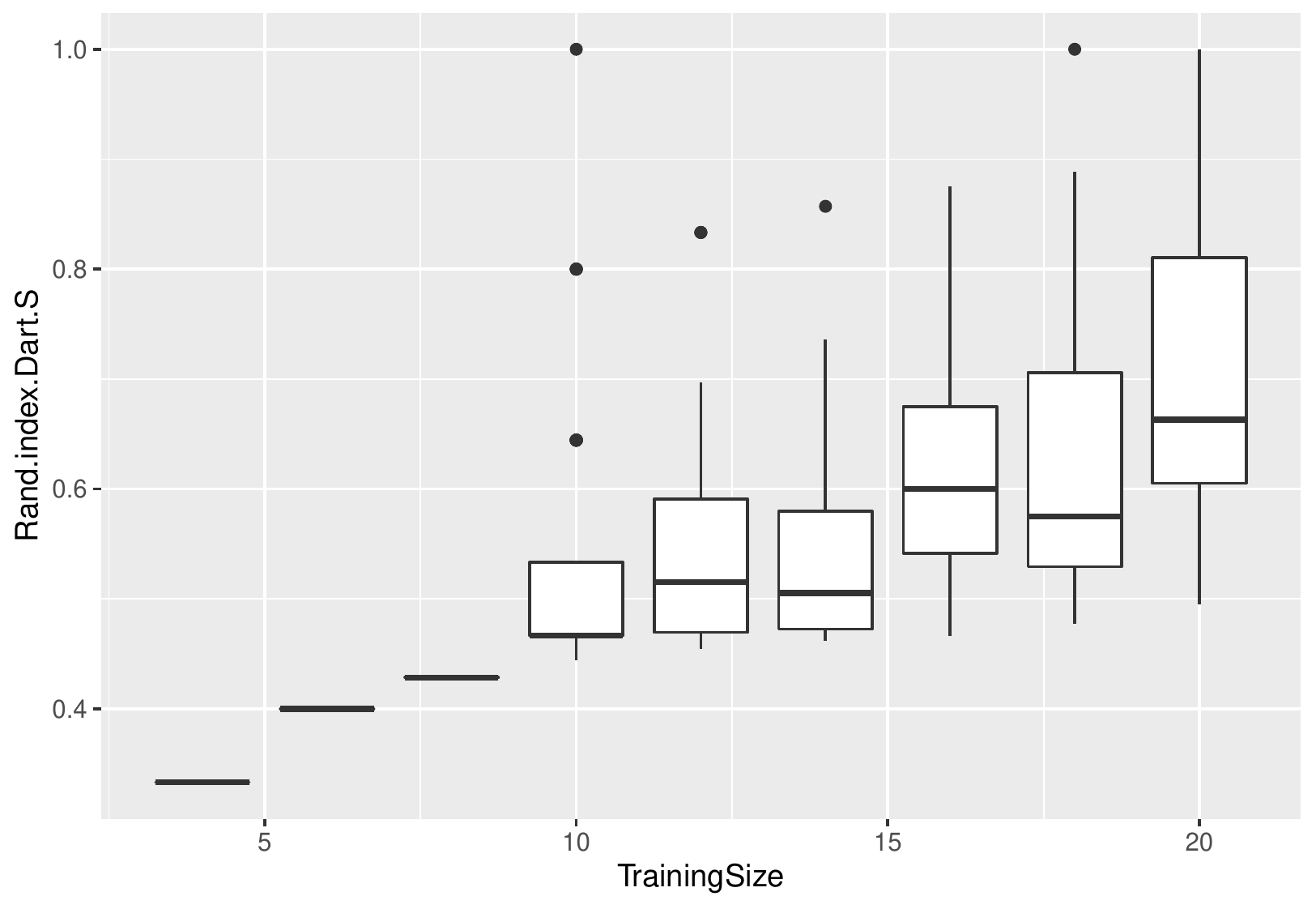}
\caption{\textit{Box plots of the Rand index of DART improved by screening.} This is in the same setting as in Figure \ref{fig:BARTacc}.  The difference here is we screen the variables with the ALB procedure before applying DART. We choose variables such that all generated ALBs are larger than 0. The power of the classification method has improved after screening for variables, despite DART being fully capable of automatically performing variable selection automatically.}
\label{fig:DARTScreenacc}
\end{figure}

\begin{figure}[h!]
\centering
\includegraphics[scale=.5]{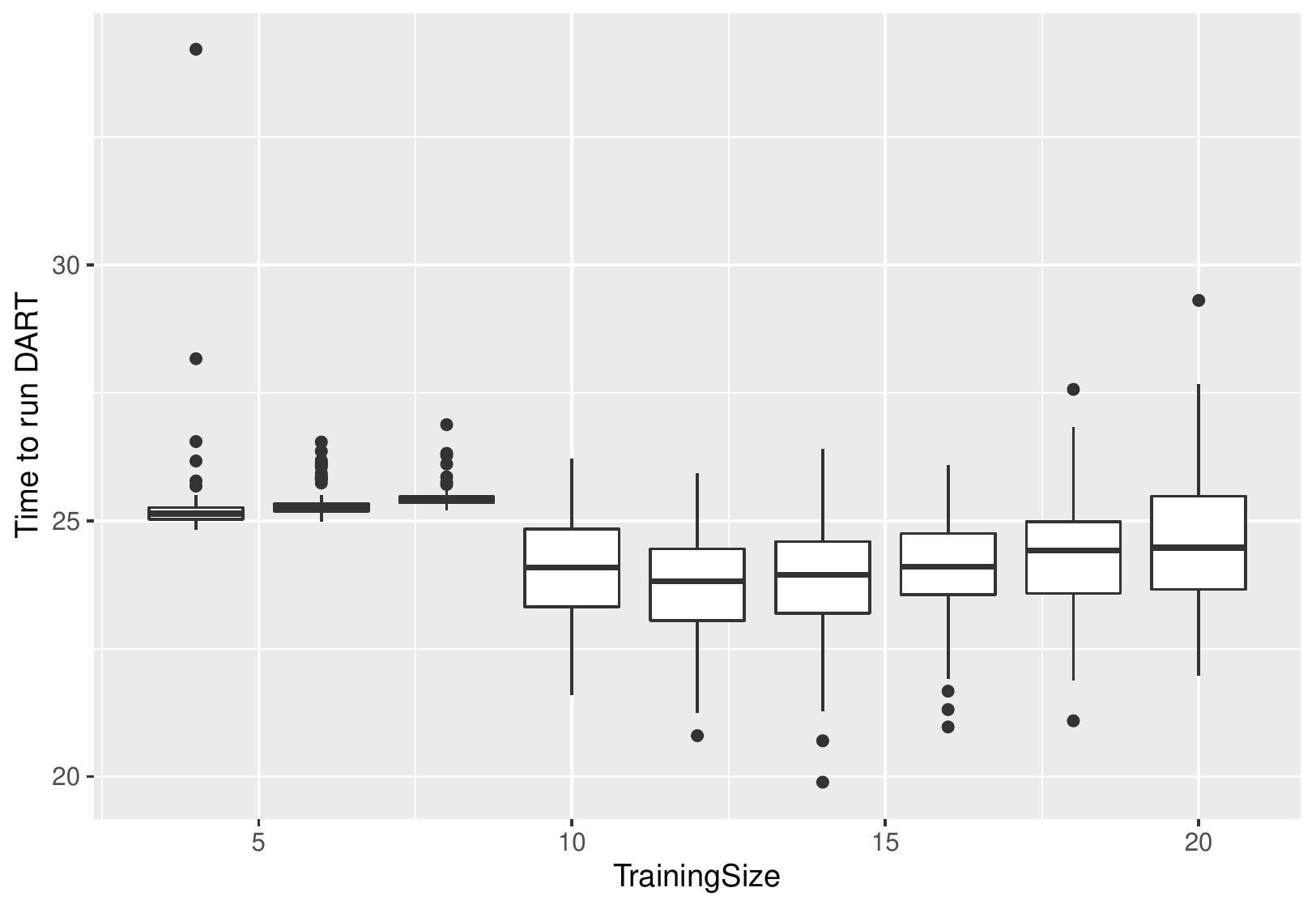}
\caption{\textit{Box plots of the time it took to run DART.} The simulation is in the same setting as in Figure \ref{fig:BARTacc}.}
\label{fig:DARTRT}
\end{figure}

\begin{figure}[h!]
\centering
\includegraphics[scale=.5]{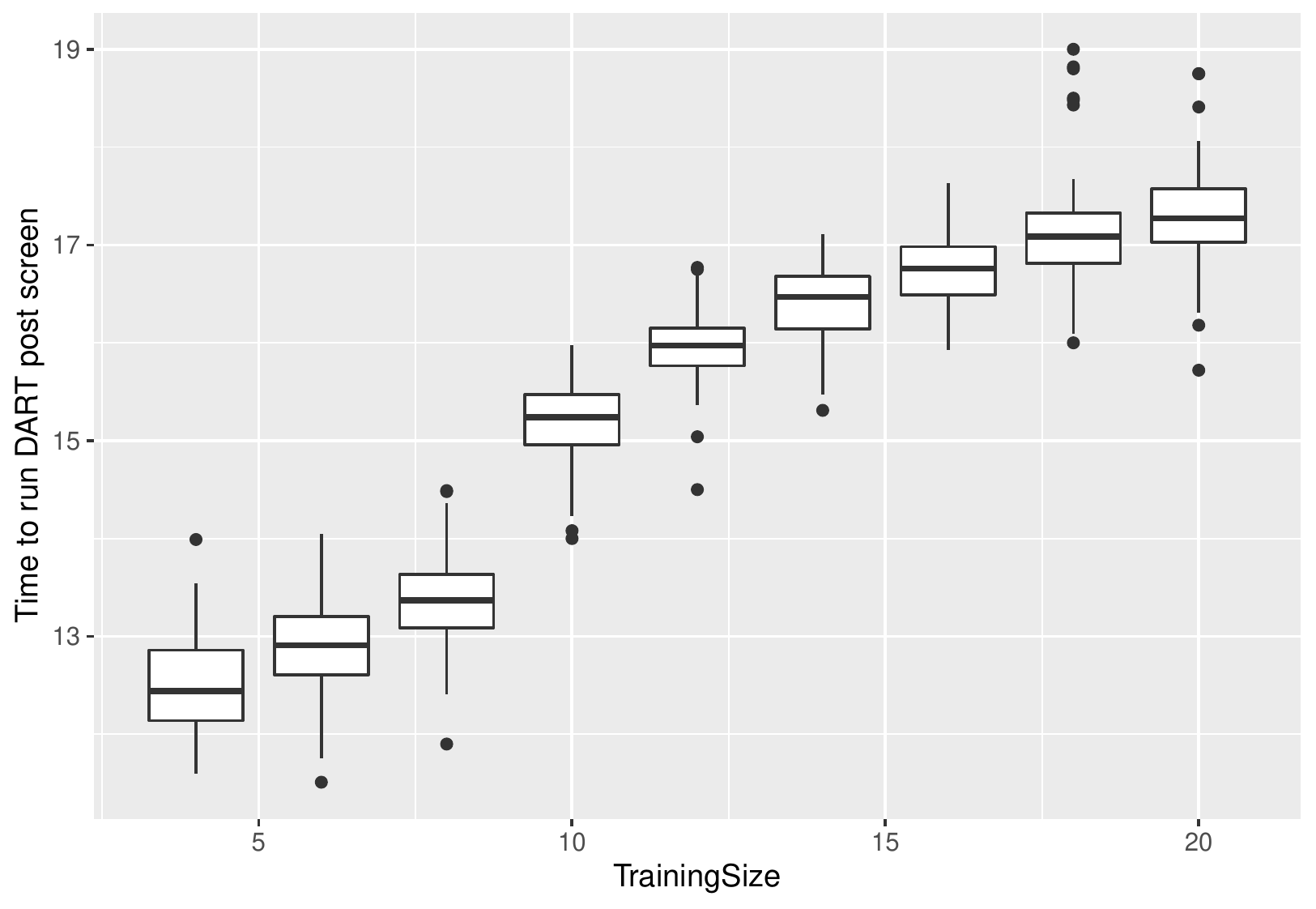}
\caption{\textit{Box plots of the time it took to run DART post screening.} This is in the same setting as in Figure  \ref{fig:BARTacc}. The difference here is we screen the variables with the ALB procedure before applying DART. The screening procedure itself takes much less than half a second, and as a result of removing a large number of irrelevant variables, greatly improves the amount of time it takes for DART to run.}
\label{fig:DARTScreenRT}
\end{figure}

Figures \ref{fig:BARTacc} and \ref{fig:DARTacc} show how accurate BART and DART alone are in these settings and Figures \ref{fig:BARTScreenacc} and \ref{fig:DARTScreenacc} show how the methods do when variables are screened for importance beforehand. There is a notable gain in the Rand index as the training set size gradually increases for both methods. Of greater note is that the time it takes to run both procedures is decreased. Figure \ref{fig:DARTRT} shows the amount of time it takes the BART method to run before screening and Figure \ref{fig:DARTScreenRT} shows how long the method takes after screening. Screening on average shaves off at least 10 seconds of computation time while increasing the average accuracy. This is an interesting result, as the methods themselves, DART especially, tend to be robust to irrelevant variables. 
%\tcr{\it When you say that the methods tend to be robust to irrelevant variables, does this include BART? I thought BART had issues in this regard.} \tcb{\it{DART is more robust than BART, but BART is still somewhat resilient to irrelevant variables if enough data is present}}.
%and we think they are. 
However, the figures suggest that a larger sample size is required to achieve that robustness. 
%Importantly,  accuracy is gained in this example by screening prior to use of either BART or DART. 
%\tcr{\it I think the last sentence is fair.  Do you agree?} \tcb{\it{I think this is fine.}}

\subsection{A simple Bayesian classifier}

Suppose our goal is to simply leverage the differences between variables, regardless of the type of difference, and that we assume independence between variables. We can construct  a simple Bayesian method for classification in the following fashion. For each variable, we compute  two kernel density estimates, one for each class. For each  variable $i$, let $\hat f_i$ and $\hat g_i$  be kernel density estimates using all variable $i$ data from classes 1 and 2, respectively.
%for the density of a variable under each class. 
The prior probability that a variable arises from a class is assumed to be proportional to the number of observations for that class.  Let $\x = (x_1,...,x_p)$ be an observation to be classified. If the underlying densities are known, then the conditional probability that $\x$ came from class 1 is
\begin{equation}
%\label{eqn:BNEWestimator}
p(\x)=P(Y = 1 | \x) = \frac{\frac{n}{m+n}\prod_{i\in D}f_i(x_i)}{\frac{n}{m+n}\prod_{i\in D}f_i(x_i) + \frac{m}{m+n}\prod_{i\in D}g_i(x_i)},
\end{equation}
where $D$ is the set of indices $i$ such that $f_i\not\equiv g_i$. 
Of course, the densities and $D$ are unknown, but $p(\x)$ can be estimated using kernel density estimates, and $D$ can be replaced by $\widehat D$, the set of indices such that the corresponding variables survive screening:
\begin{equation}
\label{eqn:BNEWestimator}
\hat p(\x)= \frac{\frac{n}{m+n}\prod_{i\in \widehat D}\hat{f_i}(x_i)}{\frac{n}{m+n}\prod_{i\in \widehat D}\hat{f_i}(x_i) + \frac{m}{m+n}\prod_{i\in \widehat D}\hat{g_i}(x_i)}.
\end{equation}
\begin{comment}
The statistic (\ref{eqn:BNEWestimator})
estimates:
\begin{equation}
\label{eqn:BNEWresultantestim}
\exp\left(\frac{n}{n + m}KL(f,\frac{n}{n + m}f+(1-\frac{n}{n + m})g)\right)  \\ 
\end{equation}

where
$$
f(x_1,...,x_p) = \prod_{i \in D}f_i(x_i), \\ 
$$

and
$$
g(x_1,...,x_p) = \prod_{i \in D}g_i(x_i), 
$$
\end{comment} 

A classifier based on (\ref{eqn:BNEWestimator}) can be a powerful tool for capturing marginal differences in distributions, but is incapable of leveraging differences that may lie in the dependence structure of the variables. 
%The classifier to which it leads is  simple and benefits greatly from our screening procedure. However, it has restricted utility when the \tcr{difference between classes is mainly with respect to dependence between variables}. 
To use (\ref{eqn:BNEWestimator}), we say that an observation $\x$ belongs to class 1 if $\hat p(\x)>n/(m + n)$ and to class 2 otherwise. 
%and this is the matter in which we will use it. 
We have found this classifier to have strong accuracy when dealing with independent variables, or with settings where the difference in multivariate distributions is dominated by marginal differences.
%\tcr{\it I deleted a lot of the discussion here since it seemed redundant.}

\begin{figure}[h!]
\centering
\includegraphics[scale=.5]{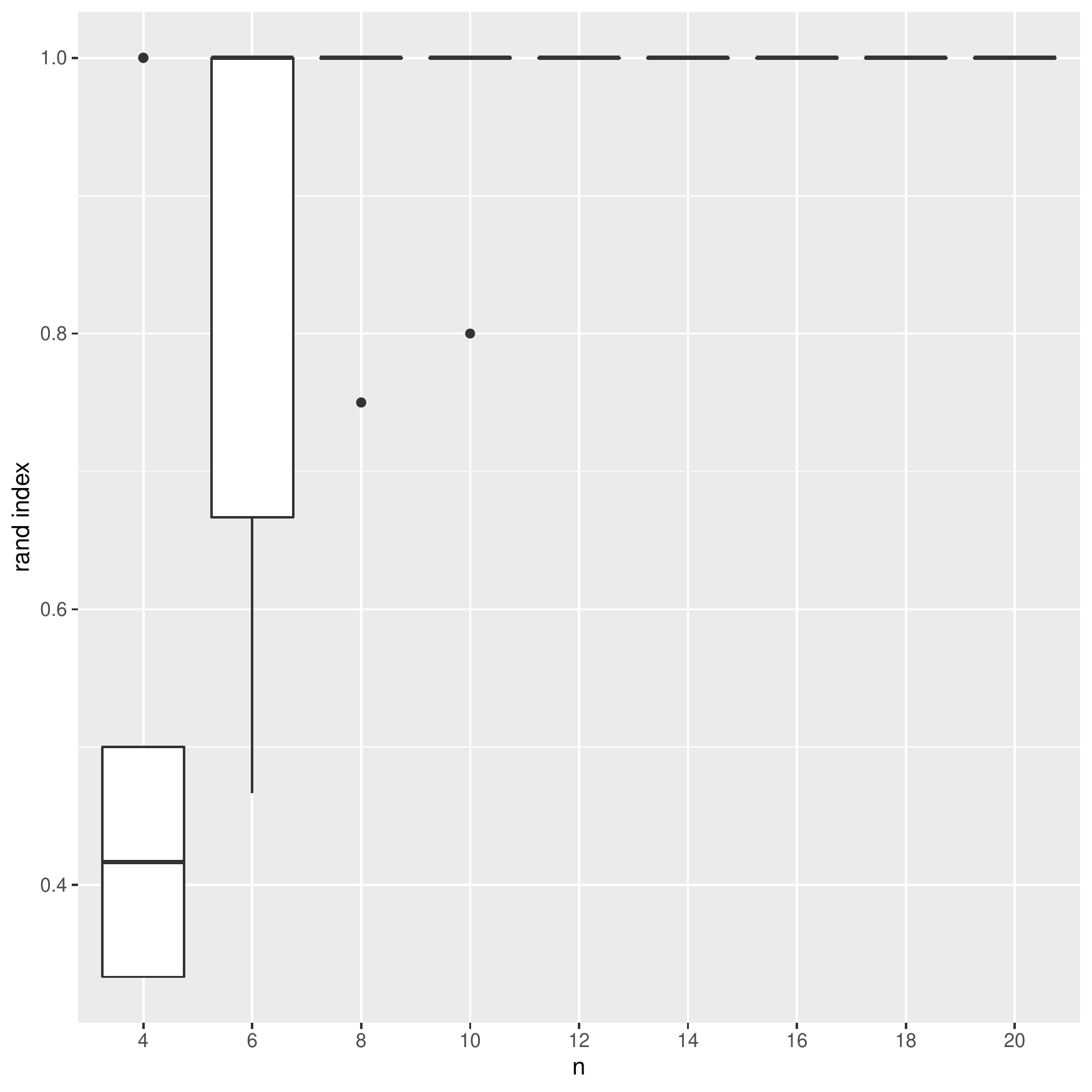}
\caption{\textit{A box plot of the Rand index of the Bayesian classifier}. The classifier is defined in terms of  (\ref{eqn:BNEWestimator}). We generate data in the same context as in Figure \ref{fig:mn20}, but vary $m$ and $n$ so that $m = n$, and the training size denotes $m + n$. We choose variables such that all generated ALBs are larger than the intepretable cutoff of $0$.}
\label{fig:BNEWestimatorresBP}
\end{figure}

To illustrate the effectiveness of the classifier based on (\ref{eqn:BNEWestimator}), we perform a simulation similar to the one referenced in Figure \ref{fig:mn10} but under a variety of different sample sizes. We repeat this procedure 5 times. For each sample size, we generate a balanced testing set of the same size, and compute a Rand index. A plot of the Rand indices against the sample size is found in Figure \ref{fig:BNEWestimatorresBP}. The power of the classification method grows to be quite large at even small $m + n$ values. Since the interpretive cutoff tends to be conservative, power of the approach is likely to be even larger if a permutation based cutoff is utilized instead. At a training size of just 9 samples in each group, the classifier makes perfect predictions over 75\% of the time. 

%\tcr{\it I'm not asking you to use a different cutoff in the sims for Fig. 14, but I still prefer the idea of using 0. The beauty of this cutoff is that it's as small as you would ever use, but still has the interpretation of being "conservative" in that the type I error probability for this cutoff tends to 0 with sample size.}
%\tcb{\it This is fair, should I add a note that another cutoff could have been used like 0, although I kind of think its okay to leave it as is.} \tcp{\it Just leave it.}
\section{Application on simulated data sets}

We want to compare our method to $t$-test screening and also compare performance of the different choices of ALB cutoff. There are at least two ways to go about this. One is to compare the percentage of variables that survive screening from both procedures, and the other is to apply a classification method after screening and see which method has a better classification rate. These procedures are carried out in three cases:
\begin{itemize}
    \item [] Case 1 -- {\it Location differences.} There are 600 variables, and those that are important arise from a case where there is a mean difference between classes. If the variable is important, one class has a standard normal distribution and the other a normal distribution with mean 1 and standard deviation 1. We let roughly $5\%$ of the variables be important by generating 600  independent Bernoulli variables, each with success probability $0.05$. 
    %The chance the bernoulli variable indicates we use the variable is $.05$. 
    This way of determining important variables is used in Cases 2 and 3 as well. Also, in this case and the following two the unimportant variables have standard normal distributions. The classifier we will use to compare performance in this case is SVM without the kernel trick. %\tcr{\it What makes it simple?} 
    \item [] Case 2 -- {\it Scale differences.} There are 600 variables, and those that are important arise from a case where there is a variance difference between classes. If the variable is important, one class has a standard normal distribution and the other a normal distribution with mean 0 and standard deviation 3. We let roughly $20\%$ of the variables be important, and the classifier used is the support vector machine with a kernel trick. 
    %\tcr{\it So, I guess the simple SVM doesn't use kernel trick. I'm not very familiar with SVM. Is the simple SVM with two classes different than a linear discriminant rule?}
    \item [] Case 3 -- {\it Shape differences.} There are 600 variables, and those that are important arise from a case where the class distributions have different shapes.  If the variable is important, one class has a  standard $t$-distribution with 4 degrees of freedom and the other a bimodal mixture of two normal distributions with means -2.5 and 2.5 and the same standard deviation of 1. Roughly $10\%$ of the variables are important, and the classifier used is the support vector machine with a kernel trick.
%    \item [Remark:] In all cases, for the sake of simplicity, if the variable is unimportant, the variable is generated from both classes as a standard normal variable.
\end{itemize}

\begin{figure}[p]
    \vspace*{-2.5cm}
    \makebox[\linewidth]{
        \includegraphics[width=.6\textwidth]{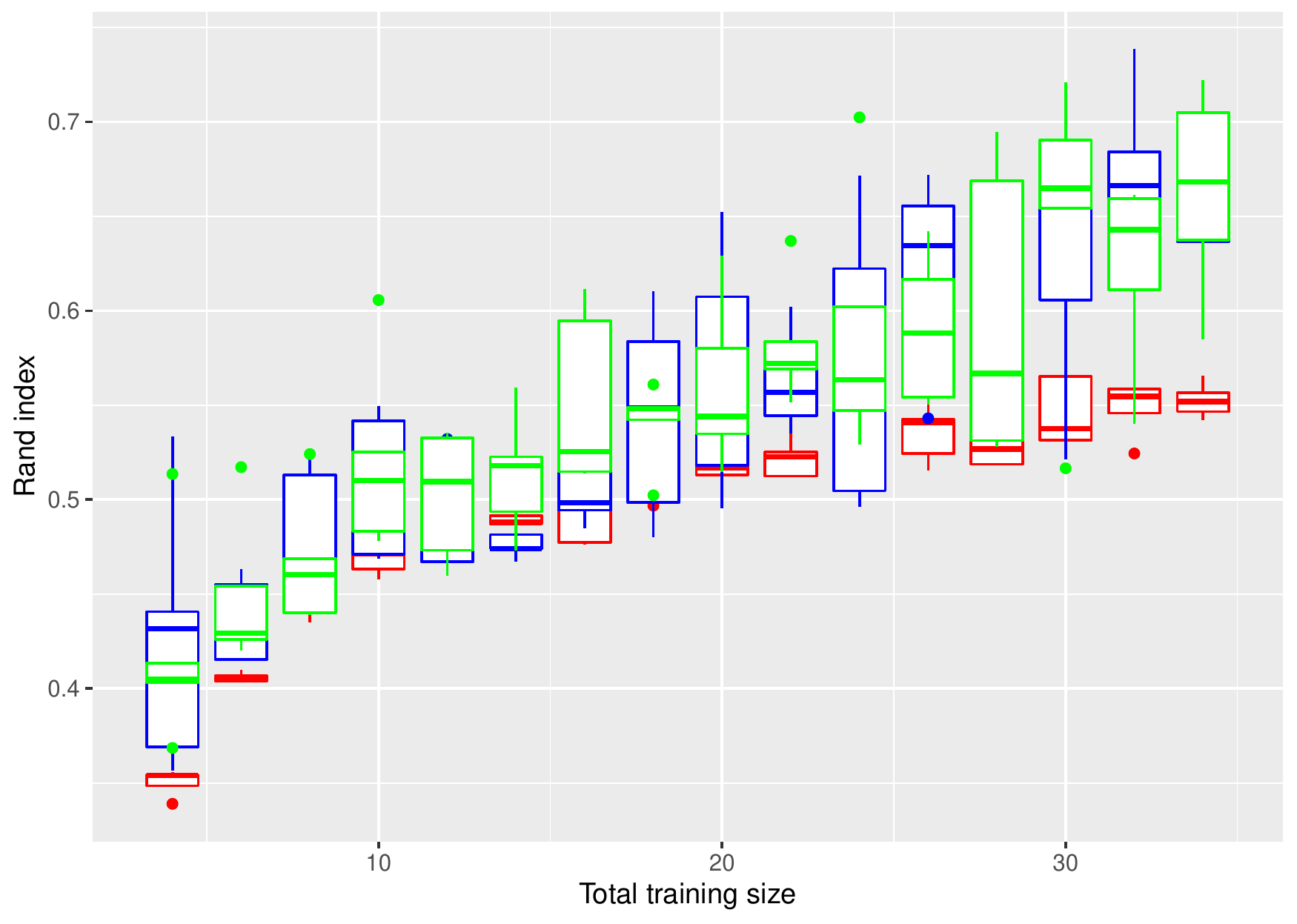}\hfill
    \includegraphics[width=.6\textwidth]{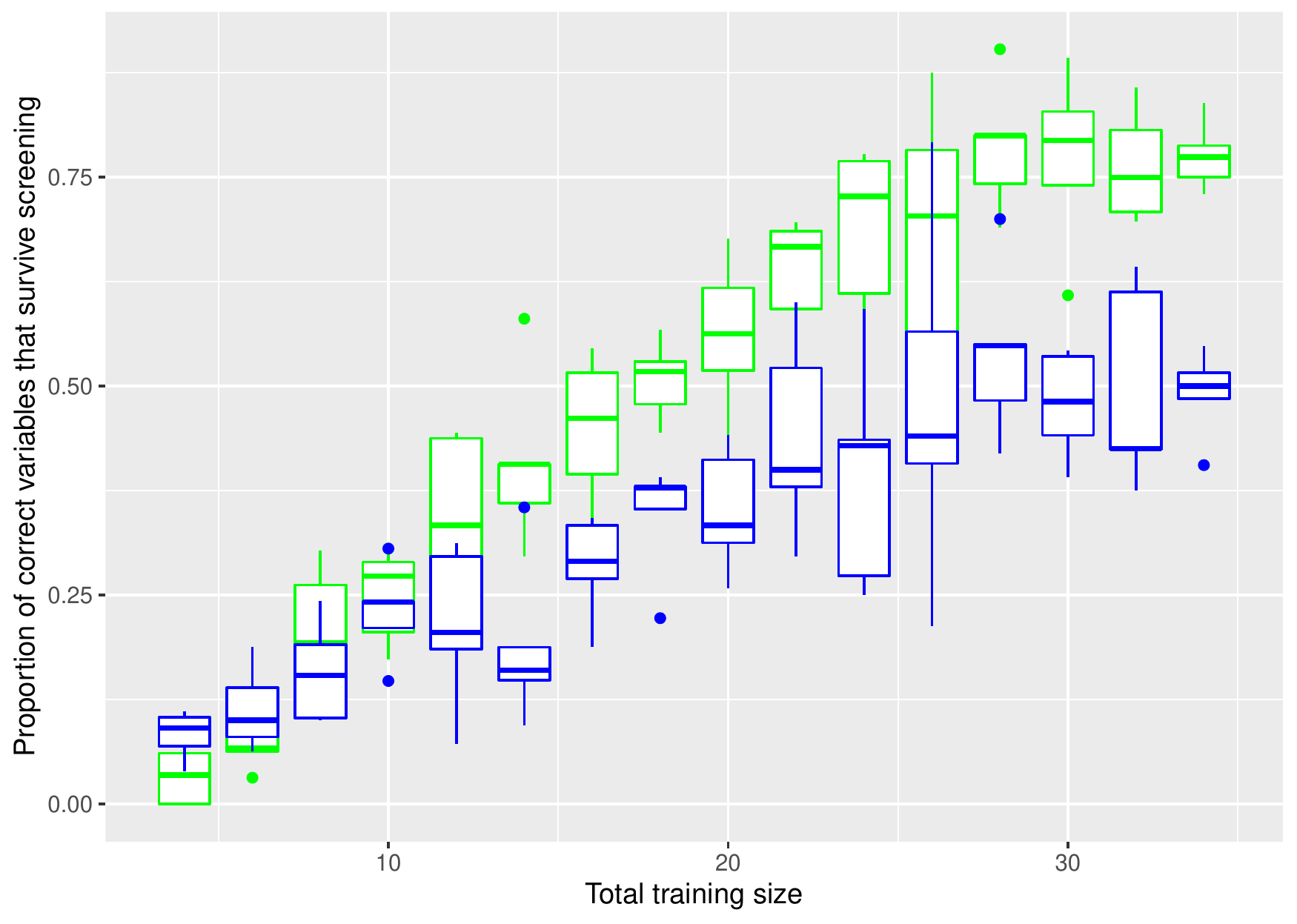}\hfill
    }
    \\[\smallskipamount]
    \makebox[\linewidth]{
    \includegraphics[width=.6\textwidth]{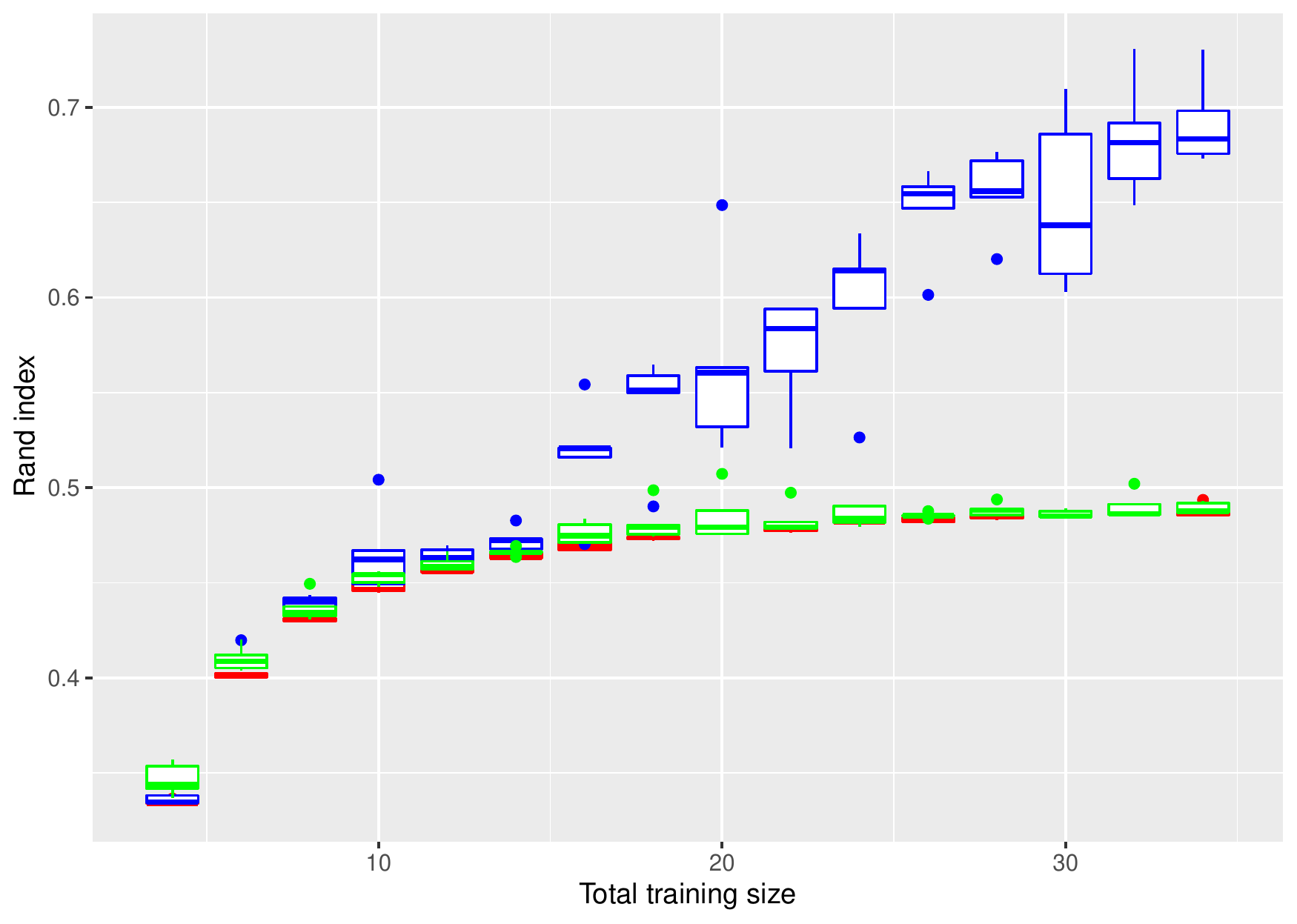}\hfill
    \includegraphics[width=.6\textwidth]{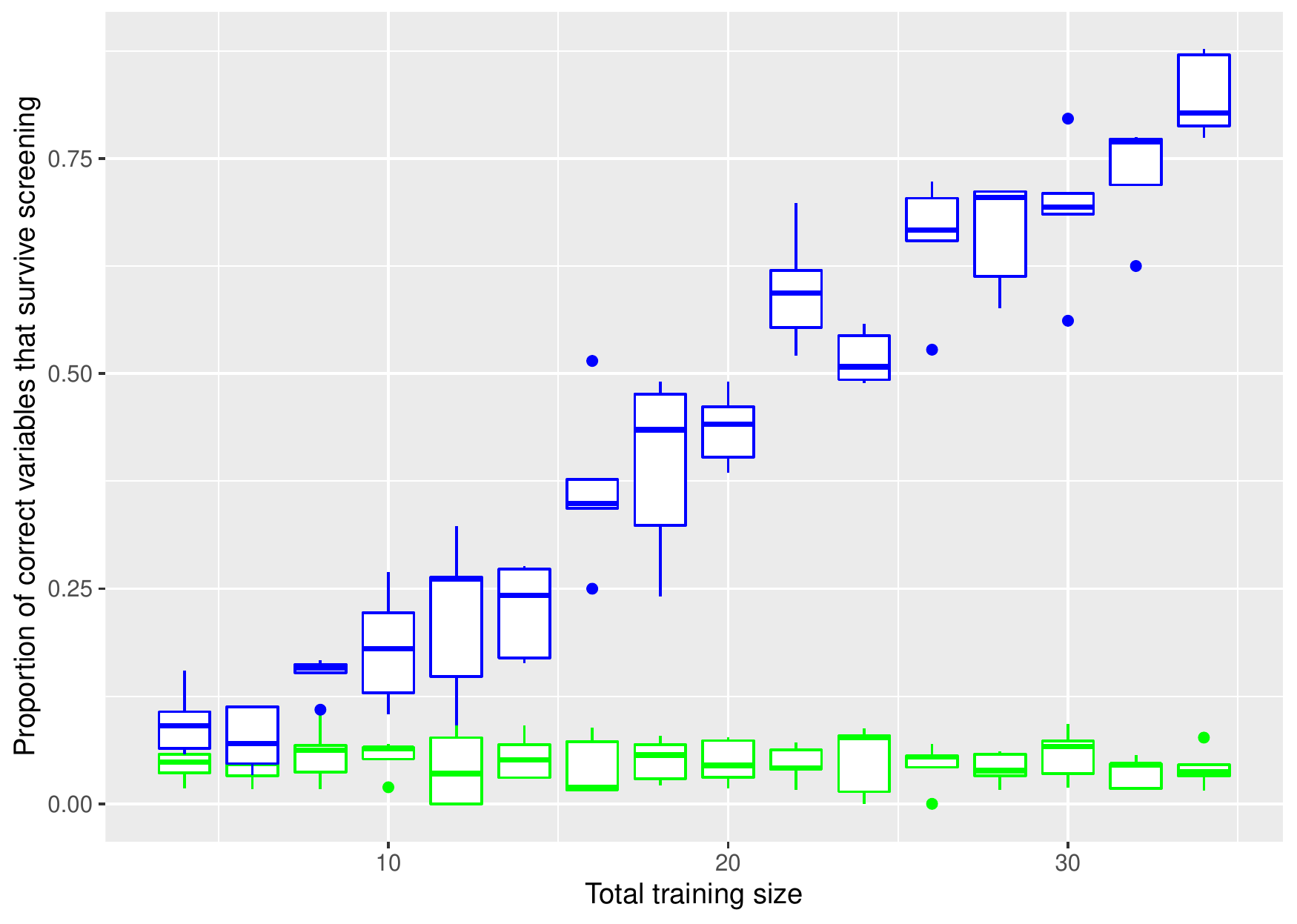}\hfill
    }
    \\[\smallskipamount]
    \makebox[\linewidth]{
    \includegraphics[width=.6\textwidth]{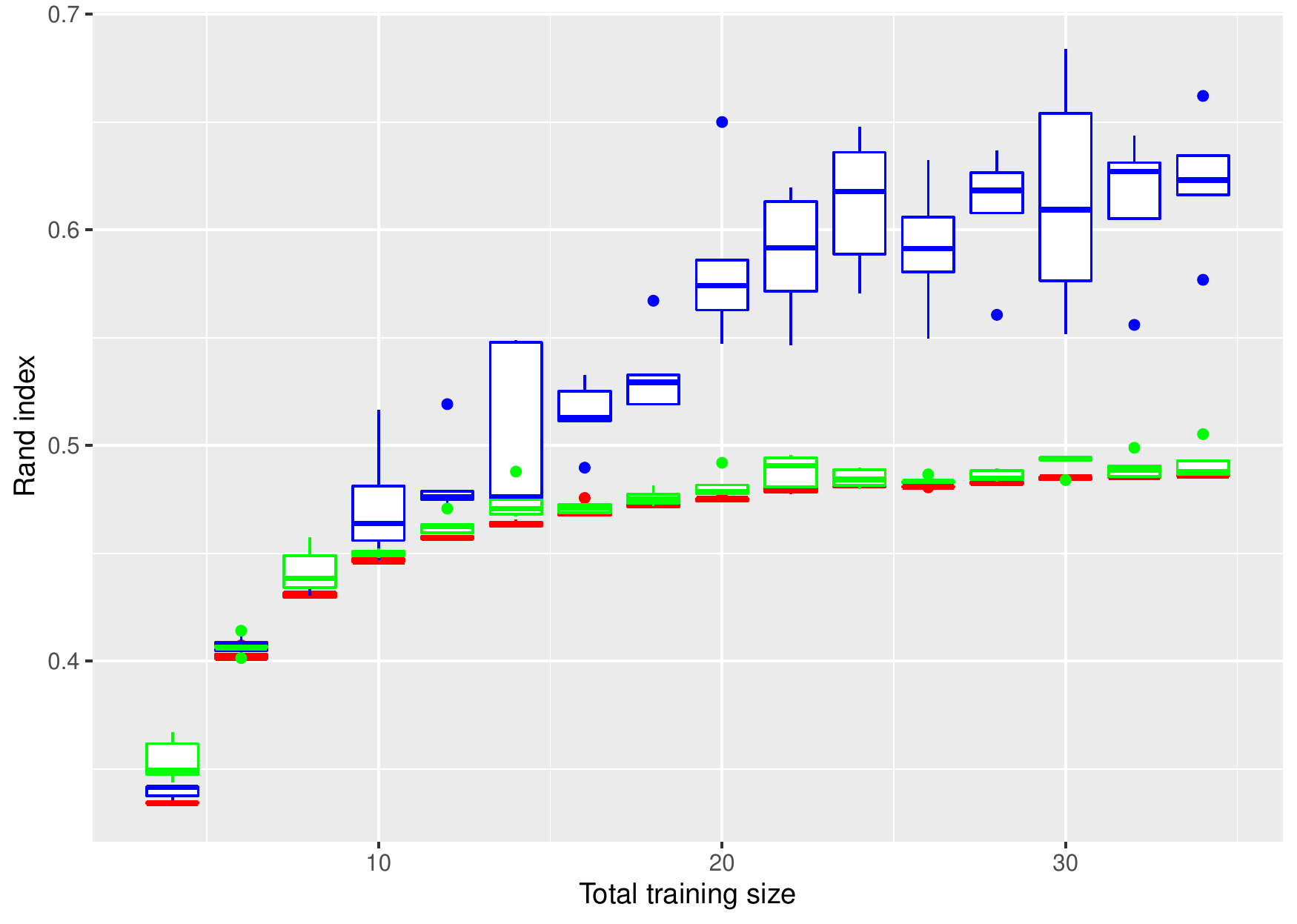}\hfill
    \includegraphics[width=.6\textwidth]{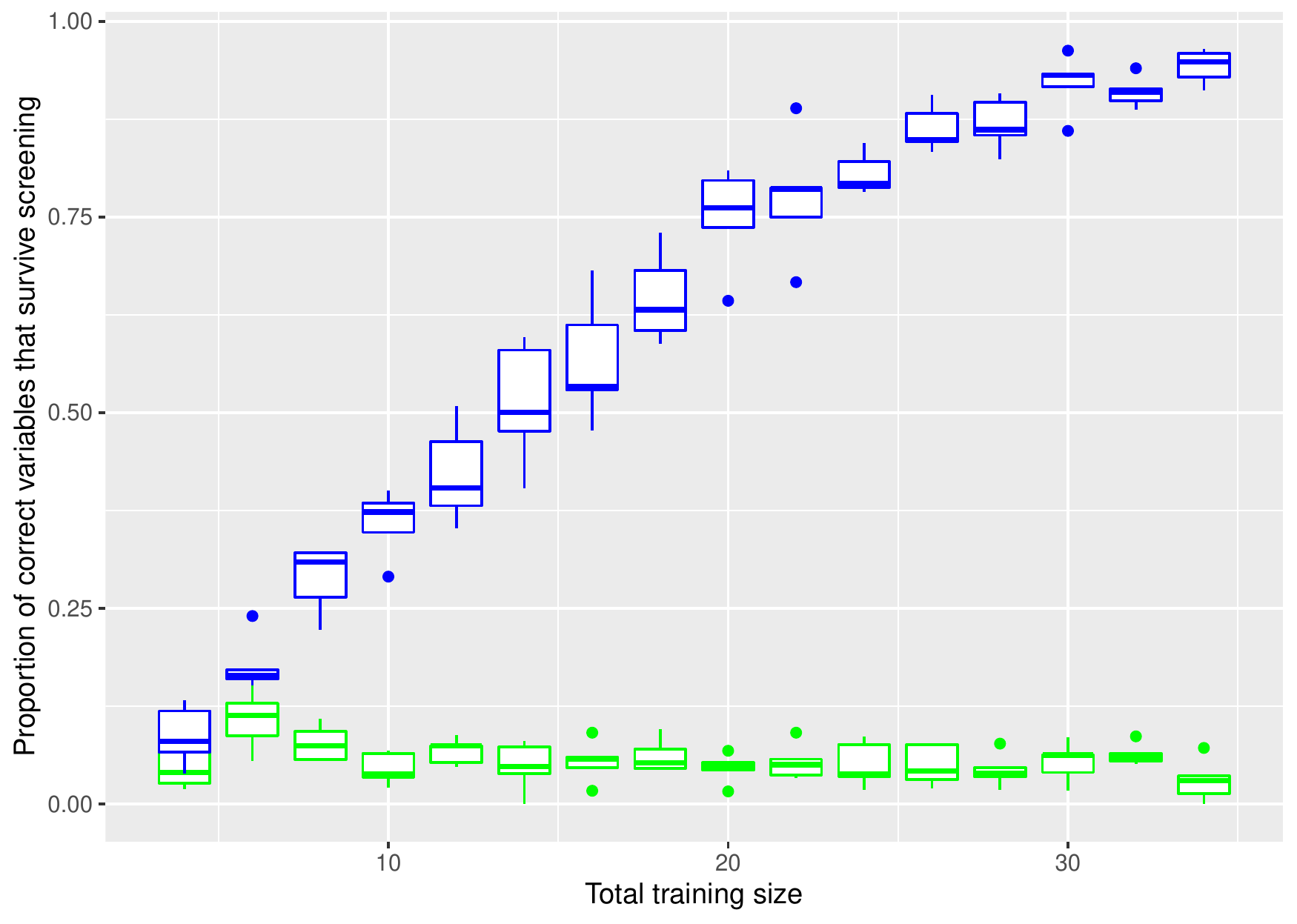}
    }
    \caption{{\it Simulation results for $t$-test screening and ALB screening that uses A3.} 
    %\tcr{Is A5 correct?}. \tcb{\it{A5 is correct}}}} 
    We generate $ALB^*$ values by permuting the data labels and computing $ALB$ for the resulting data. We do this for every variable twice. The ALB cutoff is the $95$th percentile of $ALB^*$. The $t$-statistic cutoff is chosen so that the $P$-value is less than $0.05$. Red, green and blue box plots are for no screening, $t$-test screening, and ALB  screening, respectively. 
    %Testing set sizes are always equal to training set size. 
    %\tcr{\it  Did you use A4 or A5? Rather than describing in detail how the permuting was done, just say which type you did, A4 or A5, and say that you used the 95th percentile. You give info in the captions that should be given in text, and vice versa. The fine details of the sim, such as whether you use A4 or A5, should be in text. In the captions you should include the following: (a) what's the difference between the three rows? (b) What's the difference between left and right?} 
    The first, second and third row of plots correspond to cases 1, 2 and 3, respectively.
    }
    \label{fig:PermutedScreening}
\end{figure}

\begin{figure}[p]
    \vspace*{-2.5 cm}
    \makebox[\linewidth]{
    \includegraphics[width=.6\textwidth]{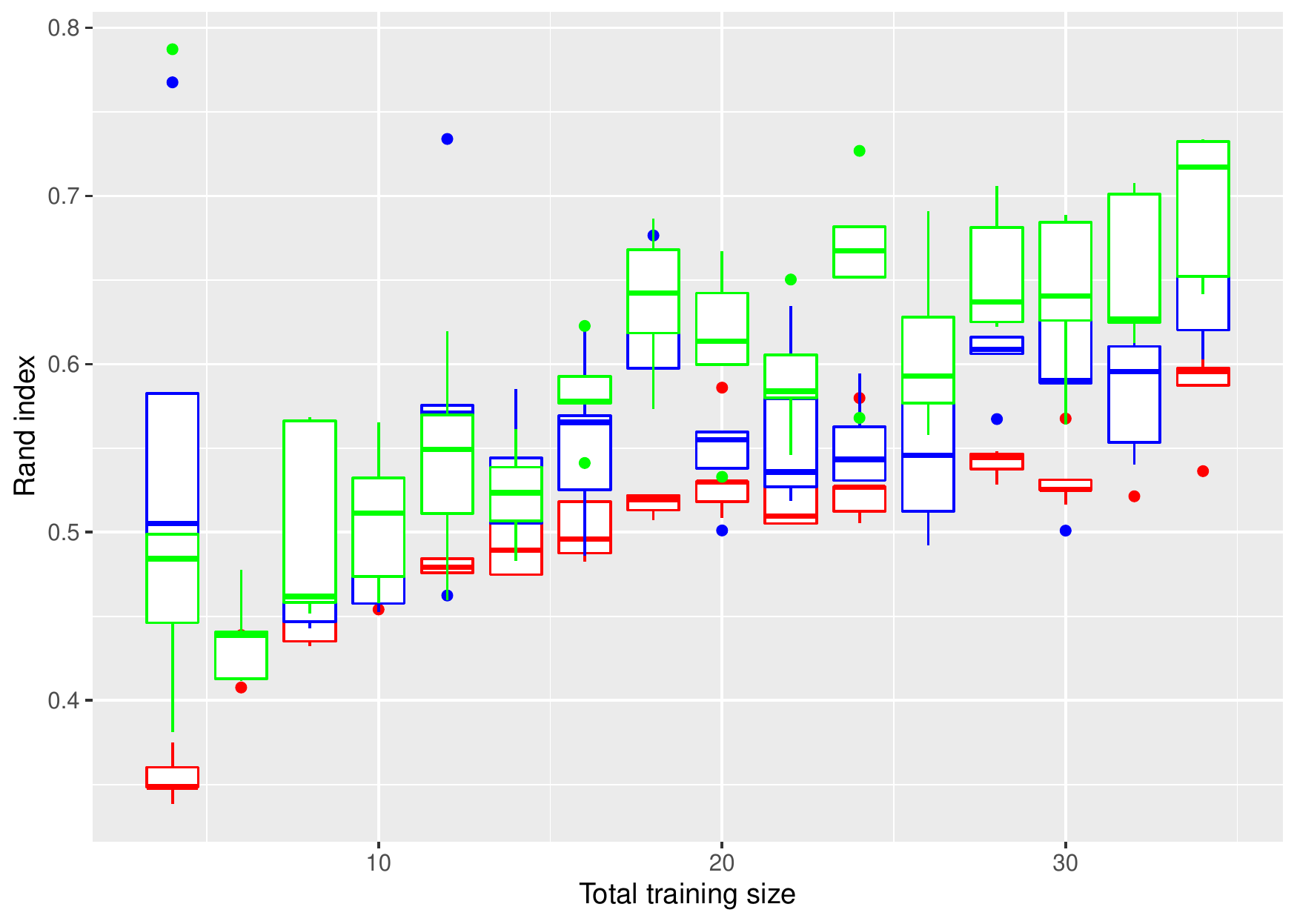}\hfill
    \includegraphics[width=.6\textwidth]{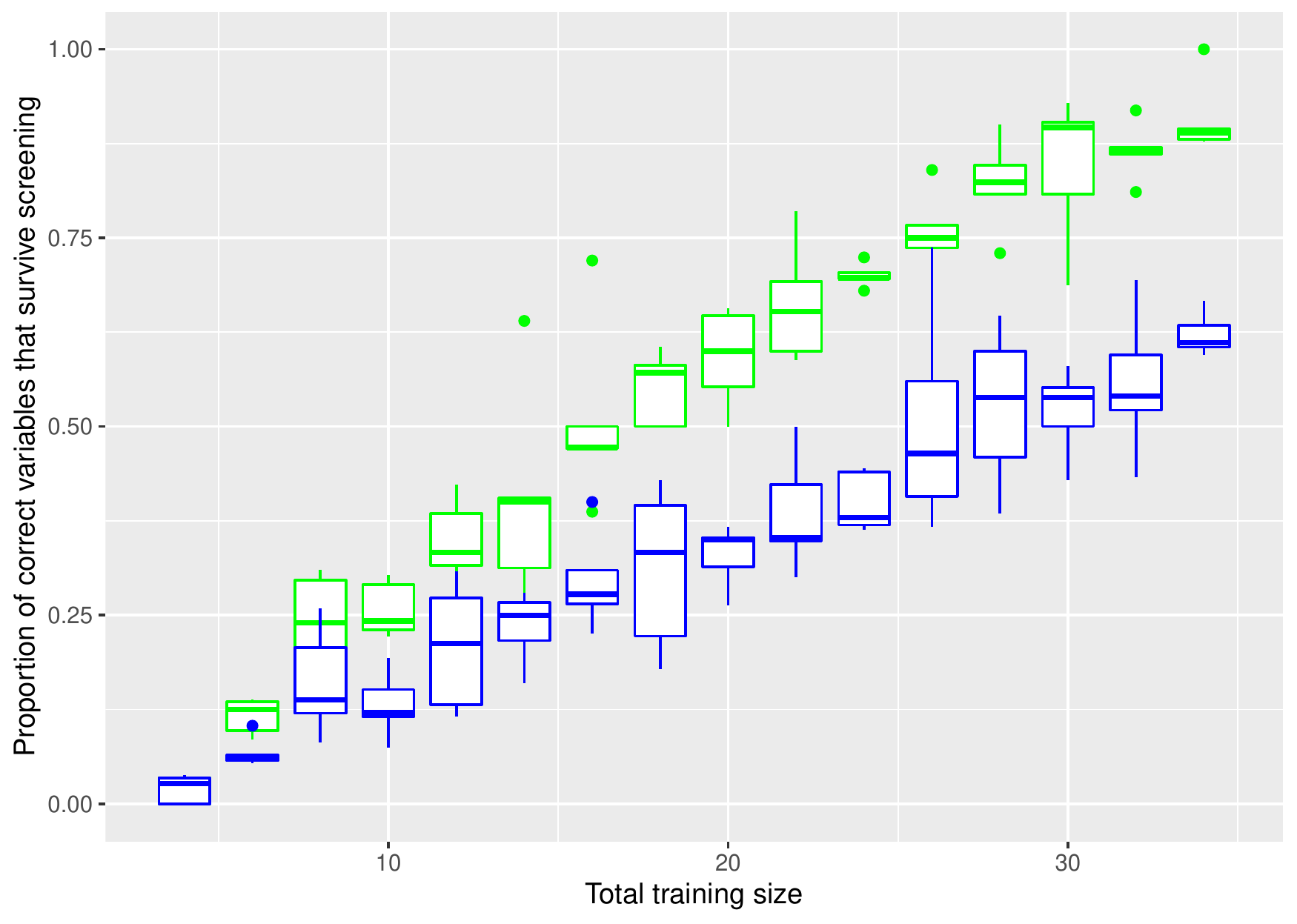}\hfill
    }
    \\[\smallskipamount]
    \makebox[\linewidth]{
    \includegraphics[width=.6\textwidth]{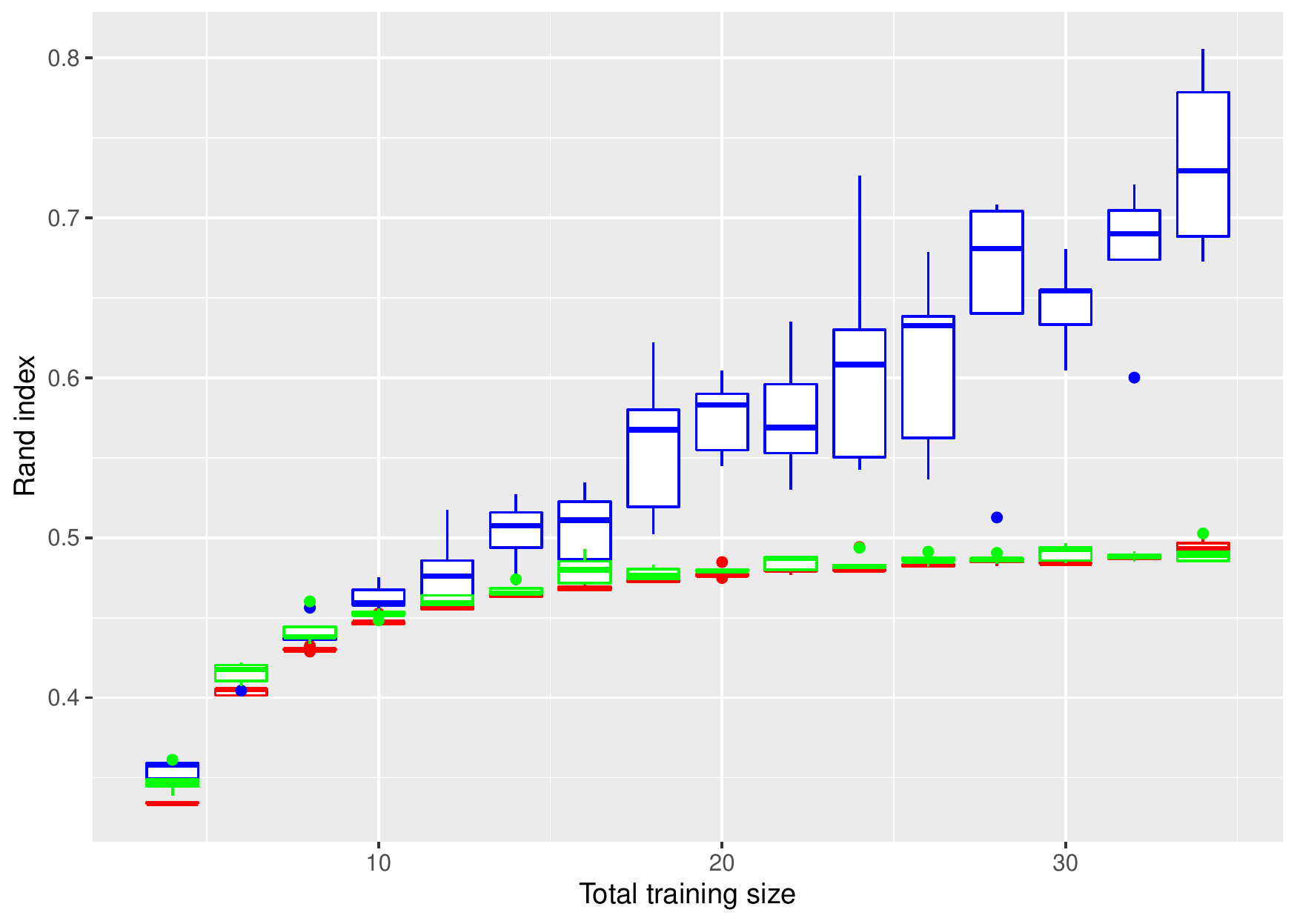}\hfill
    \includegraphics[width=.6\textwidth]{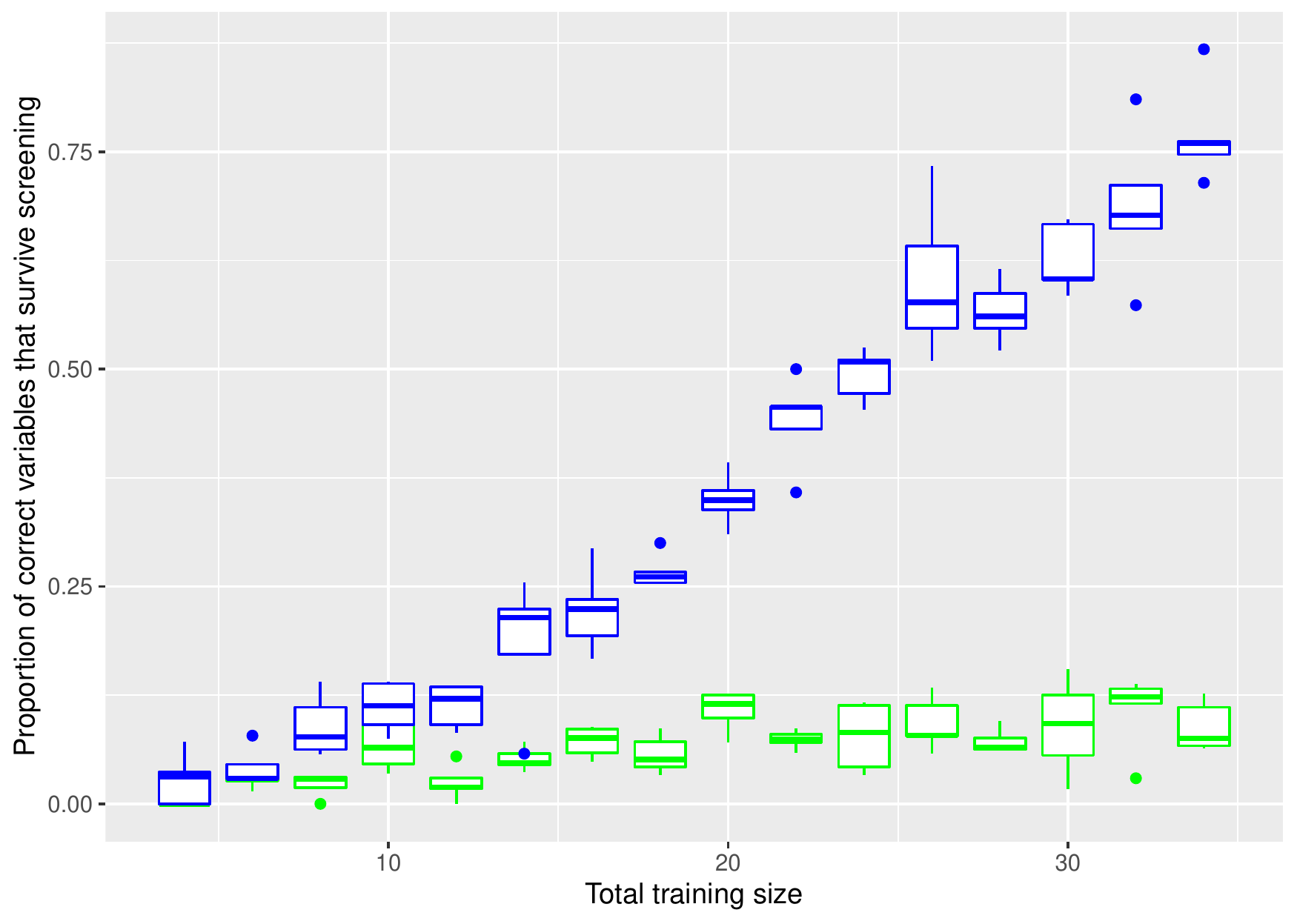}\hfill
    }
    \\[\smallskipamount]
    \makebox[\linewidth]{
    \includegraphics[width=.6\textwidth]{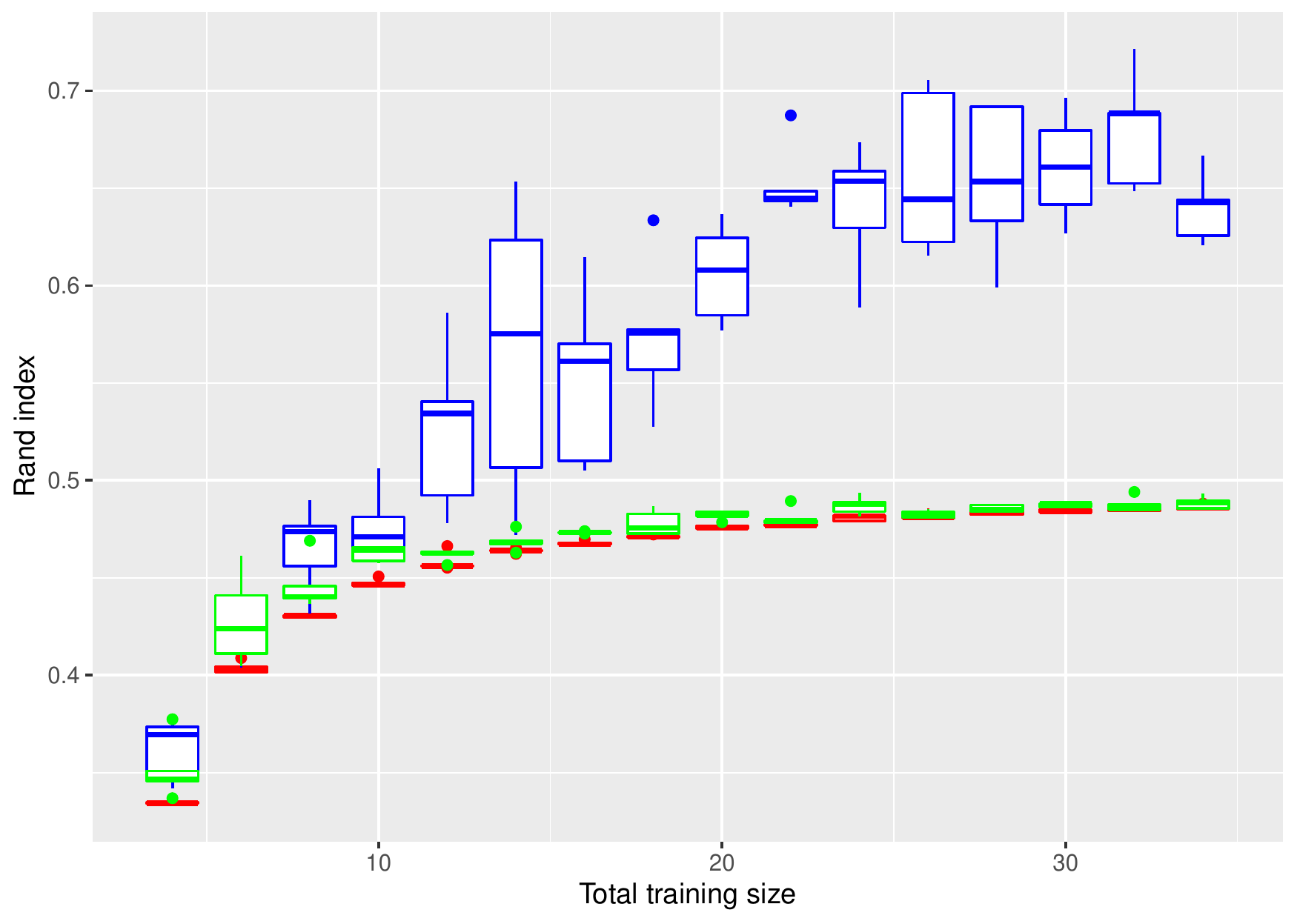}\hfill
    \includegraphics[width=.6\textwidth]{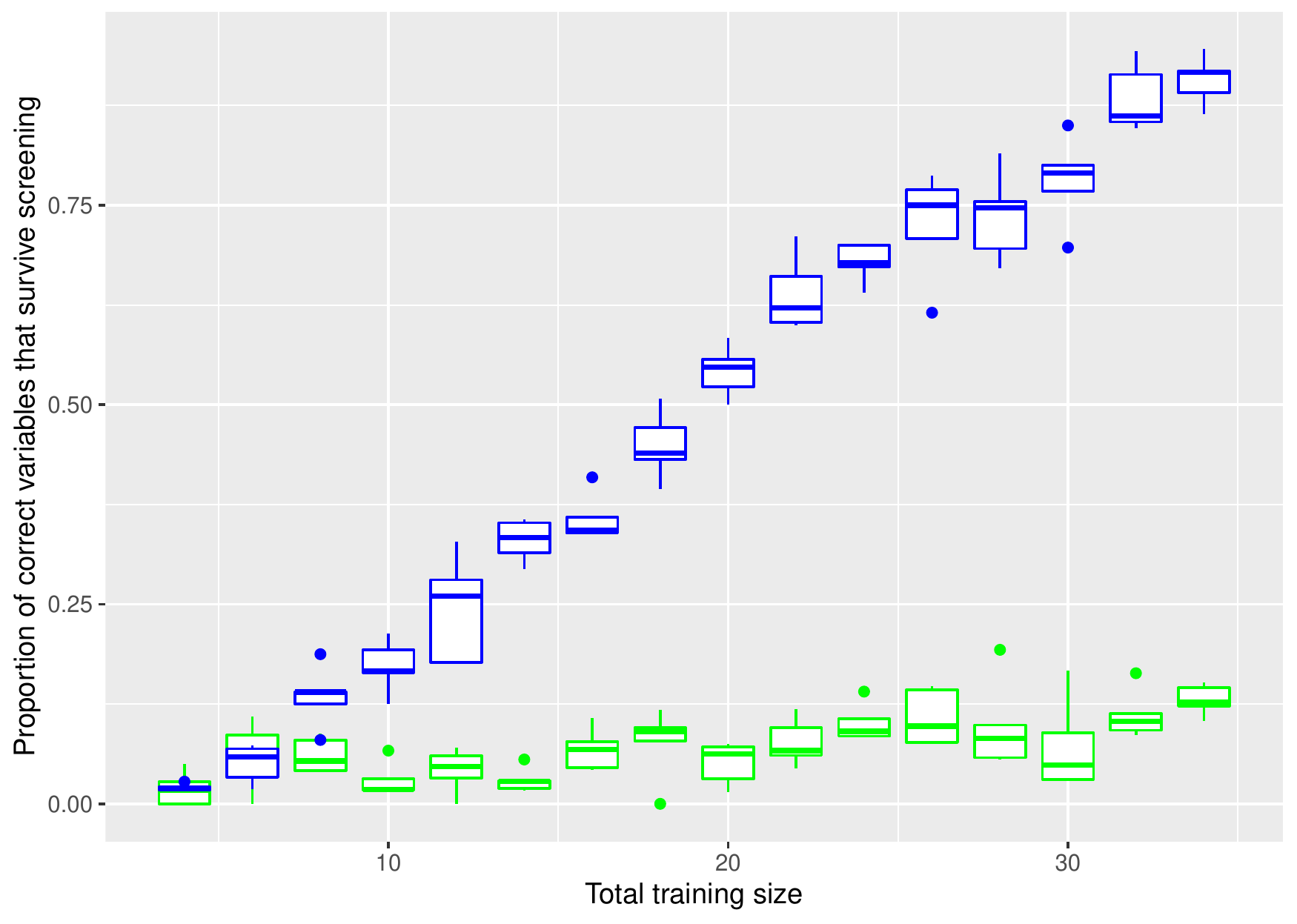}
    }
    \caption{{\it Simulation results for $t$-test screening and ALB screening that uses variables with $n + m$ largest values of ALB.} 
    %The cutoff for the $t$-statistics is generated in the same way. 
    %\tcr{\it Cutoff for $t$-stats will be stated in the setup.} 
    The colors of the box plots have the same meaning as they do in Figure \ref{fig:PermutedScreening}.}\label{fig:QuantileScreening}
\end{figure}

\begin{figure}[p]
    \vspace*{-2.5 cm}
    \makebox[\linewidth]{
    \includegraphics[width=.6\textwidth]{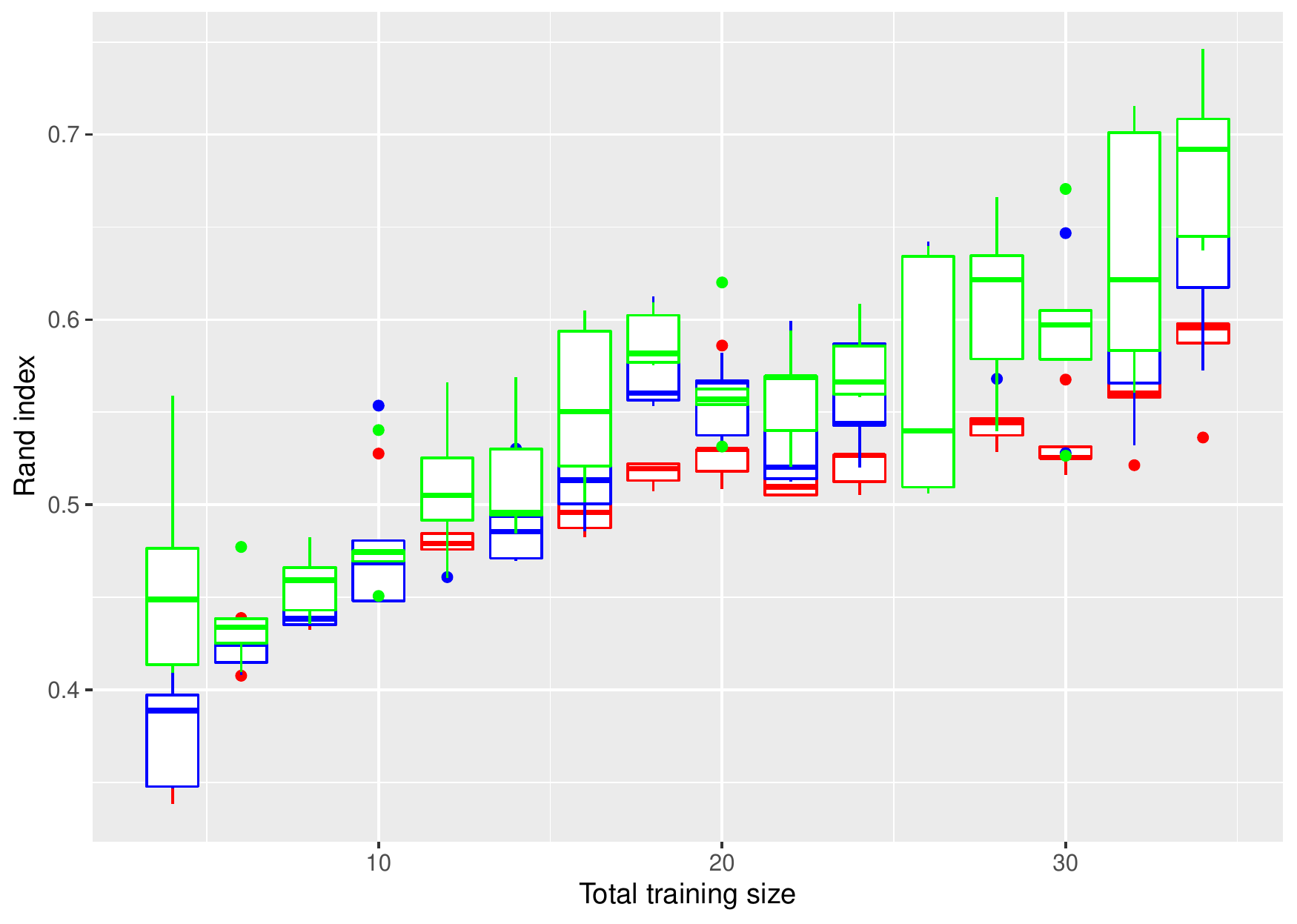}\hfill
    \includegraphics[width=.6\textwidth]{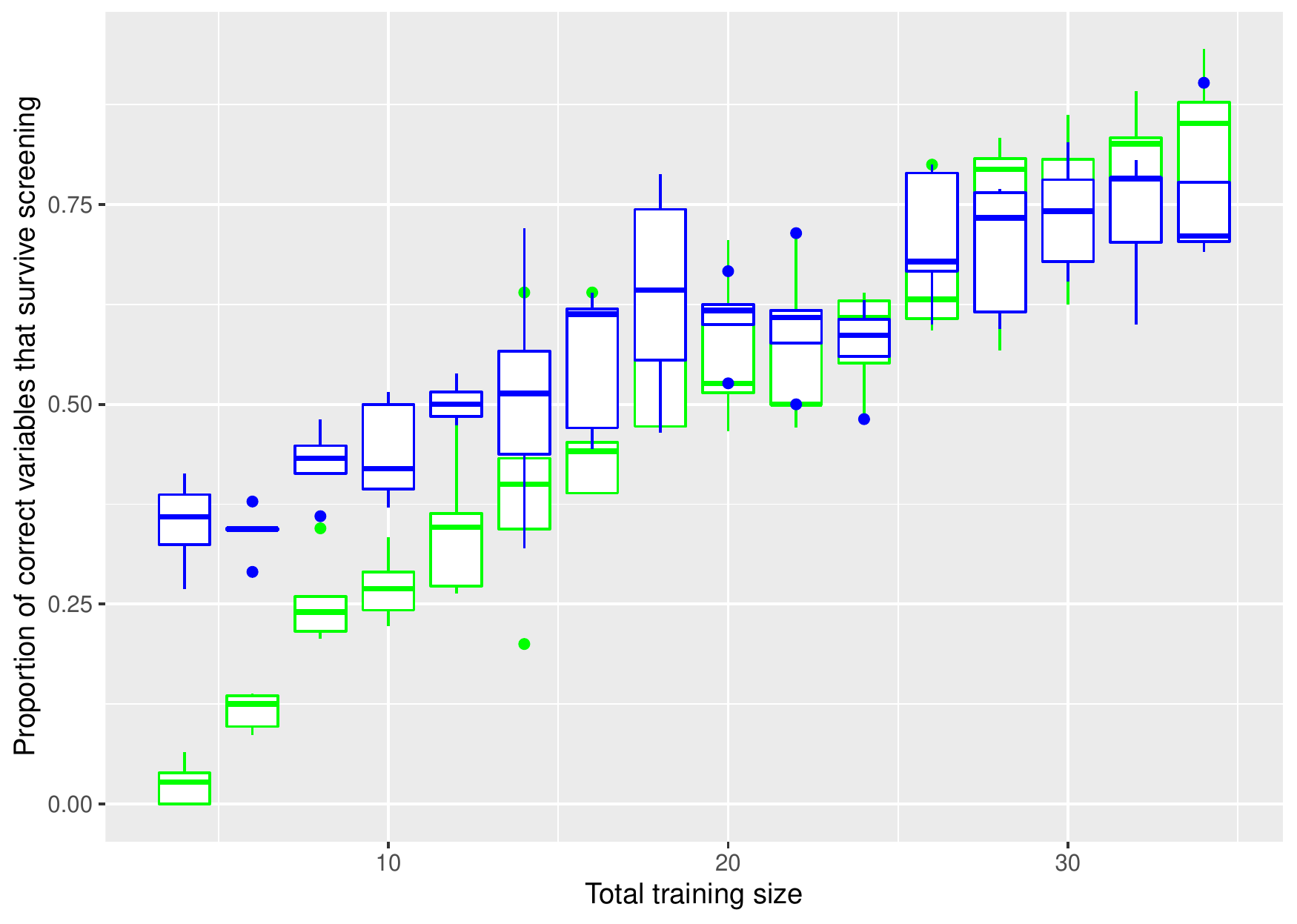}\hfill
    }
    \\[\smallskipamount]
    \makebox[\linewidth]{
    \includegraphics[width=.6\textwidth]{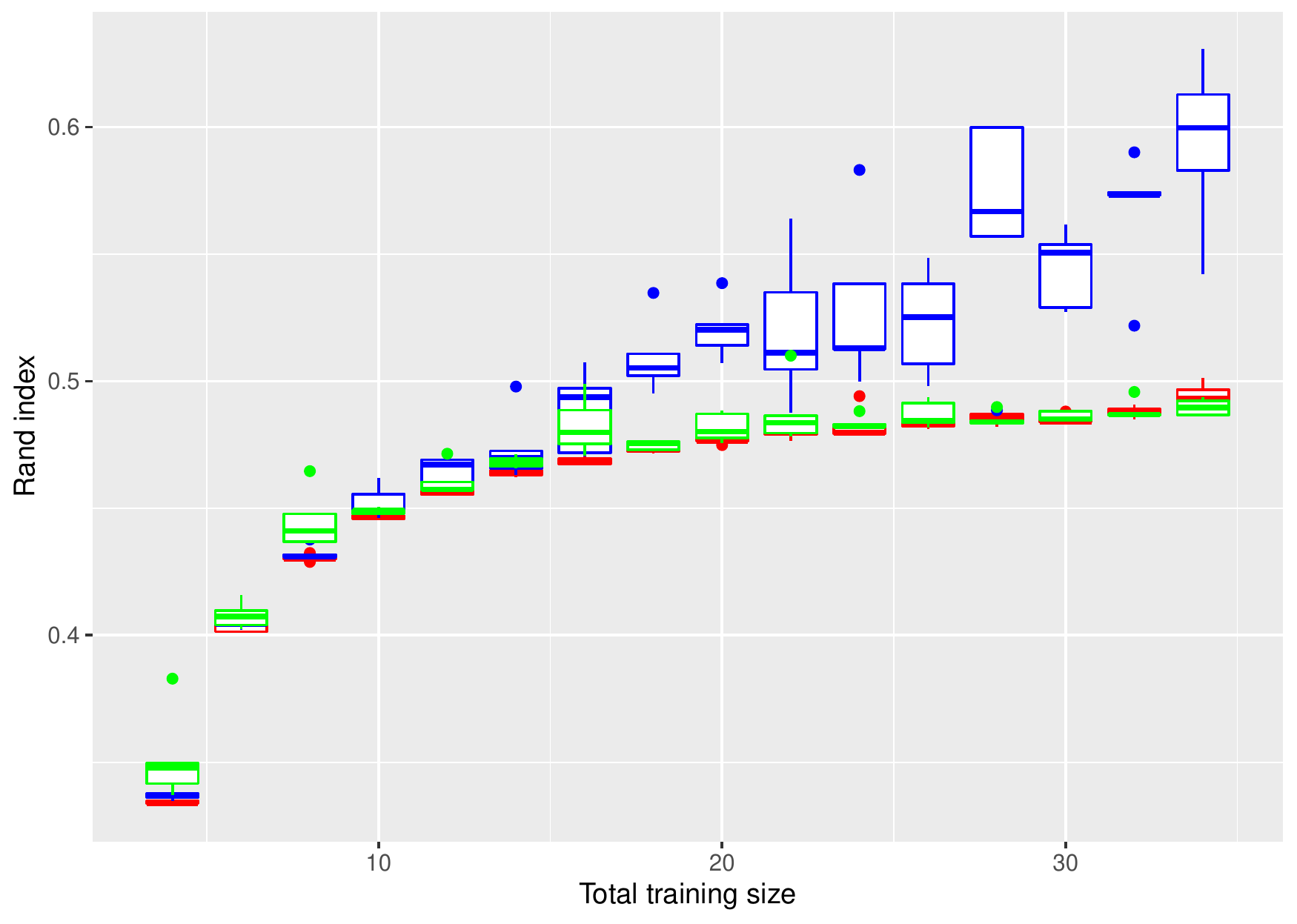}\hfill
    \includegraphics[width=.6\textwidth]{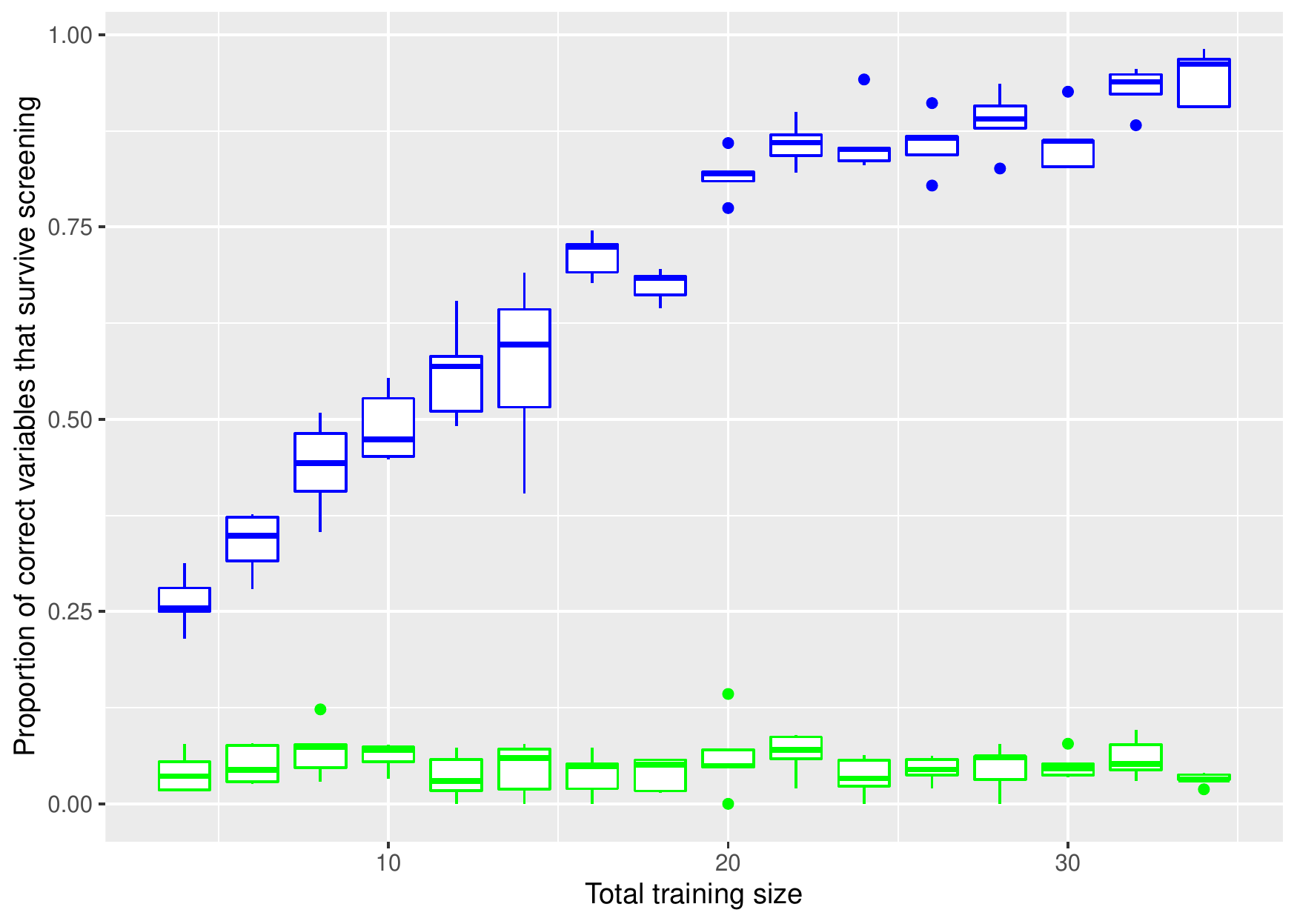}\hfill
    }
    \\[\smallskipamount]
    \makebox[\linewidth]{
    \includegraphics[width=.6\textwidth]{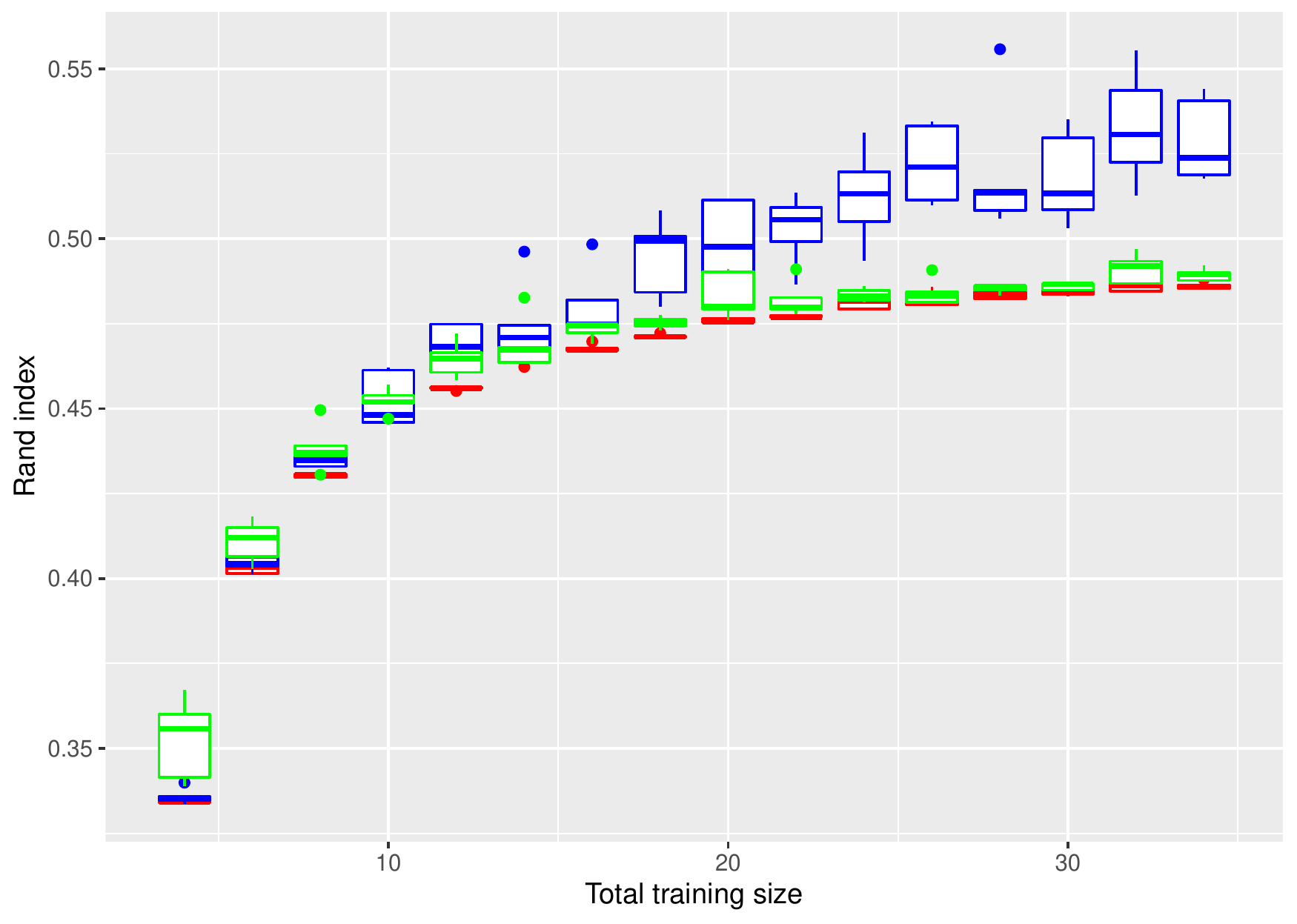}\hfill
    \includegraphics[width=.6\textwidth]{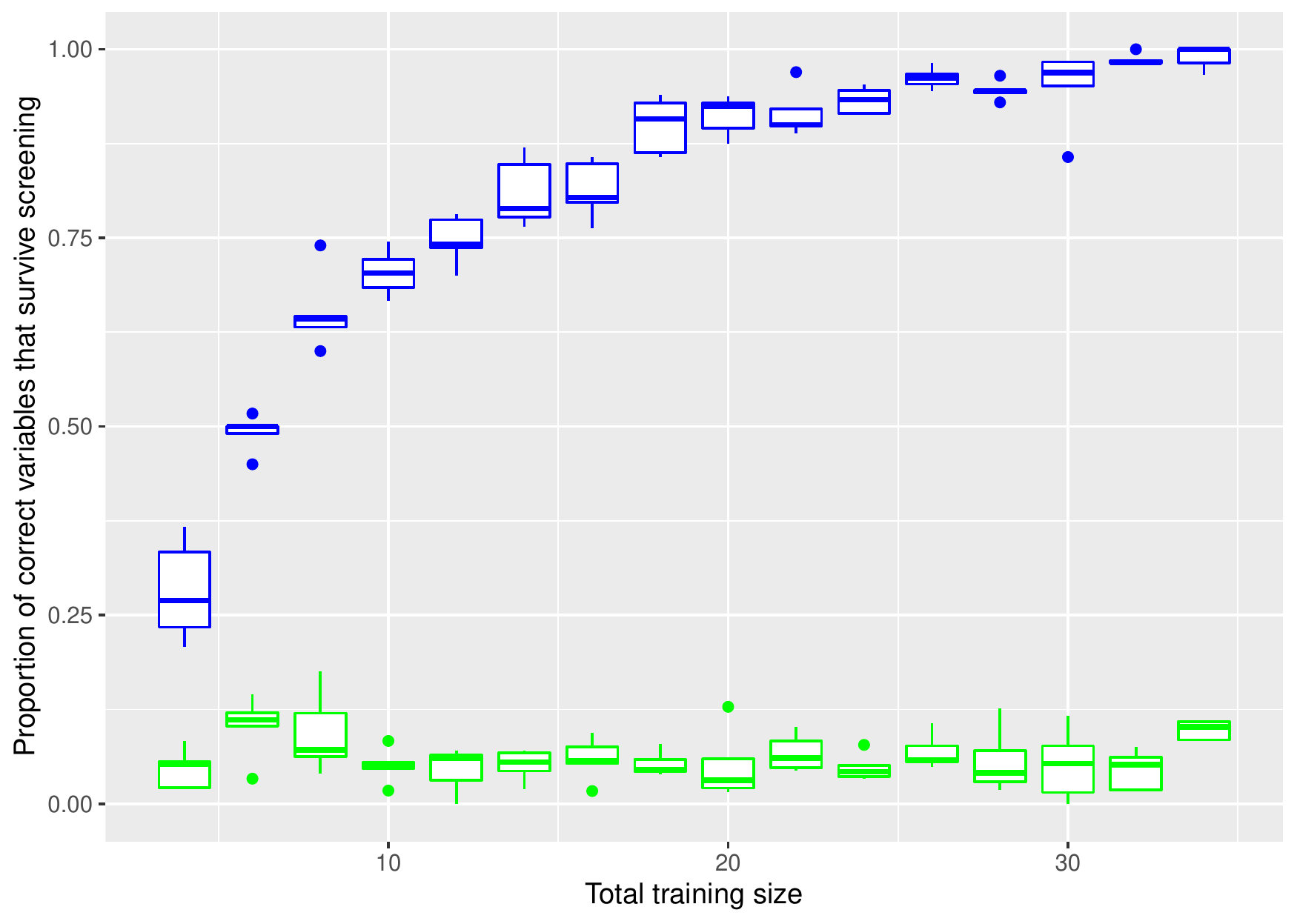}
    }
    \caption{{\it Simulation results for $t$-test screening and ALB screening with cutoff $0$.} 
    %The cutoff for the $t$-statistics is generated in the same way as  \ref{fig:QuantileScreening}. 
    The colors of the box plots have the same meaning as they do in Figure \ref{fig:PermutedScreening}. }\label{fig:InterpScreening}
\end{figure}

\noindent
For all three cases, screening was done and the classifier built from a training set of $m+n$ observations on each variable, where $m=n$. The classifier so built was applied to predict $m+n$ observations, and the resulting Rand index was calculated. In $t$-test screening, variables were selected when their $P$-values were smaller than $0.005$. We repeated this procedure 100 times for each sample size. 
%\tcr{\it As I asked before, do you randomly choose the class for each observation in the validation set? Also, how many reps were performed?} \tcb{\it{ I set it so that training and validation set sizes were equal in this case (no coin was flipped). I added the number of reps.}}
Figures \ref{fig:PermutedScreening}-\ref{fig:InterpScreening} %\ref{fig:QuantileScreening}, and
 show, respectively, how well the methods performed for three ways of choosing an ALB cutoff: a ``fixed type I error rate" approach,  the largest $n+m$ values of ALB, and a cutoff of $0$. 
 %The size of testing set (from which the Rand index was computed) was $m+n$, the same as the training set size for three cases used to compute the Rand index is always $m+n$, the same size as the training set, and $m = n$. 

The results of these simulations suggest that ALB screening is effective at detecting location differences, as in Case 1, but not to the same degree as $t$-test screening. In Case 1, the performance of the SVM with ALB screening is better than with no screening, but worse than with $t$-test screening. The proportion of variables that survive ALB screening steadily increases as sample size increases, but at a slower rate than with $t$-test screening. In Cases 2 and 3, $t$-test screening does no better than no screening in terms of classification accuracy. Regarding preservation of important variables, $t$-test screening does not improve as the sample size increases, but ALB screening does.

\section{Application to the GISETTE data}

The GISETTE data are obtained from  $m=3000$ and $n=3000$ handwritten images of the digits 4 and 9, respectively. For each of the 6000 images, $p=5000$ variables are measured, some of which are irrelevant probes, and the others pixel intensities. 
%\tcr{\it If true, say right here that the 5000 variables are intensities of 5000 pixels, and the intensity may differ depending on whether the digit being written is a 4 or a 9.} \tcb{This isn't completely correct. The data set contains variables where this is the case, and some variables where this isn't the case (the variable is a "probe" or useless). } 
%Some variables are probes and are irrelevant, others correspond to the pixel intensity of some of the image measurements. 
We will perform classification on these data, using different screening methods to choose different subsets of the variables. We will rely on DART to be the primary classification method and will explore how it performs when aided by different screening methods.

The data set was randomly split into two halves. The first half was treated as the training data. We trained our classifier and computed $t$-statistics and ALB statistics on these data. The second half was treated as the validation set, and we computed the Rand index from these data. 
%\tcr{\it This doesn't jibe with the table, which says a quarter of the data were used for training.}

Five screening methods were compared. We implemented A3, setting $B = 1000$ and $d = 1$ and choosing variables such that $ALB$ was larger than the $99.5$th percentile of $ALB^*$ values. 
%\tcr{\it This doesn't sound right.  How about ``We randomly permute labels, choose a variable, compute $ALB$, and repeat 1000 times."} \tcb{\it{I made changes to just refer to A4.}} 
To compare this with $t$-test screening, we picked variables whose $t$-test $P$-values were less than 0.005. We tried screening method A1 and compared it to it's t-test counterpart that picks the same number of variables. We also checked to see how the method did with no variable screening. 
%\tcr{\it If these methods are versions of A1-A7, just say which ones (and with what settings) without going into the previous detail.}\tcb{\it{I implemented this.}} 
We compare the Rand indices of these methods with that when no screening of variables is used. Before proceeding with any of the methods, we removed each variable for which all 6000 data values were the same. As a result there were only 4835 variables in the full data set rather than 5000. %\tcr{\it Not sure I understand what this means. Do you mean that all data values for a variable were the same?}\tcb{\it{This is correct, I updated the description.}}. 
A summary of Rand indices is given in Table \ref{table:RandIndScreenGISETTE}.

\begin{table}
\begin{center}
\begin{tabular}{||c | c | c||} 
 \hline
  Rand Index & Screening Method & Number of variables chosen  \\  
 \hline
% .748 &  top $\text{log}(n + m)$'th ALBs chosen &  8\\ 
% \hline
  0.947 &  $ALB> T^*_{0.005}$ & 1321 \\
   \hline
  0.942 &  $ALB> 0$ & 1946 \\
 \hline
% 0.750 & $ALB> \text{log}(1.2)$ & 4 \\ 
% \hline
%  .603 & $\text{log}(n + m)$'th t-statistic  chosen  \\ 
% \hline
  0.947 & $P_t< 0.005$ & 1540 \\ 
 \hline
 0.935 & No screening & 4835 \\
 \hline
\end{tabular}
\end{center}
\caption{{\it Classification and screening results for GISETTE data.} All methods used a balanced training and testing set that both consisted of 3000 observations. The quantities $P_t$ and $T^*_{0.005}$ are, respectively, the $P$-value of a $t$-test and the 99.5th percentile of permuted $ALB$s.} %\tcr{\it Say what the actual sizes were. Also, see my query about size of training set. Out of curiosity, what was the 99.5th percentile?} \tcb{\it{.001451}}
\label{table:RandIndScreenGISETTE}
\end{table}

\begin{figure}[h!]
\centering
\includegraphics[width=.6\textwidth]{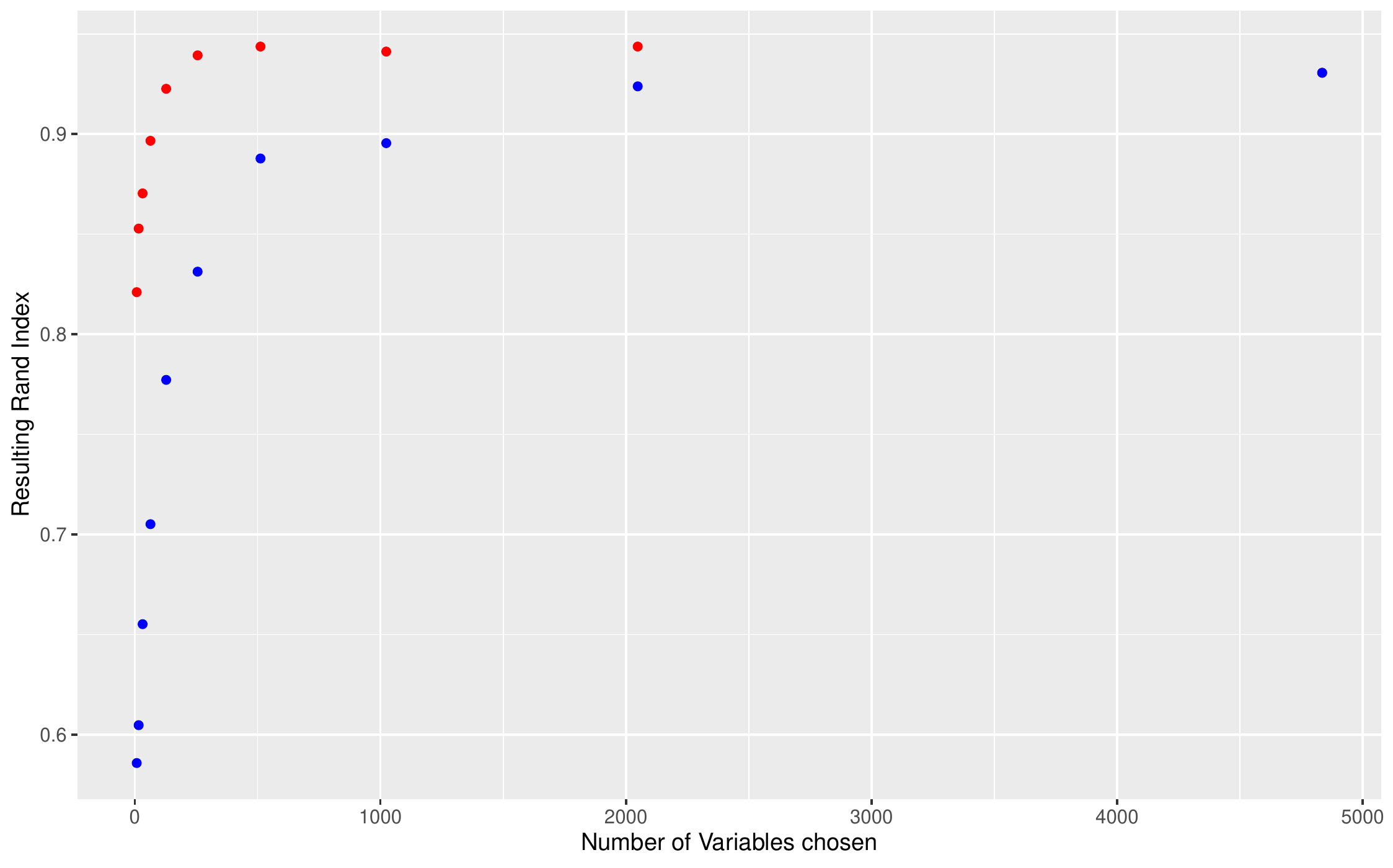}
\caption{\textit{ Rand index for the GISETTE data as a function of number of variables used.} The red points correspond to ALB screening and blue points to $t$-statistic screening. Rand index is computed for a DART classifier using only those variables having the largest $2^j$ $t$-statistics or the largest $2^j$ values of $ALB$, where $j=3,\ldots,11$. }
\label{fig:RandIndGISETTETvsALBQ}
\end{figure}

For these data, $t$-test screening 
%based on a $P$-value cutoff of $0.005$
does as well as 
%the analogous method with 
ALB screening based on A4 and the 99.5th percentile. 
%where we pick variables that are bigger than the $99.5\%$ of the permuted ALBs.
%\tcr{\it Repeating the settings for each method is redundant.  Establish the cutoffs in the setup and then just say ``$t$-test screening does as well as permuted ALB screening," or something like that.} 
The difference between the two methods is that ALB retains the same accuracy while picking roughly 200 fewer variables. The interpretable cutoff rule does the worst, but 75\% accuracy using only four ``pixel" measurements is a very interesting result.  
%\tcr{\it This suggests the possibility of high correlations between variables.  Have you checked this?} \tcb{\it{I think some of the variables are correlated, and it makes sense, one pixel being colored or large naturally increases the chance that some surrounding ones are. The correlation matrix is a bit hard to show though.}}
We believe this is a setting where most variables, or "pixels," that are marginally important are ones that are colored in for one of the two numbers (4 or 9) but not the other. 
This can be interpreted as a location difference, since the intensity of a colored-in pixel is larger than the intensity of a pixel that is rarely touched. We also believe this is the reason why $t$-test screening finds more important variables than does ALB screening based on the same type I error rate. Despite being a setting where mostly location differences exist, ALB ends up doing as well as $t$-test screening in terms of Rand index, at least when using a type I error rate of $0.005$.  In general we believe that choosing a cutoff based on a type I error rate or using 0 as a cutoff are good strategies for data sets where $n$ and $p$ are both large.
A large amount of data allows us to choose fairly small significance thresholds while still maintaining good power.
Choosing variables that have the largest $n + m$ values of ALB in this case results in no variables being screened, and so we elect not to explore that avenue. A cross-validation procedure for selecting a cutoff is expensive to perform due to the large values of $p$ and $n+m$. 
%\tcr{\it I broke the sentence up as I don't see the connection between CV and the type I error rate issue.}
%it being time consuming to fit large models.
 %\tcr{\it I'm not sure I follow. How are you defining "type I error rate issue?"  Are you saying that you can still choose a small cutoff (which is good for power) while still having a small type I error rate?} \tcb{\it{I'm trying to say that you can choose a small cutoff that corresponds to a low type I error rate by the permutation idea, and it's all right because the large amount of data implies that we have high power. Normally doing per test with small type I error rates is ideal but there's the issue that the familywise error rate is probably large. This is normally resolved by using a sparse classifier afterwards (BART / DART), since we can pick smaller thresholds, this puts "less work" on the sparse classifier to get the correct variables right. I dropped the argument a little bit in the end.}}

To explore the impact of choosing a cutoff based on quantiles, we explored the Rand indices when the cutoff corresponded to using variables with the largest $k$ statistics. We considered values of $k$ that increased geometrically: $k=8,16,\ldots,2048$.  The results can be seen in Figure \ref{fig:RandIndGISETTETvsALBQ}. 
%\tcr{\it Is this really true? It seems like the largest Rand indices in Fig.~18 are about the same as they are for the first two ALB-based procedures in Table 1.} \tcb{\it{I'll add when we pick only a few variables instead, I agree with your thoughts.}} 
%It is seen in Figure \ref{fig:RandIndGISETTETvsALBQ} that, 
At each number of variables used, ALB-based screening has a larger Rand index than does $t$-test screening. Clearly, ALB and $t$-test screening are not choosing the same variables, and the ones chosen by ALB are more effective. 
%We believe this is because our method has a harsher standard that's needed to accept a variable as relevant.
%\tcr{\it I don't understand what the last sentence means. What do you mean by harsher standard? Could it be that the variables chosen by ALB are just better, maybe because ALB is using more of the information (i.e., not just location information)? Actually, it would be pretty interesting to see what kind of overlap (or lack thereof) there is between the sets of variables chosen by the two screening methods.} \tcb{There isn't perfect overlap, some of the variables found are the same, but a fair portion are not. My understanding is we discover location differences that are larger than that the t-test screening picks up, as well as other differences that are similar but not of a location type difference. In other words, if we pick the top number of variables, I think we will get some of the variables that differ because of location differences but fail to pick up ones with smaller location differences (t-test screening picks these up instead). Instead, we have other variables that differ in other ways.}

\section{Application to Leukemia data } \label{Leukemia}

The Leukemia data set contains observations on 72 patients, 47 of which have one type of leukemia and the remainder another type. We have observations of 7129 variables on each patient to build a classifier that will help decide which type of leukemia a future patient has. This is a case where $p$ is much larger than $n$ and $m$. We will apply various screening methods to this problem in conjunction with DART and compare the accuracy of the methods via Rand indices.

To do this in a fair fashion, we split the data set randomly into two halves. The first half are validation data, and the second half are training data. 
%\tcr{\it Didn't you do the same thing with the GISETTE data?  If so, that would be where you'd want to make the ``fair" comment, and here you could say we did as in previous section.}\tcb{\it{I think I only did this for the leukemia data set, I'm not sure why I wrote that a quarter of the data set was used for training in the GISETTE part.}} 
We then split the training data in half again to make two smaller training sets. 
%from the big training data set. 
We do this for two reasons. First, we wish to assess the effect of training set size on accuracy of the methods. One of the two smaller training sets will be used to build classifiers, each one corresponding to a different screening method, and then all the training data will be used to build another set of classifiers. Both sets of classifiers will be used to predict the data in the validation set.  The second reason for dividing the training set in half is that it makes possible a cross-validation approach for selecting a cutoff. We can train the model on one of the smaller training data sets, and choose a cutoff that gives the best classification accuracy on the other training data set. We can then train this best model on the full training set and apply it to the validation set. This is applying strategy A2 in the methodology, which is feasible because of the small sizes of $m$ and $n$.

To carry out our analysis we did the following. We performed four ALB based screening methods and two $t$-test based screening methods. The four ALB methods were A1 with the $n+m$ largest $ALB$s being selected, A2 with the cutoff of $0$, and A3 with a type I error rate of $0.05$, $d = 3$ and $B = 7129$. 
\begin{comment}
the method where we declare the top $n + m$ ALBs as important variables, the method where we permute the labels of the data and observe the ALB under a permutation, and declare all ALBs greater than $95\%$ of the permuted ALBs as relevant, and the method where we use cross validation to declare an ALB as relevant.
\tcr{\it Again, make the description more concise by just referring to AX.}\tcb{\it{This is done.}} To make cross-validation more efficient timewise, we used the Bayesian classifier based on (\ref{eqn:BNEWestimator}). We applied the first three ALB-based screening methods, first to one of the small training sets, and again to the full training data set. We applied the cross-validation approach using one of the small training sets to train the classifier based on (\ref{eqn:BNEWestimator}) and then determined which cutoff worked best using the other training set. Finally, we used this best cutoff to perform ALB screening using the full training set.
\end{comment}

Two methods of $t$-test screening were used, one using the variables with the $m + n$ largest $t$-statistics, and the other using variables whose $t$-test $P$-values were smaller than $0.05$. The latter version of $t$-test screening makes it comparable to choosing a variable using method A3 with significance level $0.05$. 
Once we determined relevant variables via screening, DART based on those variables was used to compute a Rand index from the validation set. Tables  \ref{table:RandIndScreen} and \ref{table:RandIndScreen2} summarize the results.

\begin{table}
\begin{center}
\begin{tabular}{||c | c | c ||} 
 \hline
  Rand Index & Screening Method Used & Number of variables\\  
 \hline
 0.599 &  $n + m$ largest $ALB$s& 19 \\ 
 \hline
 0.529 &  $ALB>T^*_{0.05}$ & 617 \\
 \hline
% 0.549 & $ALB>\text{log}(1.2)$ & 233 \\ 
% \hline
  0.599 & $n + m$ largest $t$-statistics & 19 \\ 
 \hline
  0.529 & $P_t<0.05$& 847 \\ 
 \hline
 0.501 & No screening & 7129 \\
 \hline
\end{tabular}
\end{center}
\caption{\textit{Classification and screening results for leukemia data when training set is a quarter of full set.} All results are based on a validation set size that was roughly half that of the full data set. The quantities $P_t$ and $T^*_{0.05}$ are, respectively, the $P$-value of a $t$-test and the 95th percentile of permuted $ALB$s. The classifier used was DART.}
\label{table:RandIndScreen}
\end{table}

\begin{table}
\begin{center}
\begin{tabular}{||c | c | c ||} 
 \hline
  Rand Index & Screening Method & Number of variables\\ 
  \hline
  0.742 &  $ALB>T_{{\rm CV}}$ & 12 \\
 \hline
% 0.834 &  $ALB>\text{log}(1.2)$ & 150 \\
% \hline
 \hline
 \hline
 \hline
 0.742 &  $ALB>T_{{\rm CV}}$ & 12 \\
 \hline
 
 0.786 &  $n + m$ largest $ALB$s & 38 \\ 
 \hline
 0.572 &  $ALB>T^*_{0.05}$ & 1302 \\
 \hline
% 0.742 & $ALB>\text{log}(1.2)$ & 150 \\ 
% \hline
  0.598 & $n + m$ largest $t$-statistics & 38 \\ 
 \hline
  0.572 & $P_t<.05$& 1694 \\ 
 \hline
 0.549 & No screening & 7129 \\ 
 \hline
\end{tabular}
\end{center}
\caption{\textit{Classification and screening results for leukemia data when training set is half of full set.} All results are based on a validation set size that was roughly half that of the full data set. The first two rows of the table correspond to use of the classifier based on (\ref{eqn:BNEWestimator}), and subsequent rows to use of DART. The quantity $T_{CV}$ is the best cutoff as chosen by cross-validation, and $P_t$ and $T^*_{0.05}$ are as in Table 2. See Section \ref{Leukemia} for an explanation of how cross-validation was implemented.} 
%\tcr{\it See if you can put a little more space after first two rows to make the distinction more obvious.}
\label{table:RandIndScreen2}
\end{table}

All screening methods performed similarly when the training set size was a quarter of $m+n$. 
%the size, and the permutation based method for an ALB cutoff always did roughly the same as the t-test based method with a p-value based cutoff. 
However, $ALB$ screening that chose a cutoff as in $A1$ or $A3$ improved remarkably when the sample sizes were doubled, faring much better than the $t$-test based screening methods.
%\tcr{\it Is the quantile method the top $n+m$ or the one based on percentile of permuted ALBs? Again, this is why it would be good to have a concise method of referring to methods. What seems most logical is to refer back to AX.} \tcb{\it{I editted it to refer to the method instead}} 
While DART has been shown to be an effective classifier, our experience is that
%the sample sizes being as small as they are there is 
it may fail to recover the structure of the classification problem when the sample size is small. 
%\tcr{\it Do you have a reference for this claim?  What makes it an ensemble method and why is this bad for small sample sizes?} \tcb{\it{I don't really have a reference for it being bad for small sample sizes, I have a reference for it being an ensemble method (it is an average of many "weak learners"), I could just remove this since its just speculation, my experience with tree based methods is you need enough data to get a good tree.}} 
We thus tried the Bayesian classifier based on (\ref{eqn:BNEWestimator}) with the CV-based method to choose a cutoff, and this classifier was able to achieve decent classification accuracy, surpassing DART if a different cutoff is chosen. 
%\tcr{Prior to this point, give the classifier a name, such as Bayesian classifier, so that you don't have to describe it every time.} 
We believe this is the case due to its simplicity, and there should be enough data to construct reasonable kernel density estimates of the underlying distributions.  
%\tcr{\it I wonder if it's because the variables are near independent in this case, which would make a classifier based on (4) near optimal.}\tcb{\it{I'm not completely sure what the full reason is to be honest. I think even if there are correlated variables its hard to notice them and leverage it in the model due to how big the samples are.}} 

\begin{figure}[h!]
\centering
\includegraphics[width=.8\textwidth]{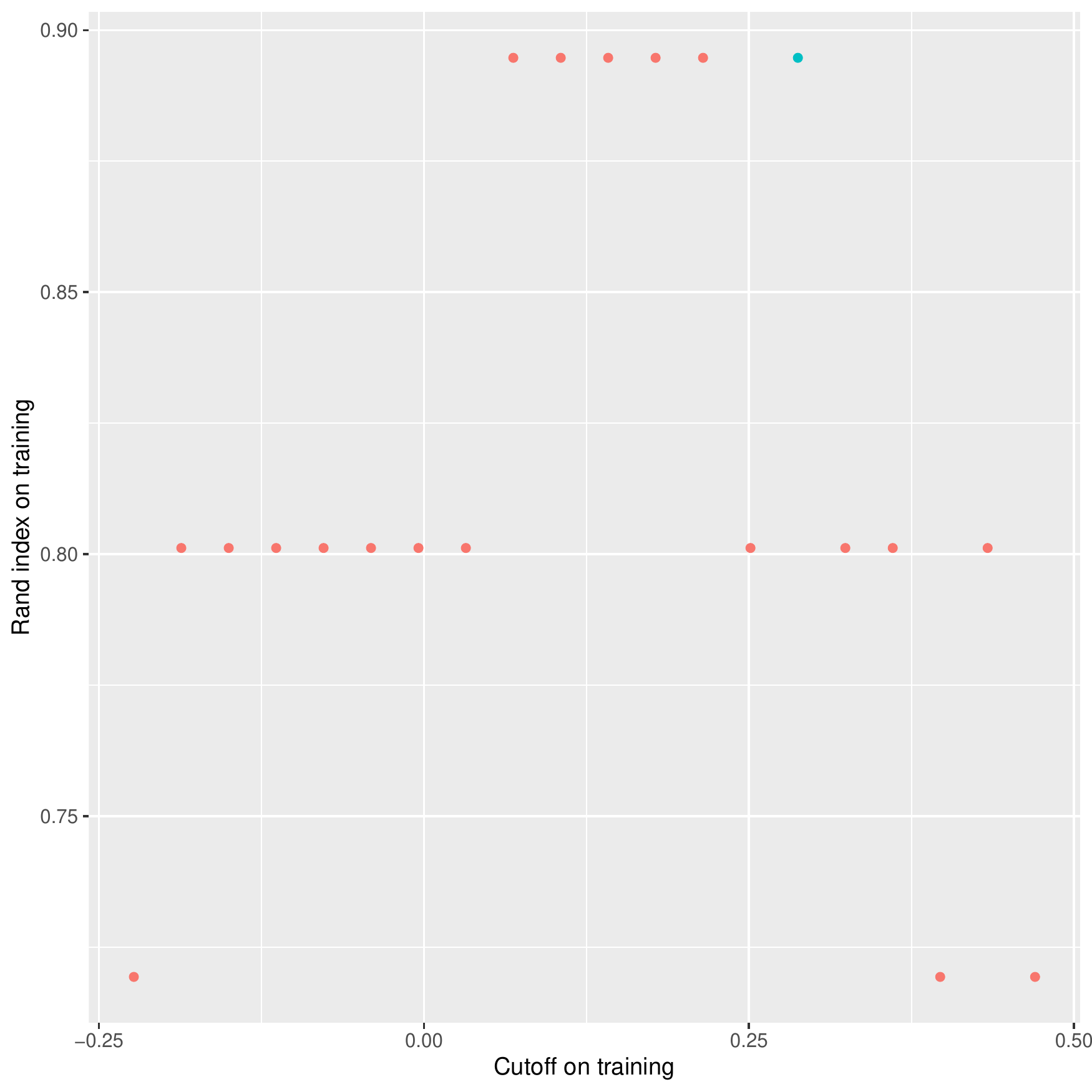}
\caption{\textit{Cross-validation performance of the Bayesian classifier on the training set of the leukemia data.} A plot of the Rand index of the method corresponding to  (\ref{eqn:BNEWestimator}) against the number of variables chosen by the method. The Rand indices are the performance of the classifier on one of the training sets, applied to the other training sets. We chose the cutoff that works best by picking the cutoff (in blue) that preserves the fewest variables in a sequence that maximizes the Rand index.}
\label{fig:RandIndNW}
\end{figure}

\begin{figure}[h!]
\centering
\includegraphics[width=.8\textwidth]{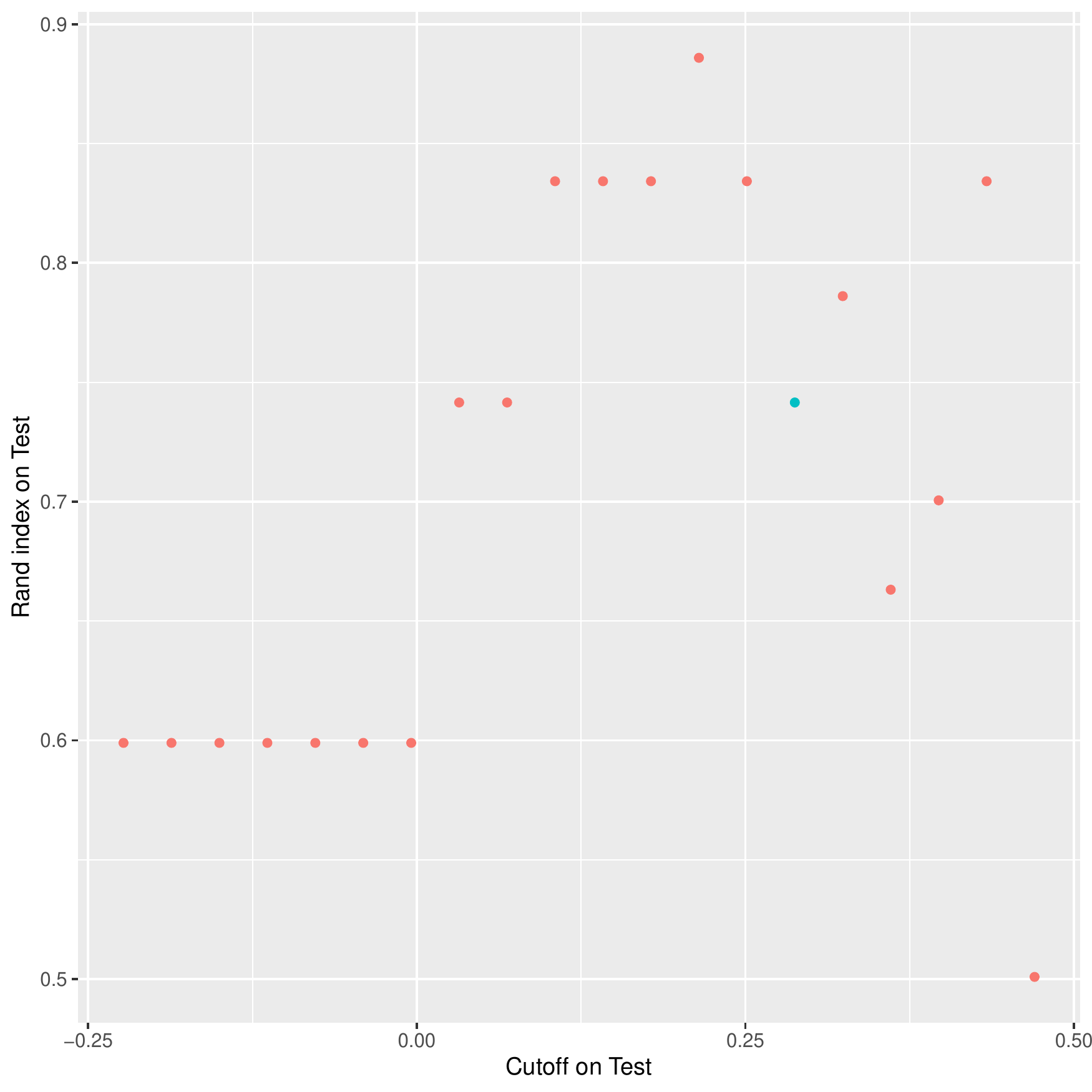}
\caption{\textit{Cross-validation performance of the Bayesian classifier on the testing set of the leukemia data.} The Rand indices are on the validation set. The cutoff we chose for cross validation is in blue \tcr{??} and was selected by examining Figure \ref{fig:RandIndNW}.}
\label{fig:RandIndNWC}
\end{figure}

Figure \ref{fig:RandIndNW} shows the cross-validated Rand indices of the Bayesian classifier as a function of cutoff.  It turns out that the largest cutoff maximizing the Rand index was .288. (The largest cutoff was chosen since this corresponds to the smallest number of variables maximizing the Rand index.)  Figure
%\ref{fig:RandIndNW} and  
\ref{fig:RandIndNWC} shows Rand indices of the Bayesian classifier that was trained on the full training set (i.e., the training set using half of the full data set). This figure shows how well the CV cutoff fared when it was used to screen variables in the full training set. Figure \ref{fig:RandIndNW} shows how the number of variables chosen is related to cutoff. Finally, Figure \ref{fig:RandIndNWC} also shows how sensitive the Rand index is to the selected cutoff. 

\begin{figure}[h!]
\centering
\includegraphics[width=.8\textwidth]{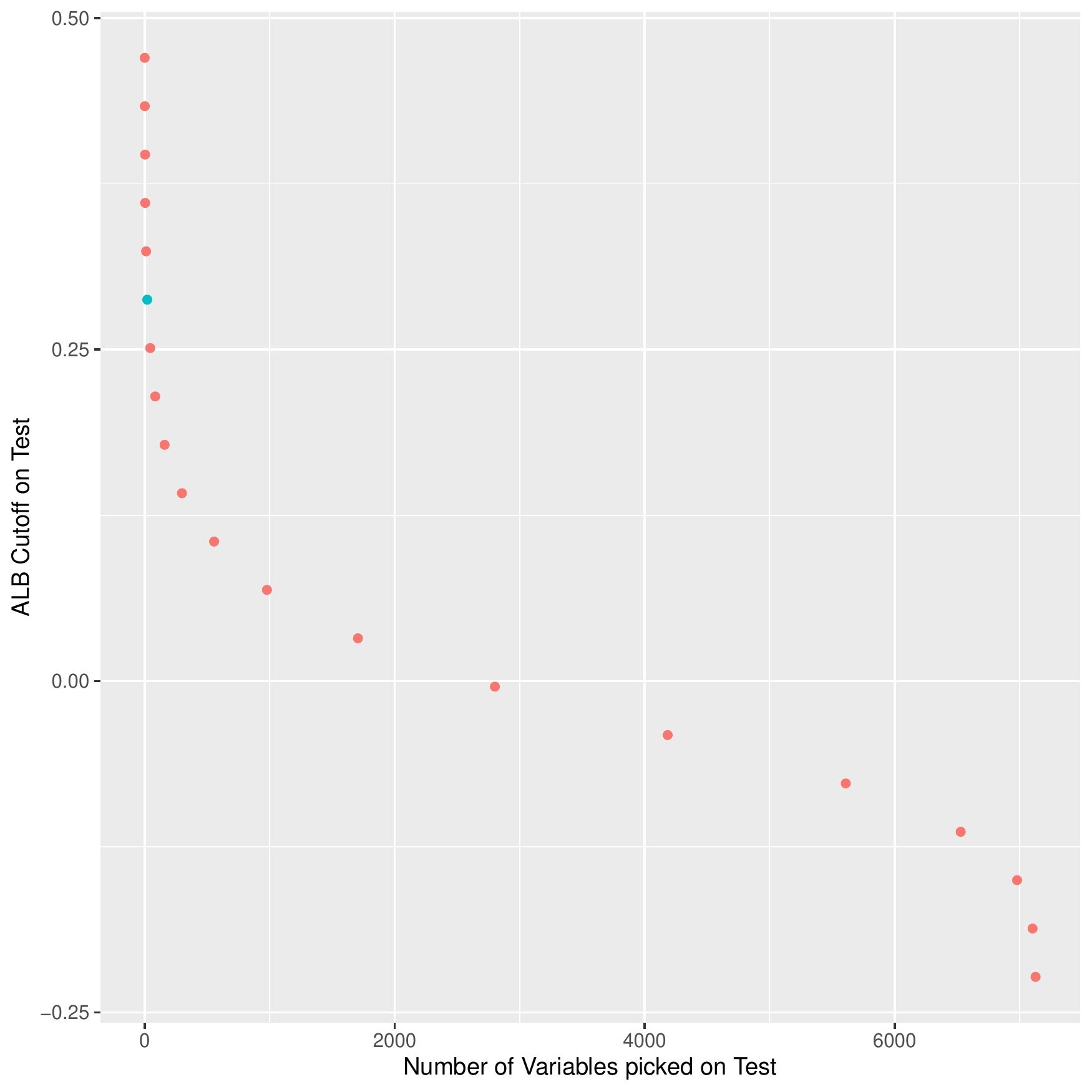}
\caption{\textit{Number of variables picked versus the cutoff chosen using the ALB screening method on the Leukemia test data.}}
\label{fig:RandIndNWT}
\end{figure}

We believe the Bayesian classifier has potential in other settings where there may not be sufficient data to train ensemble methods and there may exist differences between classes that are not of location type. We note that, like other methods, it's best to apply screening before using the Bayesian classifier, as all of their performances will degrade if there are a large amount of variables that are not useful. It is encouraging that the cutoff of $0$ was the largest cutoff before large improvements in Rand index occurred for the Bayesian classifier, but somewhat discouraging that the cross-validation approach could not pick a cutoff that resulted in the best performance for that classifier.
%in \ref{eqn:BNEWestimator}. 
The problem here is that the cross-validation approach chose an optimal cutoff when the classifier was constructed from one quarter of all the data, whereas we actually needed to know the optimal cutoff when half of all the data were used.
Future research can focus on how an optimal cutoff depends on training set size.  If the dependence is simple enough, it may be possible to estimate the optimal cutoff for a given training set size from cross-validation results based on a smaller  training set size.
%improved ways to select the ideal data set to be used with the classifier in \ref{eqn:BNEWestimator}.

\begin{figure}[h!]
\centering
\includegraphics[width=\textwidth]{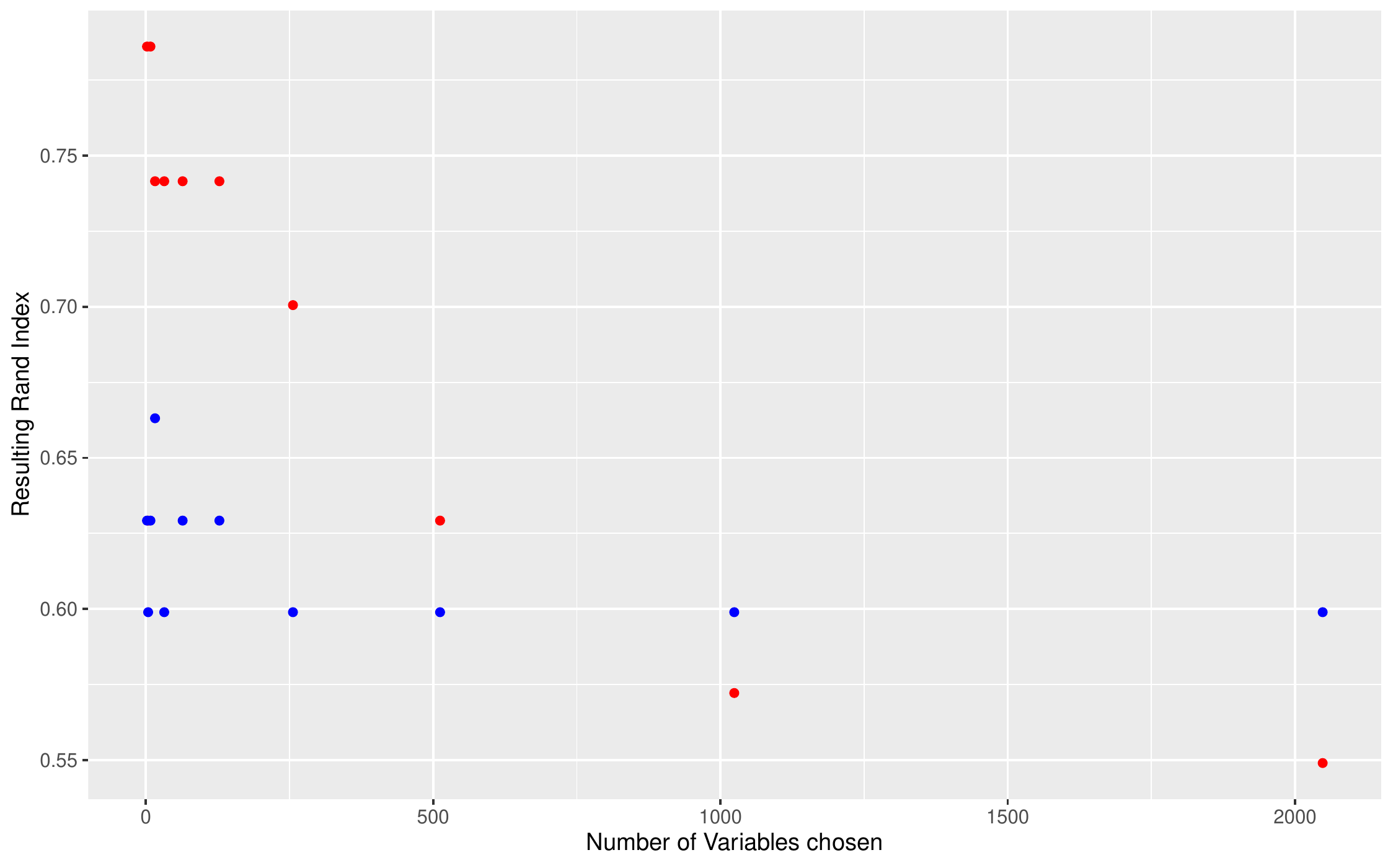}
\caption{\textit{DART Rand indices versus number of variables chosen for the Leukemia data set.} We plot the Rand index of the DART methods where we choose relevant variables corresponding to the top number of ALBs or $t$-statistics. We vary the number of variables chosen. The red points denote the Rand index of DART models using the top number of ALBs, while the blue points denote the Rand index of DART models using the top number of $t$-statistics.}
\label{fig:RandIndLeukTvsALBQ}
\end{figure}

ALB screening tends to do better when using the $k$ variables having the $k$ largest $ALB$s. Figure \ref{fig:RandIndLeukTvsALBQ} shows the Rand index resulting from applying DART after using this method of screening. When using fewer than 500 variables, ALB screening does a better job than the analogous way of performing $t$-test screening. 
%and it does pick the variables that produce the best rand index. 
After the number of variables included is large enough, $t$-test screening does better than ALB screening, but at this point the number of variables included is large enough that the Rand index becomes suboptimal for both types of screening. 
%It can be argued that ALB screening is more sensitive to the cutoff chosen in this case, but we argue that if the cutoff chosen is poorly for both methods, it is probably not the best way to apply the screening method anyway. \tcr{\it You don't always have to try to put the competing method in a good light. I don't see why sensitivity to cutoff is a bad thing. In fact, I might argue that it's a good thing because it makes it easier to pinpoint optimal cutoff when you're using something like CV.}

\section{Conclusion and future work}

We have proposed a new screening method that searches for differences other than those of location type. For this method to be more effective than $t$-test screening it needs to be paired with classification methods that can leverage these differences. In simulations, we pair ALB screening with BART, DART and a Bayesian classifier and show that it performs better than $t$-test screening in situations where class differences are not of location type.  The Bayesian classier outperformed DART when applied to a leukemia data set. Even if the data contain primarily location differences, ALB screening performs well, although, as expected, not as well as $t$-test screening.

Future work includes efforts to increase the speed of computing ALB statistics and their permutation distributions, especially for large data sets. An iterative approach to the screening method is available for SIS, and future research could involve investigating an ALB procedure that could capture differences in joint distributions. The simulated data in this paper all leveraged independent data, and how sensitive the method is to independence is also a property to explore. The interaction between ALB screening and random projection or sketching methods of dealing with settings where $n$ and/or $p$ are very large is also a promising direction for future research.

\bigskip\noindent
%\tcr{Do we really want to underline journal titles, or use italics? Also, what do the red numbers after references mean?}
\normalem
\bibliography{references}

\begin{thebibliography}{}

\bibitem[Boser et~al., 1992]{boser1992training}
Boser, B.~E., Guyon, I.~M., and Vapnik, V.~N. (1992).
\newblock A training algorithm for optimal margin classifiers.
\newblock In {\em Proceedings of the fifth annual workshop on Computational
  learning theory}, pages 144--152.

\bibitem[Brieman et~al., 1984]{brieman1984classification}
Brieman, L., Friedman, J.~H., Olshen, R.~A., and Stone, C.~J. (1984).
\newblock Classification and regression trees.
\newblock {\em Wadsworth Inc}, 67.

\bibitem[Cui et~al., 2015]{cui2015model}
Cui, H., Li, R., and Zhong, W. (2015).
\newblock Model-free feature screening for ultrahigh dimensional discriminant
  analysis.
\newblock {\em Journal of the American Statistical Association},
  110(510):630--641.

\bibitem[Fan and Lv, 2008]{fan2008sure}
Fan, J. and Lv, J. (2008).
\newblock Sure independence screening for ultrahigh dimensional feature space.
\newblock {\em Journal of the Royal Statistical Society: Series B (Statistical
  Methodology)}, 70(5):849--911.

\bibitem[Fan et~al., 2010]{fan2010sure}
Fan, J., Song, R., et~al. (2010).
\newblock Sure independence screening in generalized linear models with
  {NP}-dimensionality.
\newblock {\em The Annals of Statistics}, 38(6):3567--3604.

\bibitem[Friedman, 2002]{friedman2002stochastic}
Friedman, J.~H. (2002).
\newblock Stochastic gradient boosting.
\newblock {\em Computational statistics \& data analysis}, 38(4):367--378.

\bibitem[Hill et~al., 2020]{hill2020bayesian}
Hill, J., Linero, A., and Murray, J. (2020).
\newblock Bayesian {A}dditive {R}egression {T}rees: A {R}eview and {L}ook
  {F}orward.
\newblock {\em Annual Review of Statistics and Its Application}, 7.

\bibitem[Linero and Yang, 2018]{linero2018bayesian}
Linero, A.~R. and Yang, Y. (2018).
\newblock Bayesian regression tree ensembles that adapt to smoothness and
  sparsity.
\newblock {\em Journal of the Royal Statistical Society: Series B (Statistical
  Methodology)}, 80(5):1087--1110.

\bibitem[Mai and Zou, 2012]{mai2012kolmogorov}
Mai, Q. and Zou, H. (2012).
\newblock The {K}olmogorov filter for variable screening in high-dimensional
  binary classification.
\newblock {\em Biometrika}, 100(1):229--234.

\bibitem[Merchant and Hart, 2022]{leaveoneout}
Merchant, N. and Hart, J. (2022).
\newblock A {B}ayesian motivated two-sample test based on kernel density
  estimates.
\newblock {\em Entropy}, 24:1071.

\bibitem[Tang et~al., 2014]{tang2014feature}
Tang, J., Alelyani, S., and Liu, H. (2014).
\newblock Feature selection for classification: A review.
\newblock {\em Data classification: Algorithms and applications}, page~37.

\bibitem[Zhu et~al., 2011]{zhu2011model}
Zhu, L.-P., Li, L., Li, R., and Zhu, L.-X. (2011).
\newblock Model-free feature screening for ultrahigh-dimensional data.
\newblock {\em Journal of the American Statistical Association},
  106(496):1464--1475.

\end{thebibliography}
\bibliographystyle{apalike}
\end{document}